\documentclass[10pt,a4paper]{article}

\usepackage{geometry}
\usepackage[utf8]{inputenc}
\usepackage[T1]{fontenc}
\RequirePackage{amsmath,amssymb}
\RequirePackage{mathrsfs}
\RequirePackage{amsfonts,bm}
\usepackage{tikz,pgfplots}
\pgfplotsset{compat=newest}
\RequirePackage{dsfont}
\RequirePackage{braket}
\usepackage{array}
\RequirePackage{tabularx}
\usepackage{multirow}
\usepackage{longtable}
\usepackage{ltcaption}
\RequirePackage{nicefrac}
\RequirePackage{graphicx}
\RequirePackage{booktabs}
\RequirePackage{colortbl}
\RequirePackage{xcolor}
\RequirePackage[symbol]{footmisc}
\usepackage{mathtools}
\usepackage{comment}
\usepackage{blkarray}
\usepackage{stackengine}
\usepackage{float}
\usepackage{rotating}
\usepackage{hhline}

\usetikzlibrary{tikzmark,arrows.meta,backgrounds,calc,positioning}
\usepackage{physics}
\usepackage{multirow}
\usepackage{bbm}
\usepackage{bbold}
\usepackage{latexsym}
\usepackage[section]{placeins}
\usepackage{hyperref}
\usepackage{cleveref} 
\usepackage{arydshln}
\usepackage{adjustbox}

\usepackage[backend=biber,natbib=true,maxbibnames=99,giveninits=true,sorting=none,style=numeric-comp]{biblatex}
\usepackage{xurl}
\definecolor{colSG}{HTML}{1F4E8C}
\definecolor{colRS}{HTML}{157A5B}
\definecolor{colT}{HTML}{8A5A00}
\definecolor{colV}{HTML}{A83A20}
\definecolor{colH}{HTML}{6B3A8C}
\definecolor{colStripe}{HTML}{F0EFEA}
 
\tikzset{
  mult/.style={draw,rounded corners=1.5pt,line width=.5pt,inner sep=1pt,
               minimum width=7mm,minimum height=4.4mm,font=\scriptsize,anchor=center},
 SG/.style={mult,draw=colSG,fill=colSG!8,text=colSG},
RS/.style={mult,draw=colRS,fill=colRS!8,text=colRS},
Tm/.style={mult,draw=colT,fill=colT!8,text=colT},
Vm/.style={mult,draw=colV,fill=colV!8,text=colV},
Hm/.style={mult,draw=colH,fill=colH!8,text=colH},
  thy/.style={font=\small,inner sep=1pt},
  ax/.style={font=\scriptsize,text=black!55},
  red/.style ={-{Stealth[length=3.4pt,width=2.4pt]},line width=.45pt,black!62},
  tru/.style ={-{Stealth[length=3.4pt,width=2.4pt]},line width=.45pt,black!62,densely dashed},
  mx/.style={font=\tiny,inner sep=.6pt,fill=white,text=black!60},
}

\renewcommand{\thefootnote}{\fnsymbol{footnote}}

\def\beqn{\begin{eqnarray}}
\def\eeqn{\end{eqnarray}}

\usepackage{empheq}

\newcommand{\CC}[2]{C{\genfrac{[}{]}{0pt}{}{#1}{#2}}}

\newcommand{\ba}{\begin{eqnarray}}
\newcommand{\ea}{\end{eqnarray}}

\newcommand{\one}{\bm{1}} 
\newcommand{\Sv}{\bm{S}}
\newcommand{\bSv}{\bm{\bar{S}}} 

\newcommand{\bv}{\bm{b}}

\newcommand\x{\bm{x}}

\DeclareMathSymbol{\mg}{\mathrel}{symbols}{"1D}

\newcommand{\ga}{\alpha}
\newcommand{\gb}{\beta}

\newcommand{\gd}{\delta}

\newcommand{\gm}{\mu}

\newcommand{\gth}{\theta}

\newcommand{\gs}{\sigma}
\newcommand{\gt}{\tau}

\newcommand{\gp}{\pi}
\newcommand{\gps}{\psi}
\newcommand{\get}{\eta}
\newcommand{\gch}{\chi}
\newcommand{\gX}{\Xi}

\newcommand{\gTh}{\Theta}
\newcommand{\gO}{\Omega}

\newcommand{\gPs}{\Psi}

\newcommand{\cD}{{\cal D}}

\newcommand{\cN}{{\cal N}}

\newcommand{\cS}{{\cal S}}

\newcommand{\Id}{\mathbbm{1}}

\newcommand{\ra}{\rightarrow}

\newcommand{\der}{\partial}

\newcommand{\beq}{\begin{equation}}
\newcommand{\eeq}{\end{equation}}
\newcommand{\barr}{\begin{array}}
\newcommand{\earr}{\end{array}}
\newcommand{\equ}[1]{\begin{gather} #1 \end{gather}}

\newcommand{\items}[1]{\begin{itemize} #1 \end{itemize}}
\newcommand{\enums}[1]{\begin{enumerate} #1 \end{enumerate}}
\newcommand{\tabu}[2]{\begin{tabular}{#1} #2 \end{tabular}}
\newcommand{\arry}[2]{\begin{array}{#1} #2 \end{array}}

\newcommand{\non}{\nonumber}
\newcommand{\sfrac}[2]{\mbox{$\frac{#1}{#2}$}}
\newcounter{oldcounter}

\newcommand{\bder}{\bar\partial}
\newcommand{\bi}{{\bar \imath}}
\newcommand{\bj}{{\bar \jmath}}

\newcommand{\bmm}{{\bar m}}
\newcommand{\bn}{{\bar n}}

\newcommand{\bp}{{\bar p}}

\newcommand{\bs}{{\bar s}}

\newcommand{\bw}{{\bar w}}

\newcommand{\byy}{{\bar y}}
\newcommand{\bz}{{\bar z}}

\newcommand{\bM}{{\overline M}}
\newcommand{\bN}{{\overline N}}

\newcommand{\bga}{{\bar \alpha}}
\newcommand{\bgb}{{\bar\beta}}

\newcommand{\bgth}{{\bar\theta}}

\newcommand{\bgt}{{\bar\tau}}

\newcommand{\bgps}{{\bar\psi}}
\newcommand{\bget}{{\bar\eta}}
\newcommand{\bgch}{{\bar\chi}}
\newcommand{\bgTh}{{\overline\Theta}}

\newcommand{\bgPs}{{\overline\Psi}}

\newcommand{\Bga}{{\boldsymbol \alpha}}
\newcommand{\Bgb}{{\boldsymbol \beta}}
\newcommand{\Bgg}{{\boldsymbol \gamma}}
\newcommand{\Bgd}{{\boldsymbol \delta}}

\newcommand{\BgX}{{\boldsymbol \Xi}}

\newcommand{\Intr}{\mathbbm{Z}}

\newcommand{\Real}{\mathbbm{R}}
\newcommand{\Ratl}{\mathbbm{Q}}

\newcommand{\brkt}[2]{\bigl[ ^{#1}_{#2} \bigr]}

\newcommand{\sm}{{\,\mbox{-}}}

\numberwithin{equation}{section}

\begin{document}
\begin{titlepage}
\samepage{
\setcounter{page}{1}
\vspace{1cm} 

\begin{center}
  {\Large \bf{
  Classification of order--two T--duality orbifolds  \\\medskip
   at the $\boldsymbol{SO(12)}$ free fermionic point
  }}
\vspace{1cm}

{
Alon E. Faraggi$^{1}$\footnote{E-mail address: alon.faraggi@liverpool.ac.uk}, Stefan Groot Nibbelink $^{2,3}$\footnote{E-mail address: s.groot.nibbelink@hr.nl} and 
Benjamin Percival$^{4}$\footnote{E-mail address: b.percival@mmu.ac.uk}}
\vspace{1cm}

{\it $^{1}$ Dept.\ of Mathematical Sciences, University of Liverpool, Liverpool
L69 7ZL, UK\\}
\vspace{.08in}

{\it $^{2}$ Department of Electrical Engineering, School of Engineering and Applied Sciences, Rotterdam University of Applied Sciences,
G.J. de Jonghweg 4 -- 6, 3015 GG Rotterdam, the Netherlands\\}
\vspace{.08in}

{\it $^{3}$ HR Datalab EAS, School of Engineering and Applied Sciences, Rotterdam University of Applied Sciences,
G.J. de Jonghweg 4 -- 6, 3015 GG Rotterdam, the Netherlands\\}
\vspace{.08in}

{\it $^{4}$ Dept.\ of Natural Sciences, Manchester Metropolitan University, Manchester 
M1 5GD, UK\\}
\end{center}
\vspace{1cm}

\begin{abstract}
\noindent
Asymmetric orbifolds provide concrete examples of non--geometric constructions dubbed T--folds. 
All $\Intr_2$ point groups of six dimensional asymmetric order--two orbifolds are identified. 
The order--two T--fold configurations on the $SO(12)$ lattice are classified at the fermionic point of type II string theories for all $\Intr_2$ point groups.
The spectra of the models on these configurations are presented parametrically, in terms of certain generalised GSO phases. 
The minimal effective Hodge numbers were found to be $(h_{11},h_{12})=(1,1)$ for six T--fold configurations. 
Asymmetric generalisations of the mirror symmetry map are conjectured. 
The orientifoldable configurations using the basic worldsheet parity were identified within the classification. 
Twisted sectors corresponding to pure asymmetric twists may contain Rarita--Schwinger multiplets with spin--$3/2$ states. 
Throughout the paper, generalised GSO projections are chosen to preserve the maximal amount of supersymmetry possible. 
Relaxing this, the order--two point groups were identified for which non--supersymmetric T--folds can be constructed. 
Since some of them may enhance to $\cN=1$ or, even, $\cN=2$ supergravities, we argue that appearance of spin--$3/2$ states necessarily implies that the spectrum has reorder itself in a supersymmetric fashion and hence that the vacuum energy vanishes.

\end{abstract}

\smallskip}

\vfill 

\end{titlepage}

\setcounter{footnote}{0}
\renewcommand*{\thefootnote}{\arabic{footnote}}

\tableofcontents

\section{Introduction} 
\label{sc:Introduction}

One of the major obstacles in the quest for semi--realistic models from superstring compactifications is the so--called moduli problem: typical string constructions leave many scalar fields unfixed. This is often considered problematic as they determine the detailed phenomenology of such models. Non--geometric compactifications known as T--folds~\cite{Dabholkar:2005ve,Grana:2008yw} may be able to freeze many, or even all \cite{moduli, Tfoldwebs}, (geometric) moduli. A comprehensive review of various aspects of such non--geometric backgrounds can {\em e.g.}\ be found in~\cite{nongeomreview}. Constructions of such T--folds is often complicated and can only be described using effective supergravity means. 
A more recent framework to gain insights in such non--geometric backgrounds is double field theory~\cite{Dabholkar:2005ve,Hull:2006va,Hull:2007jy,Hull:2009sg}; for a review see {\em e.g.}~\cite{Aldazabal:2013sca}. 
The main objective of this work is to provide a larger class of explicit and simple examples of T--folds that admit exact string descriptions, which are, therefore, definitely part of the string landscape of consistent supergravity vacua. 

Asymmetric toroidal orbifolds~\cite{narain_87,asymmorbs,Ibanez:1988pj} provide examples of non--geometric superstring compactifications as exact conformal field theories. They act differently on the left-- and right--moving string coordinates by quotienting with generalised T--duality rotations~\cite{Duff:1989tf}. In particular, examples of elusive non-geometric fluxes in truly non-geometric constructions have a concrete description as certain asymmetric orbifolds~\cite{Condeescu:2012sp,Condeescu:2013yma}. Phenomenological aspects of asymmetric orbifolds have been mostly investigated in the heterotic string context in the past, see {\em e.g.}~\cite{slm2, Imamura:1991yf,Imamura:1992bz,Imamura:1992np,Kakushadze:1996hi}, and revised more recently~\cite{Bianchi:2012xz,Beye:2013moa,Beye:2013ola,Tan:2015nja,FMP1, Aldazabal:2025zht, DFP1}. A duality covariant approach to asymmetric orbifolds partially inspired by double field theory has been laid out in refs.~\cite{SGNPKV,GrootNibbelink:2020dib}

Asymmetric orbifolds, and T--folds in general, can be considered as backgrounds for type II or heterotic strings. The objects of central interest in this work are the order--two asymmetric orbifolds as prime examples of T--folds, not the possible phenomenological models that may be built using them. Considerations of possible gauge embeddings in the heterotic strings would only lead to unnecessary distractions of their main features, therefore 
we study T-folds of the type II superstring.  

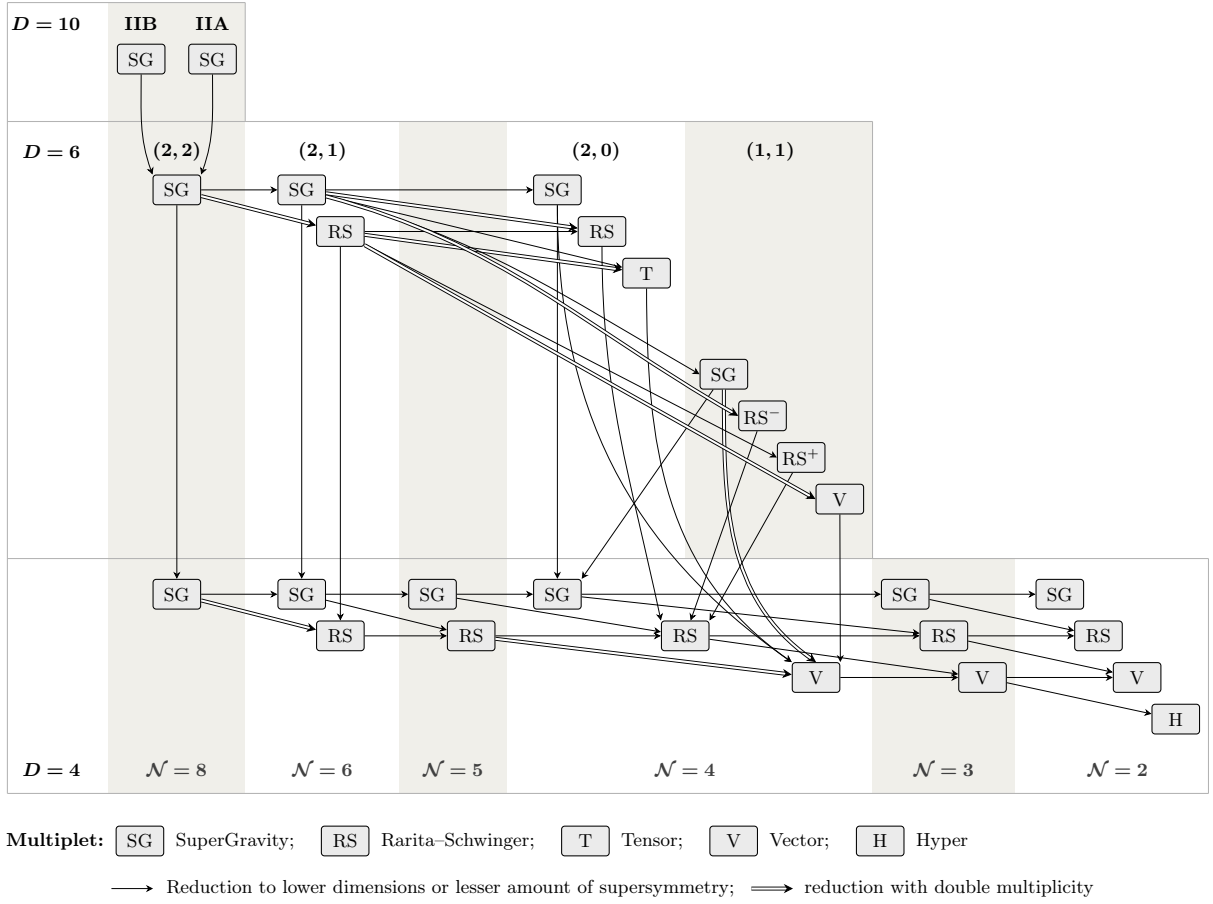
\begin{figure}[t]
\centering
\adjustbox{max width=\textwidth}{%
\begin{tikzpicture}[double distance=1pt, 
  mbox/.style={draw,rounded corners=1.5pt,line width=.5pt,inner sep=1pt,
               minimum width=8mm,minimum height=5mm,font=\small},
  mSG/.style={mbox,draw=black,fill=black!8,text=black},
  mRS/.style={mbox,draw=black,fill=black!8,text=black},
  mT/.style={mbox,draw=black,fill=black!8,text=black},
  mV/.style={mbox,draw=black,fill=black!8,text=black},
  mH/.style={mbox,draw=black,fill=black!8,text=black},
  thy/.style={font=\small,inner sep=1pt},
  axl/.style={font=\small,text=black},
  redu/.style={-{Stealth[length=3.4pt,width=3.4pt]},line width=.45pt,black},
  redu2/.style={-{Stealth[length=4.4pt,width=4.4pt]},double,line width=.45pt,black},
  trun/.style={-{Stealth[length=3.4pt,width=3.4pt]},line width=.45pt,black},
  trun2/.style={-{Stealth[length=4.4pt,width=4.4pt]},double,line width=.45pt,black},
  mx/.style={font=\scriptsize,inner sep=.6pt,fill=white,text=black},
]

\begin{scope}[on background layer]
  \foreach \a/\b in {-1.15/1.15}
    \fill[colStripe] (\a,0.95) rectangle (\b,-1.05);  
  \foreach \a/\b in {-1.15/1.15, 3.75/5.55, 8.55/11.7}
    \fill[colStripe] (\a,-1.05) rectangle (\b,-8.4);
  \foreach \a/\b in {-1.15/1.15, 3.75/5.55, 11.7/14.1}
    \fill[colStripe] (\a,-8.4) rectangle (\b,-12.35);
\end{scope}
\draw[black!30,line width=.5pt] (-2.85,0.95) -- (1.15,0.95);
\draw[black!30,line width=.5pt] (-2.85,-1.05) -- (11.7,-1.05);
\draw[black!30,line width=.5pt] (-2.85,-8.4) -- (17.35,-8.4);
\draw[black!30,line width=.5pt] (-2.85,-12.35) -- (17.35,-12.35);
\draw[black!30,line width=.5pt] (-2.85,0.95) -- (-2.85,-12.35);
\draw[black!30,line width=.5pt] (1.15,0.95) -- (1.15,-1.05);
\draw[black!30,line width=.5pt] (11.7,-1.05) -- (11.7,-8.4);
\draw[black!30,line width=.5pt] (17.35,-8.4) -- (17.35,-12.35);

\foreach \x/\n in {0/8, 2.45/6, 4.65/5, 8.55/4, 12.9/3, 15.825/2}
  \node[thy,text=black!75] at (\x,-11.95) {$\boldsymbol{\mathcal{N}=\n}$};

\node[font=\small\bfseries,anchor=east] at (-1.5,0.6) {$\boldsymbol{D=10}$};
\node[font=\small\bfseries,anchor=east] at (-1.5,-1.55) {$\boldsymbol{D=6}$};
\node[font=\small\bfseries,anchor=east] at (-1.5,-11.95) {$\boldsymbol{D=4}$};

\node[thy] at (-0.6,0.6) {\bf IIB};
\node[thy] at (0.6,0.6) {\bf IIA};
\node[mSG] (sB) at (-0.6,0.0) {SG};
\node[mSG] (sA) at (0.6,0.0) {SG};

\node[thy] at (0,-1.55) {$\boldsymbol{(2,2)}$};
\node[thy] at (2.45,-1.55) {$\boldsymbol{(2,1)}$};
\node[thy] at (7.05,-1.55) {$\boldsymbol{(2,0)}$};
\node[thy] at (9.98,-1.55) {$\boldsymbol{(1,1)}$};
\node[mSG] (a22SG) at (0,-2.2) {SG};
\node[mSG] (a21SG) at (2.10,-2.2) {SG};
\node[mRS] (a21RS) at (2.75,-2.9) {RS};
\node[mSG] (a20SG) at (6.40,-2.2) {SG};
\node[mRS] (a20RS) at (7.15,-2.9) {RS};
\node[mT]  (a20T)  at (7.90,-3.6) {T};
\node[mSG] (b11SG)  at (9.20,-5.3) {SG};
\node[mRS] (b11RSm) at (9.85,-6.0) {RS$^-$};
\node[mRS] (b11RSp) at (10.50,-6.7) {RS$^+$};
\node[mV]  (b11V)   at (11.15,-7.4) {V};

\node[mSG] (n8SG) at (0,-9.0) {SG};
\node[mSG] (n6SG) at (2.10,-9.0) {SG};
\node[mRS] (n6RS) at (2.75,-9.7) {RS};
\node[mSG] (n5SG) at (4.30,-9.0) {SG};
\node[mRS] (n5RS) at (4.95,-9.7) {RS};
\node[mSG] (n4SG) at (6.40,-9.0) {SG};
\node[mRS] (n4RS) at (8.55,-9.7) {RS};
\node[mV]  (n4V)  at (10.75,-10.4) {V};
\node[mSG] (n3SG) at (12.25,-9.0) {SG};
\node[mRS] (n3RS) at (12.90,-9.7) {RS};
\node[mV]  (n3V)  at (13.55,-10.4) {V};
\node[mSG] (n2SG) at (14.85,-9.0) {SG};
\node[mRS] (n2RS) at (15.50,-9.7) {RS};
\node[mV]  (n2V)  at (16.15,-10.4) {V};
\node[mH]  (n2H)  at (16.80,-11.1) {H};

\draw[redu] (sB) to[in=110,out=-90] (a22SG.north west);
\draw[redu] (sA) to[in=70,out=-90] (a22SG.north east);
\draw[redu] (a22SG) -- (n8SG);
\draw[redu] (a21SG) -- (n6SG);
\draw[redu] (a21RS) -- (n6RS);
\draw[redu] (a20SG) -- (n4SG);
\draw[redu] (a20SG) to[in=145,out=-90] (n4V.north west);
\draw[redu] (a20RS) to[in=102,out=-90] (n4RS.north west);
\draw[redu] (a20T)  to[in=145,out=-90] (n4V.north west);
\draw[redu] (b11SG)  -- (n4SG.north east);
\draw[redu2] (b11SG) to[in=145,out=-90] (n4V.north);
\draw[redu] (b11RSm) -- (n4RS);
\draw[redu] (b11RSp) -- (n4RS.north east);
\draw[redu] (b11V) -- (n4V.north east);

\draw[trun] (a22SG) -- (a21SG);
\draw[trun2] (a22SG) -- (a21RS);
\draw[trun] (a21SG) -- (a20SG);
\draw[trun2] (a21SG) -- (a20RS);
\draw[trun] (a21SG) -- (a20T);
\draw[trun] (a21RS) -- (a20RS);
\draw[trun2] (a21RS) -- (a20T);
\draw[trun] (a21SG) to[in=150,out=-15] (b11SG.west);
\draw[trun2] (a21SG) to[in=150,out=-15] (b11RSm.west);
\draw[trun] (a21RS) -- (b11RSp.west);
\draw[trun2] (a21RS) -- (b11V.west);
\draw[trun] (n8SG) -- (n6SG);
\draw[trun2] (n8SG) -- (n6RS);
\draw[trun] (n6SG) -- (n5SG);
\draw[trun] (n6SG) -- (n5RS);
\draw[trun] (n6RS) -- (n5RS);
\draw[trun] (n5SG) -- (n4SG);
\draw[trun] (n5SG) -- (n4RS);
\draw[trun] (n5RS) -- (n4RS);
\draw[trun2] (n5RS) -- (n4V);
\draw[trun] (n4SG) -- (n3SG);
\draw[trun] (n4SG) -- (n3RS);
\draw[trun] (n4RS) -- (n3RS);
\draw[trun] (n4RS) -- (n3V);
\draw[trun] (n4V) -- (n3V);
\draw[trun] (n3SG) -- (n2SG);
\draw[trun] (n3SG) -- (n2RS);
\draw[trun] (n3RS) -- (n2RS);
\draw[trun] (n3RS) -- (n2V);
\draw[trun] (n3V) -- (n2V);
\draw[trun] (n3V) -- (n2H);

\begin{scope}[shift={(0.4,-13.15)}]
  \node[font=\small\bfseries,anchor=east] (lLegend) at (-1.5,0.0) {\bf Multiplet:};
  \node[mSG,right=2pt of lLegend] (lSG) {SG};
  \node[axl,right=2pt of lSG] (lSGt) {SuperGravity;};
  \node[mRS,right=9pt of lSGt] (lRS) {RS};
  \node[axl,right=2pt of lRS] (lRSt) {Rarita--Schwinger;};
  \node[mT,right=9pt of lRSt] (lT) {T};
  \node[axl,right=2pt of lT] (lTt) {Tensor;};
  \node[mV,right=9pt of lTt] (lV) {V};
  \node[axl,right=2pt of lV] (lVt) {Vector;};
  \node[mH,right=9pt of lVt] (lH) {H};
  \node[axl,right=2pt of lH] (lHt) {Hyper};

  \draw[redu] (-1.5,-0.8) -- (-0.8,-0.8);
  \node[axl,anchor=west] (e1) at (-0.7,-0.8)
        {Reduction to lower dimensions or lesser amount of supersymmetry;};
  \draw[redu2] ([xshift=1ex]e1.east) -- ++(0.7,0) coordinate (e2s);
  \node[axl,right=2pt of e2s] (e2) {reduction with double multiplicity};

\end{scope}

\end{tikzpicture}}
\caption{\label{fig:MultipletReductions}
The branching relations between the multiplets of various extended supersymmetries in $D=10$, $6$ and $4$ space--time dimensions relevant for this project are visualised. 
Vertically these dimensions are indicated, while horizontally the type of extend supersymmetry in that dimension. 
The multiplets, using the nomenclature introduced in tables~\ref{tab:4DMultiplets} and~\ref{tab:6DMultiplets} of \cref{app:ExtendedSUSY}, are nested according to their highest spin states. 
An arrow indicates the reduction of the originating multiplet to lower dimension or to lesser amount of supersymmetry; a double lined arrow denotes branching into two copies of the target multiplet.
}
\end{figure}

The type II superstrings come in two variants, the type IIA and IIB theories that are described as non--chiral and chiral type II supergravities in ten dimensions. Compactification on these order two (a)symmetric orbifolds lead to four dimensional effective field theories with extended supersymmetries with $\cN=2,3,4,5,6$ and $8$ supercharges \cite{Ferrara:1989nm}. 
Consequently, for a detailed understanding of some of the analysis in the present work some basic knowledge of theories with extended supersymmetries is mandatory. \Cref{app:ExtendedSUSY} furnishes an overview of the relevant supersymmetry multiplets in four and six dimensions: the SuperGravity (SG), Rarita--Schwinger (RS), Vector (V) and Hyper (H) multiplets, characterised by their highest spin components of spin--2, 3/2, 1 and 1/2, respectively. We will refer to a spin--3/2 state as a gravitino only when it is identified to be part of a supergravity multiplet; and as a Rarita--Schwinger field otherwise. 
Based on the properties of these supersymmetric multiplets, \cref{fig:MultipletReductions} visualises relations between these multiplets in ten, six and four dimensions and reductions of the amount of supersymmetry. For example, from this figure one may infer that the six dimensional (2,1) RS multiplet branches to RS+4\,V+3\,H of $\mathcal{N}=2$ supersymmetry in four dimensions. This work makes heavy use of the properties of the various supersymmetry multiplets and their relations as depicted in this figure. 

The main purpose of this work is to obtain a classification of a large collection of T--folds, which can be described as asymmetric orbifolds using the free fermionic description. Classifications of symmetric $\Intr_2^2$ orbifolds have been performed using the free fermionic formulation~\cite{gkr, fknr, Donagi2004, DW}. A detailed identification between the bosonic and fermionic descriptions of six dimensional symmetric $\Intr_2^2$ orbifolds is given in ref.~\cite{Athanasopoulos:2016aws}. A full general classification of all six dimensional (possibly non--Abelian) symmetric orbifolds can be found in ref.~\cite{Opgenorth:1998} using the CARAT--package~\cite{CARAT}. The subset of these orbifolds that preserve some amount of supersymmetry was identified in~\cite{FRTV}. This work should therefore be considered as a moderate step towards a full classification of all asymmetric orbifolds. 

It should be realised that classification of asymmetric orbifolds is more complicated than the classification of symmetric constructions. First of all, since a direct geometrical picture is missing, the construction and its interpretation is more subtle. This holds in particular for counting of twisted states and their properties under residual orbifolds symmetries. Secondly, for symmetric orbifolds one can in principle 
only focus on the geometrical aspects, while the non--geometric nature of generalised geometry immediately requires that aspects of torsion due to the B--field are taken into account. Moreover, since T--folds in general have many fewer geometrical moduli, there may be many distinct discrete configurations which, in an analogous symmetric setting, would be continuously connected by varying some combination of moduli. It is precisely for these reasons that we work with 
the free fermionic formulation, in an attempt to classify certain classes of asymmetric order--two orbifolds. 

The free fermionic formulation \cite{ABK1, ABK2, KLT} is a convenient framework to describe asymmetric orbifolds concisely, in which fermion boundary conditions are the primary input and just as naturally give rise to non--geometric spacetime compactifications as geometric ones. When working primarily with real worldsheet fermions, a natural restriction is to allow for only periodic or anti--periodic boundary conditions for the fermions as they propagate around the non-contractible cycles of worldsheet amplitudes. This restriction corresponds to allowing for order--two twists, shifts or roto--translations within the complementary bosonised description as order two asymmetric orbifolds. We will distinguish between symmetric $\Intr_2$ actions, in which the twist action is identical on the left-- and right--moving degrees of freedom, and asymmetric, $\Intr_{2L}$ or $\Intr_{2R}$, twists acting exclusively on the left-- or right--movers, respectively. The considered asymmetric toroidal orbifolds are six dimensional in the sense that the resulting effective field theories live in four dimensions. The lattice of the underlying six dimensional torus $T^6$ is taken to be the root lattice of $SO(12)$; other lattices would complicate the description as they require to mod out further (asymmetric) shifts. 

It should be emphasised that all our results are at the so--called free fermionic point, {\em i.e.}\ at very specific values of the untwisted moduli. For symmetric orbifolds this is often not very important because the moduli, like radii, are free parameters, hence results obtained at the free fermionic point can be extended to arbitrary points in the moduli space. For T--folds this is generally not the case, since many, if not all, untwisted moduli are frozen. In particular, our classification is obtained on the $SO(12)$ lattice at the free fermionic point. The possibility cannot be ignored that at other (discrete) points in the moduli space different classification results (with more or fewer configurations) of order--two T--folds may be obtained.

\subsection*{A reader's guide}

This paper can be read in different ways depending on the interest of the reader. 

The paper can of course be read in the order that it is organised: 
\cref{sc:AsymOrbifolds} gives a short introduction to asymmetric orbifolds as examples of T--folds in the bosonic language. 
This section mainly serves to provide some interpretation of these non--geometric objects in the formalism used primarily in this work, namely the free fermionic description.
\Cref{sc:FreeFermionicStrings} presents the essentials of this formalism for our classifications presented in \cref{sc:ClassificationsOrder2Tfolds}. 
The first classification provides all the order--two point groups of six--dimensional T--folds. 
The second classification gives all inequivalent T--fold configurations on the $SO(12)$ torus lattice at the free fermionic point. 
\Cref{sc:TwistedSectors} gives details of the computation of the projected twisted spectra, emphasising in particular that Rarita--Schwinger multiplets may arise in asymmetric constructions. 
\Cref{sc:ApplicationsExtensions} gives a number of applications and extensions of our classification results. 
After the conclusions, \cref{sc:Conclusions}, the paper is closed with two appendices giving some technical details on extended supersymmetries in various dimensions (\cref{app:ExtendedSUSY}) and partition functions and generalised GSO phases (\cref{app:PartitionFunctions}). 

However, readers primarily interested in our classification results can immediately jump to the order--two point group classification in \cref{tab:NarainPointGroupClass} or the full classification of all inequivalent order--two T--fold configurations on the $SO(12)$ lattice in \cref{tab:NarainRotoTransClass}. The latter provides the massless type II spectra on them in a parametric form. 

Other readers may be mainly interested in applications of our results like the search for configurations with low effective Hodge numbers (\cref{sc:LowHodgeNumbers}), possible generalisations of mirror symmetry maps on T--folds (\cref{sc:MirrorSymmetries}), orientifolds of T--folds (\cref{sc:Orientifolds}), enhanced supergravities due to twisted Rarita--Schwinger multiplets (\cref{sc:EnhancedSUGRAs}) or the appearance of Rarita--Schwinger fields in models with reduced supersymmetry (\cref{sc:ReducedSusyModels}). 

Finally, readers who wish to reproduce or extend our results may find the accompanying codes useful. All the codes used in this work are collected in ref. \cite{percival2026z2n_typeii_code} and consist of two parts. The first classifies the inequivalent T–fold configurations of each order–two point group: it enumerates the modular invariant choices of the basis vectors and reduces them modulo the equivalence relations of section 4.1 by an exhaustive search, so that the classification of table \ref{tab:NarainRotoTransClass} is obtained. The second takes the basis and the generalised GSO phases of each model and determines its massless spectrum in terms of supermultiplets, which we used to crosscheck the spectra quoted in table \ref{tab:NarainRotoTransClass}.

\section{Asymmetric orbifolds as six--dimensional T--folds}
\label{sc:AsymOrbifolds}

Asymmetric orbifolds can be thought of as discrete examples of T--folds. 
On shell, the bosonic coordinate fields $X=(X_L,X_R)$ of such backgrounds can be split into left--movers $X_L=(X_L^1,\ldots , X_L^6)$ and right--movers $X_R=(X_R^1,\ldots ,X_R^6)$ in six dimensions. 
On these doubled bosonic coordinate fields a Narain space group element $g=(\gTh;L)$ acts as 
\equ{
g\circ Y = (\gTh; L)\circ Y = \gTh\, Y + 2\gp\, (L+V_\gTh)~.
}
Here $\gTh$ denotes the twist elements which may be accompanied by a fractional lattice translation $V_\gTh$ to form a so--called roto--translation. $L$ denotes a lattice translation. The action can be split into the action on left-- and right--movers $\gTh\, Y = (\gTh_L X_L, \gTh_R X_R)$, $V_\gTh=(V_{\gTh L},V_{\gTh R})$ and $L=(L_L,L_R)$. The generators of the space group define an asymmetric orbifold as they specify all the boundary conditions and identifications. The restriction of the space group to the non--twist elements $(\Id; L)$ defines the lattice of the underlying six--dimensional doubled torus. 
The projection 
\equ{
(\gTh; L) \mapsto \gTh
}
of the space group onto the twist part is called the Narain point group. An orbifold is called symmetric if its point group is generated by symmetric twist elements, {\em i.e.}\ $\gTh_L=\gTh_R$ for all $\gTh$.

The doubled coordinates $Y$ of the six--dimensional square torus are subject to the periodic identifications
\equ{
Y \sim Y + 2\gp\, e_i~,
\qquad 
Y \sim Y + 2\gp\, e_\bi~,
}
where the unit vectors in the $i$--th left--moving direction are denoted by $e_i$ and the $i$--th right--moving vector by $e_\bi$. In terms of these unit vectors define 
\equ{
    {e}_{i_1\ldots i_p \bi_1\ldots \bi_{\bp}}= {e}_{i_1}+\ldots {e}_{i_p} + {e}_{\bi_1}+\ldots {e}_{\bi_\bp}~, 
    \qquad 
    {E} = {e}_{1\ldots 6 \bar{1}\ldots\bar{6}}~. 
}

The point groups of the Narain orbifolds considered in this work are all of order two, {\em i.e.}\ isomorphic to $\Intr_2^N$ on the $SO(12)$ lattice. The products $\gth_{1\bar{1}}, \gth_{2\bar{2}}$ define symmetric twists. The $SO(12)$ torus lattice arises by the shift identification 
\equ{
    Y \sim Y + 2\pi\, \sfrac 12 {E}~. 
}
The generating $\Intr_2$ twists act as reflections of some of the double coordinates. Supersymmetry preserving $\Intr_2$ actions reflect in four left-- or right--moving directions and overlap in two of them. To fix notation define the following six twists (where only the transforming coordinates are indicated): 
\equ{
\arry{c}{
\gth_1~: X_L^{3,4,5,6} \ra -X_L^{3,4,5,6}~,
\qquad 
\gth_{\bar{1}}~: X_R^{\bar{3},\bar{4},\bar{5},\bar{6}} \ra -X_R^{\bar{3},\bar{4},\bar{5},\bar{6}}~,
\\[2ex]
\gth_2~: X_L^{1,2,5,6} \ra -X_L^{1,2,5,6}~,
\qquad 
\gth_{\bar{2}}~: X_R^{\bar{1},\bar{2},\bar{5},\bar{6}} \ra -X_R^{\bar{1},\bar{2},\bar{5},\bar{6}}~,
\\[2ex] 
\gth_3~: X_L^{1,2,3,4} \ra -X_L^{1,2,3,4}~,
\qquad 
\gth_{\bar{3}}~: X_R^{\bar{1},\bar{2},\bar{3},\bar{4}} \ra -X_R^{\bar{1},\bar{2},\bar{3},\bar{4}}~.
}
}
Products of these twist generators are denoted as follows $\gth_{ij} = \gth_i\gth_j$, $\gth_{i\bj}=\gth_i\gth_\bj$, etc. Note that $\gth_3=\gth_{12} =\gth_1\gth_2$ and $\gth_{\bar{3}}=\gth_{\bar{1}\bar{2}}=\gth_{\bar{1}}\gth_{\bar{2}}$ can be considered to be shorthands, and that $\gth_{123} = \Id$. These twists may be accompanied with shifts to form roto--translations. For example, the space group element $g_1=(\gth_1, \sfrac 12 e_{1\bar{1}})$: 
\equ{ \label{eq:g1example}
    X_L^1 \ra X_L^1 + \gp~, 
    \qquad 
    X_R^1 \ra X_R^1+ \gp~, 
    \qquad
    X_L^{3,4,5,6} \ra -X_L^{3,4,5,6}~
}
acts as a translation in the 1--direction for both the left-- and right--moving directions and as a reflection on the left--moving directions $3,4,5,6$.

\section{Free fermionic type II superstrings}
\label{sc:FreeFermionicStrings}

This section provides the necessary basis for our investigations of asymmetric orbifolds in the free fermionic formulation of closed strings and fix the notation used throughout this work. The central interest of this paper are the properties of these asymmetric string backgrounds, though some can be exhibited more clearly in type II setting while others in the heterotic context. This section develops a description that can be directly applied to type II strings. With minor modifications it also applies to standard embedding--like heterotic constructions. Details of the associated one--loop partition functions may be found in \cref{app:PartitionFunctions}. 

The construction of any free fermionic model starts with a collection of real holomorphic and anti--holomorphic fermions, $f$ and $\bar{f}$ subject to boundary conditions as
\equ{\label{eq:feEarmichanges}
  f \mapsto -e^{i\pi \ga(f)}\, f~, 
  \qquad 
  \overline{f} \mapsto -e^{-i\pi\, \bga(\overline{f})}\, \overline{f}~, 
}
in terms of certain vectors $\Bga = (\Bga_L\,|\,\Bga_R)$, $\Bga_L = (\ga(f_1), \ga(f_2), \ldots)$ and $\Bga_R = (\bga(\overline{f}_1), \bga(\overline{f}_2), \ldots)$. The Euclidean and Minkowskian inner products $\cdot_{\pm}$ of two boundary condition vectors are defined as  
\equ{ \label{eq:InnerProducts}
\Bga \cdot_{\pm} \Bgb =
\sfrac{1}{2} \Bga_L^T \Bgb_L \pm \sfrac 12 \Bga_R^T\Bgb_R~, 
}
with halves in the description with real fermions. If the signature of the inner product is not specified, then the Minkowskian inner product is implied: $\Bga\cdot\Bgb = \Bga\cdot_{-}\Bgb$. The complex overlap of two vectors is defined as
\equ{ \label{eq:Overlap}
\Bga \cap \Bgb = [\Bga]\cdot_+ [\Bgb]~,
} 
where $[\Bga]$ denotes the reduced vector associated to $\Bga$ in which all entries of $\Bga$ are restricted to $0,1$ by adding appropriate even multiples. Because of the factor $1/2$ in the inner products \eqref{eq:InnerProducts}, it counts the number of complex fermions the two vectors overlap on.  The notion of overlap can be generalised to three or more vectors as the sum of the products of the components of the reduced vectors involved. 

These vectors  $\Bga = (\Bga_L\,|\,\Bga_R) =\sum_a n_a \Bgb_a$ are elements of an additive set $\BgX$ generated from a set of basis vectors $\Bgb_a \in \mathscr{B}$. For consistency, any fermionic model needs to contain the vector $\bm{1}$ as part of the additive set. Often $\bm{1}$ is taken to be one of the basis vectors, however, for later convenience in understanding the bosonic interpretation, we do not make this assumption. The basis vectors are subject to various modular invariance conditions 
\equ{ \label{eq:12loopModInv}
\Bgb_a^2 = 0\ \text{mod}\ 4~, 
\qquad 
\Bgb_a\cdot \Bgb_b = 0\ \text{mod}\ 2~,
\qquad
\Bgb_a \cap \Bgb_b \cap \Bgb_c \cap \Bgb_d = 0 \ \text{mod } 1~,
}
for all $a\neq b \neq c \neq d =0,..., |\mathscr{B}|$. The third condition ensures the absence of single real fermions.

The description of a given fermionic model is completed by the specification of a number of generalised GSO (GGSO) phases $\CC{\Bgb_a}{\Bgb_b}$ for the basis vectors in the one--loop partition function. Some further technical details have been collected in \cref{app:PartitionFunctions}. To each vector $\Bga$ of the additive set, there is a sector associated in the string Hilbert space. 
There are an infinite number of states associated to a given sector $\Bga$. 
The terms in the expansion of the partition function \eqref{eq:FullPartitionFun} (for complex fermions) labelled by $\bm{n}$ are in one--to--one correspondence to these string states $| \bm{n} \rangle_\Bga$ in sector $\Bga$. 
Consequently, the true vacuum state is denoted by $|\bm{0}\rangle_\Bga$. 
For the entries in $\Bga$ equal to $1$, the vacuum state is degenerate in the sense that if the corresponding entry in $\bm{n}$ equals $-1$, then both states have the same left-- and right--moving masses. 
$|\Bga\rangle$ denotes the collection of all states degenerate with the true vacuum state $|\bm{0}\rangle_\Bga$ and is, with slight abuse of notation, referred to as the vacuum state as well. 
The vacuum state $|\Bga\rangle$ defines a spinor representation of dimension $2^n$, assuming that $\Bga$ contains $2n$ real fermions, and 
is often denoted as $|\pm \sfrac 12^n\rangle$. 

Any other state $|\text{state}\rangle_\Bga$ in the sector $\Bga$ is obtained by acting with bosonic and fermionic oscillators on the vacuum state. 
The left-- and right--moving masses are determined by 
\equ{ \label{eq:MassesStates}
M_L^2 = -c_L + \sfrac18\, \Bga_L^2 + N_L~, 
\qquad 
M_R^2 = -c_R + \sfrac18\, \Bga_R^2 + N_R~, 
}
where the zero-point energies $c_L=c_R=1/2$ (in the type II theories) and $N_L$ and $N_R$ count the number of left-- and right--moving oscillations, respectively. 
Bosonic excitations, $\der X$ and $\bder X$, count as $1$ and the real fermionic excitations as $\sfrac 12$. 
States in the physical on--shell spectrum should be level matched with equal left-- and right--moving masses. 

Most excited states are massive, some massless and in principle a small number can be tachyonic. 
In particular, the vacuum state $|0\rangle_\Bga$ is typically tachyonic.
Conventionally, generalised GSO projections 
\equ{ \label{eq:GGSOprojections}
e^{i\gp \Bgb_a\cdot F }\, \big| \text{state} \big\rangle_\Bga
=
\gd_\Bga\, \CC{\Bga}{\Bgb_a}^*\, 
\big| \text{state}  \, \big\rangle_\Bga~, 
}
for all $\Bgb_a\in \mathscr{B}$ with the spin statistics operator $\delta_\Bga = (-1)^{\alpha(\psi^\mu)+\alpha(\bgps^\mu)}$ (the fermions $\gps$ and $\bgps$ are characterised in the subsection below), are chosen such that all level--matched tachyonic states are projected out. 
Here $F$ is the fermion number operator.

\subsection{Ten dimensional type II fermionic superstrings}

In ten dimensions the degrees of freedom defining the superconformal algebras on the left (holomorphic) and right (anti--holomorphic) are the real spacetime bosons $X^M$ and fermions $\gps^M, \bgps^M$
\begin{align}\label{10DDoFs}
    \text{holomorphic:}~~ \der X^M, \psi^M~, 
    \qquad 
    \text{anti--holomorphic:}~~ \bar{\der}\bar{X}^M, \bgps^M~, 
\end{align}
where, in the lightcone gauge, $M=2,\ldots,9$, and $\der, \bder$ are holomorphic and anti--holomorphic derivatives of the complex worldsheet coordinates $z,\bz$. The (1,1) worldsheet supersymmetries are generated by the currents 
\begin{align}
\label{eq:SupercurrentTypeII}
T_F = i\,\psi^M \der X_M ~, 
\qquad 
\bar{T}_F = i\,\bgps^M \bder X_M~. 
\end{align}
All free fermionic models need to contain a sector in which all fermions are periodic (Ramond) around one of the two cycles of the torus when we consider the one--loop amplitude. This corresponds to the unity boundary condition vector 
\beq 
\bm{1}=\big\{\psi^M \,\big|\, \bgps^M\big\}~.
\eeq 
This vector is associated to a sector of the theory with a Ramond vacuum for both the spacetime fermions $\ket{\bm{1}}$, \textit{i.e.}\ the R--R sector. The NS--NS sector is then associated with $\bm{1}+\bm{1}$, in which the fermions are anti--periodic. Target space spinors result from sectors generated by the basis vectors
\begin{align}
    \Sv=\big\{\psi^M\big\}~, 
    \qquad 
    \bar{\Sv} =\big\{\bgps^M\big\}~.
\end{align}
In ten dimensions the sum of these basis vectors results in the unity vector: $\bar{\Sv}+\Sv =\bm{1}$.

The $\Sv$ and $\bSv$ are more generally responsible for defining the structure of supermultiplets: on both the holomorphic and anti--holomorphic sides of the theory target space supersymmetries in ten dimensions are induced by the replacements 
\equ{ \label{eq:SusyMappings}
\gps^M \big| \Bgb \big\rangle \leftrightarrow \big| \Bgb + \bm{S}\big\rangle 
\qquad\text{or}\qquad 
\bgps^M \big| \Bgb \big\rangle \leftrightarrow \big| \Bgb + \bar{\bm{S}}\big\rangle~, 
}
provided that $\Bgb$ does not have any overlap with $\bm{S}$ or $\bar{\bm{S}}$, respectively. In particular, the gravity multiplet is then formed from the supersector built from the NS vacuum state $|\bm{0}\rangle$, according to
\equ{ \label{eq:10DsugraStates}
||\bm{0}\rangle \rangle = \{ \gps^M \bgps^N \big| \bm{0}\big\rangle~, 
\qquad 
\bgps^M \big| \Sv \big\rangle~,
\qquad
\gps^M \big| \bar{\Sv} \big\rangle~, 
\qquad
\big| \Sv +\bar{\Sv} \big\rangle~ \}~. 
}
The spin statistics $\CC{\bm{S}}{\bm{0}} = \CC{\bar{\bm{S}}}{\bm{0}}=-1$ necessitates the presence of the gravitons and the gravitini arises as the massless states associated with the $\Sv$ and $\bar{\Sv}$ sectors. 
As the states associated with $\Sv$ and $\bSv$ define the gravitini, these basis vectors may be thought of as the generators of local supersymmetry in target space. 

The standard choice of basis in the fermionic literature for these 10D type II superstrings would be $\mathscr{B}=\{\bm{1}, \Sv\}$, 
however
for the choice of basis $\mathscr{B}=\{\Sv,\bSv\}$, the conversion of the GGSO phases can be determined via \eqref{eq:Conversion}. 
We note that the mixed GGSO phases have to be taken to be 
\equ{ 
\CC{\Sv}{\bar{\Sv}}= \CC{\bar{\Sv}}{\Sv} = +1~, 
}
otherwise 
the gravitini ($\bgps^M | \Sv \rangle$, $\gps^M | \bar{\Sv} \rangle$) are projected out by the GGSO projection \eqref{eq:GGSOprojections} and the model would not be supersymmetric. 

The type IIB theory is chiral in target space due to identical GGSO projections on either side of the theory encoded in the GGSO phases 
\equ{ \label{eq:GSOtypeIIB}
\phantom{-}\CC{\bar{\Sv}}{\bar{\Sv}} = \CC{\Sv}{\Sv} = -1~. 
}
The resulting projections ensure that both gravitini arise from a spinorial vacua $|\pm \sfrac 12^4\rangle$ with an even number of minus signs and thus of the same positive chirality.  
The type IIA theory is non--chiral due to opposite GSO projections 
\equ{ \label{eq:GSOtypeIIA}
-\CC{\bar{\Sv}}{\bar{\Sv}} = \CC{\Sv}{\Sv} =  -1~. 
}
In this case, the gravitino arises from the holomorphic spinorial vacuum $\big|\pm \sfrac 12^4\big\rangle$ with an even number of minus signs, while the gravitino from the anti--holomorphic spinorial vacuum has an odd number of minus signs and hence the opposite chirality.

\subsection{Free fermionic four dimensional type II models}
\label{sc:FFin4DtypeII}

Having described the type II superstring in ten dimensions, we turn to four dimensional models. Along with the spacetime bosons and fermions of eq. (\ref{10DDoFs}) but with $\mu=1,2$ (for the light--cone gauge in four dimensions), we introduce the following Majorana--Weyl fermions associated with the internal space
\begin{align}\label{4DDoFs}
    \text{holomorphic:}~ \chi^i, \ y^i, \ w^i~, 
    \qquad
    \text{anti--holomorphic:}~ \bgch^i, \ \byy^i, \ \bw^i~, 
\end{align}
for $i=1,...,6$. From the worldsheet CFT perspective, these 18 internal free fermions contribute $c=9$ on the left and right to accompany the $c=6$ generated by the spacetime fields and ensuring the overall cancellation of the conformal anomaly. 

Worldsheet supersymmetry is realised non--linearly by the holomorphic and anti--holomorphic supercurrent
\begin{align}
\label{eq:Supercurrent}
T_F(z) = i\,\psi^\mu \partial_z X_\mu +i\,\chi^i y^i w^i~, 
\qquad 
\bar{T}_F(\bar{z}) = i\,\bgps^\mu \partial_{\bar{z}} X_\mu +i\, \bgch^i \byy^i \bw^i.
\end{align}
In four dimensions, a generic sector $\Bga$ will be of the form 
\beq 
\Bga = \{\Bga_L \ | \ \Bga_R \} = 
\{\ga(\psi^\mu),\ga(\chi^{1}),\ldots,\ga(\chi^{6}),\ga(y^1),\ldots,\ga(y^6),\ga(w^1),\ldots,\ga(w^6) \ | \ (f\rightarrow 
\bar{f})\}.
\eeq  
All models explored in this paper have $\ga(f)=0,1$, corresponding to NS and R (real) boundary conditions, respectively. Preservation of the supercurrent implies that the following components of the additive group vectors satisfy
\beq \label{PreserveSUSY}
\ga(\chi^i) + \ga(y^i) + \ga(w^i) = \ga(\gps^\gm) \text{ mod } 2~, 
\eeq
and similarly on the anti--holomorphic side. 

Possibly by adding $\Sv$ or $\bar{\Sv}$, any further basis vector $\Bga$ can be assumed to not contain any $\gps^\mu$ or any $\bgps^\mu$. Assuming that such a basis is chosen, one can distinguish between twist and lattice basis vectors. Twist basis vectors involve some $\gch$'s or $\bgch$'s, while non--twist or lattice basis vectors do not. 

\subsubsection*{Lattice basis vectors}

Lattice basis vectors only contain $y^iw^i$ or $\byy^i\bw^i$ pairs and no $\gps, \bgps, \gch, \bgch$ fermions. In particular, 
\equ{
\bm{E} = \bm{1}-\Sv-\bar{\Sv} = \{y^{1,...,6},w^{1,...,6} \ | \ \byy^{1,...,6},\bw^{1,...,6} \},
}
defines a lattice basis vector. Since in supersymmetric type II models the vectors $\bm{1},\Sv,\bar{\Sv}$ are necessarily present in the additive set, any four dimensional model contains this lattice basis vector $\bm{E}$. For later convenience, we define the lattice vectors 
\equ{
\bm{e}_i = \{y^i, w^i\}~,
\qquad 
\bm{e}_\bi = \{\byy^i, \bw^i\}~,
\qquad
\bm{e}_{i_1\ldots i_p \bi_1\ldots \bi_{\bp}}= \bm{e}_{i_1}+\ldots \bm{e}_{i_p} + \bm{e}_{\bi_1}+\ldots \bm{e}_{\bi_\bp}~,
}
so that 
\(
\bm{E} = \bm{e}_{1\ldots 6} + \bm{e}_{\bar{1}\ldots\bar{6}}\,.
\)

\subsubsection*{Twist basis vectors}

For the consideration of order--two asymmetric twists, we note that there are two consistent independent twists on each side. As such, we define the pure twist basis vectors as
\equ{ \label{eq:PureTwistBasisVectors}
\arry{cc}{
\bm{b}_1=\{\gch^{3456},y^{3456}\}~,
&\qquad 
\bm{b}_2=\{\gch^{1256},y^{1256}\}~,
\\[2ex] 
 \bm{b}_{\bar{1}}=\{\bgch^{3456},\byy^{3456}\}~, 
 &\qquad 
\bm{b}_{\bar{2}}=\{\bgch^{1256},\byy^{1256}\}~, 
} 
}
which act to preserve half of the target space supersymmetries on their respective sides. They are compatible with the worldsheet supercurrents and combined with $\bm{E},\Sv,\bar{\Sv}$ satisfy the modular invariance requirements given above. The elements are chosen to only involve $y$'s or $\byy$'s but no $w$'s or $\bw$'s. For later convenience, we introduce the following notation for the combination of pure twist basis vectors 
\equ{
\arry{c}{
\bm{b}_3 = [\bm{b}_1+\bm{b}_2] = \{\gch^{1234},y^{1234}\}~,
\\[2ex] 
\bm{b}_{\bar{3}} = [\bm{b}_{\bar{1}}+\bm{b}_{\bar{2}}] = \{\bgch^{1234},\byy^{1234}\}~, 
} 
\qquad
\bm{b}_{\ga\bgb} = \bm{b}_\ga + \bm{b}_\bgb~, 
}
with $\ga, \bgb=1,2,3$. (Here, $[\Bgb]$ means that all entries of the vectors $\Bgb$ are considered mod 2.) In particular, the symmetric twist basis can be defined as 
\equ{ \label{eq:PureTwistBasisVectorsSym}
\arry{c}{
\bm{b}_{1\bar{1}} = \bm{b}_1 + \bm{b}_{\bar{1}} = \{\gch^{3456},y^{3456}\ | \ \bgch^{3456},\byy^{3456}\}~, 
\\[2ex]
\bm{b}_{2\bar{2}} = \bm{b}_2 + \bm{b}_{\bar{2}} = \{\gch^{1256},y^{1256}\ | \ \bgch^{1256},\byy^{1256}\}~.
}
}
%

\subsubsection*{Na\"{\i}ve bosonic interpretation}

The basis vectors $\bm{b}_1, \bm{b}_2$ are in one--to--one correspondence with the point group elements $\gth_1, \gth_2$, leaving the first and second two--torus inert, respectively, while reflecting in the remaining directions on the holomorphic side. Similarly, $\bm{b}_{\bar{1}}, \bm{b}_{\bar{2}}$ correspond to $\gth_{\bar{1}}, \gth_{\bar{2}}$ and $\bm{b}_{1\bar{1}}, \bm{b}_{2\bar{2}}$ to $\gth_{1\bar{1}}, \gth_{2\bar{2}}$, respectively. In light of this bosonic interpretation, these basis vectors are referred to as pure twist basis vectors\footnote{There are subtleties in this na\"{\i}ve interpretation: The pure basis vectors $\bm{b}_{1\bar{1}}, \bm{b}_{2\bar{2}}$ define in \eqref{eq:PureTwistBasisVectorsSym} overlap in the 5,6--directions and therefore do not give chiral projection. This is contrary to the convention employed in the so-called ``NAHE'' set \cite{Ferrara:1987jr, Antoniadis1988, NAHE2}. where there is no overlap of the two twist basis vectors. For the main classification purposes of this work, this is of no concern as all cases will be enumerated anyway.}. Through addition of lattice basis vectors to pure twist basis vectors, generic twist basis vectors are obtained, which can then be thought of as roto--translations. For example, the twisted basis vector $\bm{b}_1+\bm{e}_{1\bar{1}}$ corresponds to the space group element $g_1=(\gth_1,\sfrac 12 e_{1\bar{1}})$ which acts as a rototranslation~\eqref{eq:g1example}.

\subsubsection*{Basis vectors for T--folds on the $\boldsymbol{SO(12)}$}

This work considers four dimensional models defined on the $SO(12)$ lattice only, hence there are no other lattice basis vectors besides $\bm{E}$. In addition, there might be $n$ twist basis vectors $\Bgb_a$. The models under investigation contain $N=3+n$ basis vectors,
\equ{ \label{eq:BasisOfBasisVectors}
\bm{S}~, \qquad \bar{\bm{S}}~, \qquad \bm{E}~, \qquad \Bgb_a~,\quad  a=1,\ldots, n~, 
}
in total. A generic fermionic model with $N$ basis vectors is only fully specified by defining its GGSO phases. In light of \eqref{eq:CCinterchange}, there are only $\sfrac 12N(N-1)+1$ independent phases. For the basis \eqref{eq:BasisOfBasisVectors} the independent phases can be chosen to be the $3$ diagonal phases with $\bm{S}, \bar{\bm{S}}, \bm{E}$, the $n$ phases of $\Bgb_a$ with $\bm{E}$
\begin{subequations} \label{eq:IndependentPhases}
\equ{
\CC{\bm{S}}{\bm{S}}~, \qquad 
\CC{\bar{\bm{S}}}{\bar{\bm{S}}}~, \qquad 
\CC{\bm{E}}{\bm{E}}~, \qquad 
\CC{\Bgb_a}{\bm{E}}~, \qquad a=1,\ldots, n~, 
}
the $2n+1$ phases relevant for target space supersymmetry 
\equ{
\CC{\bm{S}}{\bar{\bm{S}}}~, \qquad \CC{\bm{S}}{\Bgb_a}~, \qquad \CC{\bar{\bm{S}}}{\Bgb_a}~, ~a=1,\ldots, n~, 
}    
and, finally, the $\sfrac 12n(n-1)$ off--diagonal twist vector phases
\equ{
\CC{\Bgb_a}{\Bgb_b}~, \quad a < b~. 
}
\end{subequations}
In fact, the phases with $\bm{E}$, and its diagonal phase in particular, fix the chiralities of internal spins associated with the fermions $y,w,\byy, \bw$. As these internal chiralities are physically irrelevant, these phases are all set to unity in the following 
\equ{ \label{eq:ConventionCC_E}
\CC{\Bgb_a}{\bm{E}} = \CC{\bm{E}}{\bm{E}} = +1~. 
}
By using the expansion \eqref{eq:ExpansionGGSO} this leads to the relation
\equ{
    \CC{\bm{1}}{\bm{1}} = - \CC{\bm{\bar{S}}}{\bm{\bar{S}}}~.
}
For the IIB theory this matches with our convention, while for the type IIA the sign is opposite.

\subsection{Four dimensional type II models of maximal supersymmetry}
\label{sc:typeII4d}

The amount of supersymmetry preserved in four dimensions depends on specific choices for the generalised GSO phases involving $\Sv$ and $\bar{\Sv}$. Throughout this work (with the exception of \cref{sc:ReducedSusyModels}) these choices are such that the maximal amount of supersymmetry possible is preserved. The form of the minimal surviving gravitini in four dimensions depends on certain conventions. The choices for these made here are: the surviving spinorial vacua should be of the form $\big|\pm(\sfrac12,\sfrac12^3)\big\rangle$ in the case of the type IIB models. For type IIA models the same spinorial vacua are assumed on the holomorphic side, while on the anti--holomorphic side $\big|\pm(\sm\sfrac 12,\sfrac12^3)\big\rangle$ ({\em i.e.}\ in the internal directions the spinor component are all taken to be equal). With these conventions, supersymmetric type II models in four dimensions have 
\equ{ \label{eq:4DGSOtypeII}
\CC{\Sv}{\Sv}=  \CC{\Sv}{\Bgb_a} = \CC{\bar{\Sv}}{\Bgb_a} = -1~, 
\qquad
\CC{\Sv}{\bar{\Sv}}= \CC{\bar{\Sv}}{\Sv} = +1~, 
\qquad
\CC{\bar{\Sv}}{\bar{\Sv}} = \pm 1~, 
}
for any twist basis vector $\Bgb_a$ that does not involve the fermions $\gps^\gm$ or $\bgps^\gm$. 
The final phase $\CC{\bar{\Sv}}{\bar{\Sv}}$ again distinguishes the type IIA $(+1)$ and type IIB models $(-1)$, respectively. 
Note that \eqref{eq:MixedPhasesE} applied to $\Sv$ and $\bar{\Sv}$ imply that 
\equ{ 
\CC{\Sv}{\bm{E}}= \CC{\bar{\Sv}}{\bm{E}} = -1~, 
}
which one needs to enforce to preserve supersymmetry anyway.

\subsection[Four dimensional ${\cN=8}$ supergravity]{Four dimensional $\boldsymbol{\cN=8}$ supergravity}

The particular case with no additional twist basis vectors with the above generalised GSO phase choices leads to $\cN=8$ supergravity in four dimensions. The massless states from the supersector: $\{\bm{0}, \Sv, \bSv, \Sv+\bSv\}$ corresponding to the usual NS--NS, R--NS, NS--R and R--R sectors, respectively. The massless states from these sectors are summarised in \cref{tab:N=8typeII}. They can be understood as the dimensional reduction of the ten dimensional states in \eqref{eq:10DsugraStates}. The massless states fit naturally into representations of $SU(4)_L\times SU(4)_R\times U(1)$, where the additional $U(1)$ arises from complexifying the R--R scalars from the $\Sv+\bSv$ sectors. By embedding this group into $SU(8)$ the massless states are exactly those of $\mathcal{N}=8$ pure supergravity, such that the spin--$(0,1/2,1,3/2,2)$ fields transform as $(70,56,28,8,1)$, in agreement with the $\mathcal{N}=8$ multiplet of table \ref{tab:4DMultiplets}. As delineated in \cite{Ferrara:1989nm}, we can identify the moduli space of the $\mathcal{N}=8$ type IIA/B superstring in four dimensions from the massless scalars. The axion (the dual of the $B$--field in four dimensions) and the dilaton parameterise the coset space
\equ{
    \mathcal{M}_{\rm AD} = \frac{SU(1,1)}{U(1)}~. 
}
The 36 NS--NS scalars parametrise the geometrical moduli space 
\equ{ \label{eq:GeomModuliSpace}
    \mathcal{M}_{\rm NS} = \frac{SO(6,6)}{SO(6)\times SO(6)}
}
of deformations of a six--dimensional torus, whilst the 16 complex R--R scalars parametrise coset 
\equ{ \label{eq:RRModuliSpace}
    \mathcal{M}_{\rm RR} = \frac{SU(4,4)}{SU(4)\times SU(4) \times U(1)}~.
}

\renewcommand{\arraystretch}{1.5}
\begin{table}[t]
    \centering
    \begin{tabular}{|c||c|c|c|c|}
        \hline 
        \textbf{Sector} & \textbf{Type IIB/A} & $\boldsymbol{SU(4)_L\times SU(4)_R}$ & \textbf{Interpretation}
        \\ \hline\hline
         $\bm{0}$ & $\psi^{\mu} \bgps^{\nu} \ket{\bm{0}}$  & $(1,1)$ & Graviton, dilaton, $B$--field
        \\ 
         & $\psi^{\mu} \bgch^{i}\ket{\bm{0}}, \chi^i \bgps^\mu \ket{\bm{0}}$ & $(6,1)+(1,6)$ & 12 gauge bosons
        \\ 
        & $\chi^{i} \bgch^{j}\ket{\bm{0}}$ & $(6,6)$& 36 scalars \\ \hline
        $\Sv$ & $\bgps^{\mu}\ket{\pm\sfrac{1}{2}{}^4}_{e}$  & $(4,1)$ & 4 gravitini, 4 spin--$1/2$ \\
        & $\bgch^i\ket{\pm\sfrac{1}{2}{}^4}_{e}$  & $(4,6)$ & 24 spin-$\frac{1}{2}$
        \\ \hline
        $\bSv$ &$\psi^{\mu}\ket{\pm\sfrac{1}{2}{}^4}_{e/o}$ & $(1,4/\bar{4})$ & 4 gravitini, 4 spin--$12$ 
        \\ & $\chi^i\ket{\pm\sfrac{1}{2}{}^4}_{e/o}$ & $(6,4)$ & 24 spin--$1/2$
        \\\hline
        $\Sv +\bSv$ &$\ket{\pm\sfrac{1}{2}{}^4}_{o}\otimes \ket{\pm\sfrac{1}{2}{}^4}_{o/e}$ & $(\bar{4},\bar{4}/4)$ & \tabu{c}{16 complex scalars,\\[-1ex] 16 gauge bosons}
        \\ \hline
        
    \end{tabular}
    \caption{Massless states for the $\mathcal{N}=8$ supergravity multiplet in four dimensions (see {\em e.g.}\ \cref{tab:4DMultiplets}) as arising from the basis set $\{\bm{E},\Sv, \bSv\}$ in the type IIB/A theories. The subscripts $e$ and $o$ indicate an even or odd number of minus signs, respectively. }
    \label{tab:N=8typeII}
\end{table}

\section[Classifications of order--two T--folds]{Classifications of order--two T--folds}
\label{sc:ClassificationsOrder2Tfolds}

This section outlines our classification procedure of order--two (a)symmetric orbifolds. This is done in two parts: the first step provides a classification of the Narain point groups (Narain $\Ratl$--classes) of order--two (a)symmetric orbifolds in six dimensions that preserve the maximal amount of supersymmetry possible. This classification applies to any lattice. The second part describes the necessary ingredients to obtain a full classification of all inequivalent configurations compatible with the $SO(12)$ lattice at the free fermionic point. Both parts rely on a notion of when order--two T--folds are considered to be equivalent.  

The purpose of this section is to classify all order--two T--fold configurations on the $O(12)$ lattice at the free fermionic point, the type II strings constitute the means to this end. Hence the four dimensional models resulting from type II strings propagating on the background of these configurations are studied. A configuration is then specified by a set of twist basis vectors together with the basis vector $\bm{E}$ defining the $SO(12)$ lattice. To define a full free fermionic string model on such a configuration, also the relevant GGSO phases have to be specified.

\subsection{Equivalence relations}
\label{sc:EquivalenceRelations}

A complication in classifying different order--two T--folds is that different basis vectors may give rise to the same configuration. In order to obtain a unique classification it is necessary to define a number of equivalence relations for which configurations can be considered to be physically identical. 

There are a number of equivalence relations that can relate different sets of basis vectors: 
\renewcommand{\labelenumi}{E\arabic{enumi})}
\begin{enumerate}
    \item $\mathfrak{g} \in GL(|\mathcal{B}|;\Intr)$ transformations: \label{E1}
    $\Bgb_a \ra \Bgb_a' = [\mathfrak{g}_a{}^b \Bgb_b]$.  
    \item Interchange of $y^i,w^i$ pairs in all basis vectors simultaneously; or interchange of $\byy^i,\bw^i$ pairs in all basis vectors simultaneously. 
    \item {\emergencystretch=1em
    Any permutation of the holomorphic indices $i=1,\ldots,6$; or any permutation of the anti--holomorphic indices $\bi=1,\ldots,6$.\par}
\end{enumerate}
\renewcommand{\labelenumi}{\arabic{enumi}}
Equivalence E1 uses the notation $[\Bgb]$ introduced below \eqref{eq:Overlap}. The resulting models on these configurations are then considered to be equivalent up to possible changes in the GGSO phases. 

These equivalence relations are natural in the sense that: E1 states that the physics is independent of the choices of basis vectors that define the fermionic boundary conditions. The equivalence relations E2 says that the nomenclature of the internal fermions $y^i, w^i$ and $\byy^\bi,\bw^\bi$ have no intrinsic meaning; only their products are locked in the supercurrents \eqref{eq:Supercurrent}. Finally, the equivalence relations E3 reflects that the labeling of the internal holomorphic and anti--holomorphic coordinates is arbitrary.

\begin{table}[t]
\centering
\renewcommand{\arraystretch}{1.2}
\begin{tabular}{| c| c| c| c| c|c| c|c| c|c| c|c|}
\hline 
\textbf{Point}  &  \textbf{Point} & \textbf{Basis} &  \textbf{S/} & \multicolumn{4}{c|}{\textbf{6D\,|\,4D--SUSYs}}& \multicolumn{2}{c|}{\textbf{$\boldsymbol{\#}$(Moduli)}}\\ 
\textbf{group}  &  \textbf{generators}  & \textbf{vectors} & \textbf{A} & \multicolumn{2}{c|}{\textbf{IIB}} & \multicolumn{2}{c|}{\textbf{IIA}} 
& \textbf{Geom} & \textbf{R--R} \\ 
\hline\hline 
$\Id$ & $\Id$ & -- & S & $(2,2)$ & $8$ & $(2,2)$ & $8$ & 36 & 32 \\ \hline
$\Intr_{2L}$ & $\gth_1$ & $\bm{b}_1$ & A & $(2,1)$ & 6 & $(2,1)$ & 6 & 12 & 16 \\
$\Intr_{2R}$ & $\gth_{\bar{1}}$  & $\bm{b}_{\bar{1}}$  & A & $(2,1)$ & 6 & $(1,2)$ & 6 & 12 & 16 \\
$\Intr_2$ & $\gth_{1\bar{1}}$  & $\bm{b}_{1\bar{1}}$& S & $(2,0)$ & 4 & $(1,1)$ & 4  & 20 & 8 \\ \hdashline
$\Intr_{2L}\times \Intr_{2R}$ & $\gth_1,~ \gth_{\bar{1}}$ & $\bm{b}_1,~\bm{b}_{\bar{1}}$ & A & $(2,0)$ & 4 & $(1,1)$ & 4 & 4 & 8 \\ \hline 
$\Intr_{2L}^2$ & $\gth_1,~ \gth_2$ & $\bm{b}_1,~\bm{b}_2$ & A & \multicolumn{2}{c|}{5} & \multicolumn{2}{c|}{5} & 0 & 8\\ 
$\Intr_{2R}^2$ & $\gth_{\bar{1}},~ \gth_{\bar{2}}$ & $\bm{b}_{\bar{1}},~\bm{b}_{\bar{2}}$ & A &  \multicolumn{2}{c|}{5} & \multicolumn{2}{c|}{5} & 0 & 8\\ 
$\Intr_{2L}\times \Intr_2$ & $\gth_1,~\gth_{2\bar{2}}$ & $\bm{b}_1,~\bm{b}_{2\bar{2}}$ & A & \multicolumn{2}{c|}{3} & \multicolumn{2}{c|}{3} & 8 & 4 \\ 
$\Intr_{2R} \times \Intr_2$ & $\gth_{\bar{1}},~\gth_{2\bar{2}}$ & $\bm{b}_{\bar{1}},~\bm{b}_{2\bar{2}}$ & A & \multicolumn{2}{c|}{3} & \multicolumn{2}{c|}{3}  & 8 & 4\\ 
$\Intr_2^2$ & $\gth_{1\bar{1}},~ \gth_{2\bar{2}}$ & $\bm{b}_{1\bar{1}},~\bm{b}_{2\bar{2}}$ & S &  \multicolumn{2}{c|}{2} & \multicolumn{2}{c|}{2} & 12 & 2\\ \hdashline 
$\Intr_{2L}^2\times \Intr_{2R}$ & $\gth_1,~ \gth_2,~ \gth_{\bar{1}}$ & $\bm{b}_1,~\bm{b}_2,~\bm{b}_{\bar{1}}$ & A & \multicolumn{2}{c|}{3} & \multicolumn{2}{c|}{3} & 0 & 4 \\
$\Intr_{2L}\times \Intr_{2R}^2$ & $\gth_1,~\gth_{\bar{1}},~ \bgth_{\bar{2}}$ & $\bm{b}_1,~\bm{b}_{\bar{1}},~\bm{b}_{\bar{2}}$ & A & \multicolumn{2}{c|}{3} & \multicolumn{2}{c|}{3} & 0 &4 \\
$\Intr_{2L}\times \Intr_{2R}\times \Intr_2$ & $\gth_1,~ \gth_{\bar{1}},~ \gth_{2\bar{2}}$ & $\bm{b}_1,~\bm{b}_{\bar{1}}~,\bm{b}_{2\bar{2}}$ & A & \multicolumn{2}{c|}{2} & \multicolumn{2}{c|}{2} & 4 & 2 \\ \hdashline
$\Intr_{2L}^2\times \Intr_{2R}^2$ & $\gth_1,~ \gth_2,~\gth_{\bar{1}},~ \gth_{\bar{2}}$ & $\bm{b}_1,~\bm{b}_2,~\bm{b}_{\bar{1}},~\bm{b}_{\bar{2}}$ & A & \multicolumn{2}{c|}{2} & \multicolumn{2}{c|}{2}  & 0 & 2 \\ \hline
\end{tabular}
\caption{\label{tab:NarainPointGroupClass} Classification of all order--two Narain point groups, $\Intr_{2L}^{p_L}\times\Intr_{2R}^{p_R}\times\Intr_2^p$, $p_L+p,p_R+p\leq 2$, preserving some target space supersymmetry. Here the point group generators, the basis vectors are indicated as well as whether the orbifold is A/Symmetric, the amount of supersymmetry that is preserved in six and four dimensions for the type IIA and  IIB theories. (Only when the model can be interpreted as a six dimensional theory compactified on a two--torus, the amount of six dimensional supersymmetry is indicated.) The last two columns give the number geometrical and R--R moduli preserved by the orbifold twists.}
\end{table}

\subsection{Six--dimensional T--fold order--two point groups}

In order to give a complete classification of all inequivalent  point groups of order--two T--folds, it is convenient to start with the asymmetric pure twist basis vectors $\bm{b}_\ga,\bm{b}_\bgb$ and the symmetric pure twist basis vectors $\bm{b}_{\ga\bga}$ defined in~\eqref{eq:PureTwistBasisVectors} and~\eqref{eq:PureTwistBasisVectorsSym}, respectively. This representation of the twist basis vectors is unique up to the equivalences E1 through E3.

Then starting with the unit point group, one increases the number compatible independent symmetric or asymmetric pure twist vectors to the set of basis vectors, see \cref{tab:NarainPointGroupClass}. There are three choices when there is just a single twist basis vector: either two asymmetric ones or one symmetric, up to the equivalences E1 through E3. For two twist basis vectors there are six choices: one fully symmetric, two where a symmetric element is combined with an asymmetric one on the left-- or the right--side, three fully asymmetric choices of which one acts on both left-- and right--moving sides, while the other two only on the left or only on the right. By the equivalences one can choose that when there is a symmetric element, it is $\bm{b}_{2\bar{2}}$. There are three point groups with three twist basis point groups. Two that only involve asymmetric elements, one on one side and two on the other and one set with a symmetric twist element and two asymmetric twist elements that act on both sides. Finally, there is just a single point group with four twist elements, all of them asymmetric, two acting on the left and two on the right. In summary, all the order--two Narain point groups are of the form 
\equ{
    \Intr_{2L}^{p_L} \times \Intr_{2R}^{p_R} \times \Intr_{2}^{p}~, 
    \qquad 0 \leq p_L+p,~ p_R+p \leq 2~. 
}
These results are tabulated in \cref{tab:NarainPointGroupClass} indicating whether the construction is symmetric or asymmetric and the amount of supersymmetry that is preserved.  Only three out of the fourteen cases are symmetric, all others are asymmetric. The table contains some redundancy between left-- and right--moving bosonic twists. The reason why this redundancy is not removed is to accommodate the type IIA case where the GGSO projection distinguishes the holomorphic and anti--holomorphic sides or to accommodate the extension to heterotic models. 

All the resulting type II models possess at least $\cN=2$ supersymmetry with the choices for the GGSO phases presented in the previous section. The number and types of supersymmetries preserved by different point group basis vectors are indicated in \cref{tab:NarainPointGroupClass}. Note that all extended supersymmetries $\cN=2,\ldots,6$ and $8$ are covered. The fact that with purely asymmetric orbifolds all these extended supergravities can be obtained was realised in the past~\cite{Ferrara:1989nm}. In addition, this table indicates the type of six dimensional supersymmetry that a given point group gives rise to if the twist elements only act in four internal dimensions. The resulting models can then be interpreted as strings on four dimensional T--folds times a two torus, hence the model can also be interpreted as four dimensional models with a related amount of four dimensional supersymmetry. When applicable, both six and four dimensional supersymmetries are indicated in \cref{tab:NarainPointGroupClass}, otherwise only the resulting four dimensional supersymmetry is given. 

The orbifold twist restrict the number of geometrical and RR moduli; the number of unfixed moduli are also indicated in \cref{tab:NarainPointGroupClass}. This means that the geometrical moduli space~\eqref{eq:GeomModuliSpace} is reduced to 
\equ{
    \mathcal{M}_{\rm NS} = 
    \frac{SO(\cD_L,\cD_R)}{SO(\cD_L)\times SO(\cD_R)} 
    \times 
    \Big(\frac{SO(2,2)}{SO(2)\times SO(2)}\Big)^{\cD_{LR}}~,
}
with 
\equ{
    \arry{l}{
    \cD_L= 2 + (1-p_L)(4-2p-p_L)~, 
    \\
    \cD_R = 2 + (1-p_R)(4-2p-p_R)~, 
    }
    \qquad 
    \cD_{LR} = (1-p_L)(1-p_R)p~,  
}
and the R--R moduli space~\eqref{eq:RRModuliSpace}  to
\equ{
    \mathcal{M}_{\rm RR} = 
    \frac{SU(\cN_L,\cN_R)}{SU(\cN_L)\times SU(\cN_R)\times U(1)},
}
in terms of the amounts of supersymmetry preserved on the left-- and right--moving sides,
\equ{ \label{eq:NsusyLR}
    \cN_L=\frac{4}{2^{(p_L+p)}}~, 
    \qquad 
    \cN_R=\frac{4}{2^{(p_R+p)}}~,
}
respectively; $\cN=\cN_L+\cN_R$ is the total amount of supersymmetric remaining in four dimensions, as indicated in \cref{tab:NarainPointGroupClass}.  Consequently, their real dimensions read 
\equ{
    \dim_{\mathbb R}\mathcal M_{\rm NS} = \cD_L\cD_R + 4\,\cD_{LR}~,
    \qquad 
    \dim_{\mathbb R}\mathcal M_{\rm RR} = 2\,\cN_L\cN_R~; 
}
reproducing the numbers of untwisted geometrical and R--R moduli listed in Table \ref{tab:NarainPointGroupClass}.
Note , in particular, the $\Intr_{2L}^2\times\Intr_{2R}^2$ orbifolds possess no geometrical and only two R--R moduli. Together with the dilaton and the universal axion the two R--R moduli for the four scalars of the universal hyper multiplet of the $\cN=2$ supersymmetry that these orbifolds possess.

\subsection[{$SO(12)$} lattice compatible extensions of twist basis vectors]{\boldsymbol{$SO(12)$} lattice compatible extensions of twist basis vectors}
\label{sc:ExtendedTwistBasisVectors}

To prepare for a full classification of all order--two T--folds on the $SO(12)$ lattice at the free fermionic point, it is useful to introduce some standard forms of the twist basis vectors possibly extended with shifts by making repeated use of the equivalence relations E1 through E3.

\subsubsection*{Extended symmetric twist basis vectors}

The extension of the symmetric twist elements \eqref{eq:PureTwistBasisVectorsSym} can be written as 
\equ{\label{eq:TwistBasisVectorsSym}
  \arry{ll}{
\bm{B}_{1\bar{1}} = [\bm{b}_{1\bar{1}} + n\cdot \bm{e}\,+\, \bn\cdot \overline{\bm{e}}] &= (0^21^4;n_{12},1^4;n\,|\, 0^21^4;\bn_{12},1^4;\bn)~,
\\[2ex]
\bm{B}_{2\bar{2}} = [\bm{b}_{2\bar{2}} + m\cdot \bm{e}+\bmm\cdot \overline{\bm{e}}] &= (1^20^21^2;1^2,m_{34},1^2-m_{56}; m \,|\, 1^20^21^2;1^2,\bmm_{34},1^2-\bmm_{56}; \bmm )~,
} 
}
subject to the restrictions 
\renewcommand{\labelenumi}{S\arabic{enumi})}
\begin{enumerate}
    \item $\bn = n=(n_{12},0^4)$, with $n_{12} = (n_1,n_2)=(0^2)$ or $(10)$. 
    \item $\bmm =m=(0^2,m_{3456})$, with $m_{3456}=(m_{34},m_{56})$ with $m_{34}=(m_3,m_4) = (0^2)$ or $(10)$ and $m_{56}=(m_5,m_6) = (0^2), (10)$ or $(1^2)$. 
\end{enumerate}
\renewcommand{\labelenumi}{\arabic{enumi}}
Here the notation is as follows: the first 6 entries indicate which $\gch$'s are present; the second 6 entries which $y$'s; the third 6 entries which $w$'s and similarly for the next 18 entries for the $\bgch$'s, $\byy$'s and $\bw$'s. $n,m$ and their barred versions are six dimensional vectors with either 0 or 1 entries. 

The consequences of modular invariance conditions of these symmetric twist elements require that the number of non--zero entries in $\bn$ and $n$ and in $\bmm$ and $m$ are equal. By permutations of the holomorphic and anti--holomorphic indices (equivalences E3) one can ensure that $\bm{B}_{1\bar{1}}$ acts as a shift in the 1,2 directions. Possibly by adding $\mathbf{E}$, we can ensure that $n_{12}$ has at most one non--zero entry.  By interchanges of the $y,\byy$'s and the $w,\bw$'s (equivalences E2) one can ensure that $\bm{B}_{1\bar{1}}$ only contains $y^3,\ldots, y^6$ and their holomorphic conjugates. Next, by possibly adding $\mathbf{E}$ to $\bm{B}_{2\bar{2}}$ (equivalence E1) we can ensure that $m_{34}=(0^2)$ or $(1^2)$. (Note that this does not restrict the number of non-zero entries in $m_{56}$.) Since in the 1,2 directions we have not yet used the possible interchange of the $y,\byy$'s and $w,\bw$'s, we can ensure that $\bm{B}_{2\bar{2}}$ only involves $y^1,y^2$ and their holomorphic conjugates.

\subsubsection*{Extended asymmetric twist basis vectors}

The asymmetric pure twist basis vectors \eqref{eq:PureTwistBasisVectors} can be extended to generic asymmetric twist elements that in the target space have the interpretation of roto--translations by adding lattice basis vectors to it: 
\equ{ \label{eq:TwistBasisVectors}
  \arry{ll}{
\bm{B}_1= [\bm{b}_1 + n\cdot \bm{e}\,+\, N\cdot \overline{\bm{e}}] &= (0^21^4;n_{12},1^4;n\,|\, 0^6;N;N)~,
\\[1ex]
\bm{B}_2= [\bm{b}_2 + m\cdot \bm{e}+M\cdot \overline{\bm{e}}] &= (1^20^21^2;1^2,m_{34},1^2-m_{56}; m \,|\, 0^6;M;M)~,
\\[1ex] 
\bm{B}_{\bar{1}}= [\bm{b}_{\bar{1}}+ \bn\cdot \bm{e}\,+\, \bN\cdot \overline{\bm{e}}] & =(0^6;\bN;\bN \,|\,0^21^4;\bn_{12},1^4;\bn)~, 
\\[1ex]
\bm{B}_{\bar{2}}= [\bm{b}_{\bar{2}} + \bmm\cdot \bm{e}+\bM\cdot \overline{\bm{e}}] &= (0^6;\bM;\bM\,|\, 1^20^21^2;1^2,\bmm_{34},1^2-\bmm_{56}; \bmm)~.
} 
}
The notation here is the same as in the symmetric case but in addition there are now also six dimensional vectors $N, M$  and their barred versions. In light of the equivalence relation E1, the twist basis vectors are defined modulo the addition of shift basis vectors. Free fermionic models on the $SO(12)$ lattice involve just a single shift basis vector $\mathbf{E}$. By using the equivalence relations E1 to E3 we can adjust these basis vectors to the standard form given above with the following restrictions: 
\renewcommand{\labelenumi}{A\arabic{enumi})}
\begin{enumerate}
    \item $n=(n_{12},0^4)$, with $n_{12} = (n_1,n_2)=(0^2), (10)$ or $(1^2)$ and at most 3 entries in $N$ are non--zero. 
    \item $\bn=(\bn_{12},0^4)$, with $\bn_{12} = (\bn_1,\bn_2)=(0^2), (10)$ or $(1^2)$ and at most 3 entries in $\bN$ are non--zero.    
    \item $m=(0^2,m_{3456})$, with $m_{3456}=(m_{34},m_{56})$, $m_{34}=(m_3,m_4)$, $m_{56}=(m_5,m_6) = (0^2), (10)$ or $(1^2)$ and at most 3 entries in $M$ are non--zero. 
    \item $\bmm=(0^2,\bmm_{3456})$, with $\bmm_{3456}=(\bmm_{34},\bmm_{56})$, $\bmm_{34}=(\bmm_3,\bmm_4)$, $\bmm_{56}=(\bmm_5,\bmm_6) = (0^2), (10)$ or $(1^2)$ and at most 3 entries in $M$ are non--zero.    
\end{enumerate}
\renewcommand{\labelenumi}{\arabic{enumi}}
On any first twist basis vector $\bm{B}_1$ the restriction A1 may be enforced: If $N$ has four or more non--zero entries, then by adding the shift basis vector $\mathbf{E}$ the new $N$ has two or less entries non-zero. (Three non-zero entries are mapped to three non-zero entries.) By permutation of the holomorphic indices (equivalence E3) we can enforce that in the first two directions $\bm{B}$ does not act as a reflection. Then by possible interchanges of $y^i,w^i$ for $i=3,\ldots,6$ (equivalence E2) we can ensure that $\bm{B}_1$ only involves $y^i$ and not $w^i$ for that range of indices. The argument that $M$ has at most three non--zero entries is the same as for $N$. By permutation of the holomorphic indices (equivalence E3) we can ensure that in the 3,4 directions $\bm{B}_2$ does not act as reflections. Since $\bm{B}_1$ act as a pure shift in the 1,2 directions, we may still use interchanges of $y^i,w^i$ for $i=1,2$ to ensure that $\bm{B}_2$ contains $y^1,y^2$ and not $w^1,w^2$. Only in the 5,6--directions we distinguish between a pure twist and a roto--translation in $\bm{B}_2$ for $\bm{b}_2$ since $\bm{b}_1$ already acts as a pure twist in those directions. These distinctions are made by $m_{5,6}$: for example, if $m_5=0$ $\bm{b}_2$ acts as a pure twist in the 5--th direction ($y^5$ present but not $w^5$), if $m_5=1$ as a roto--translation ($w^5$ present but not $y^5$). (Similar arguments may be used to arrive at the standard forms A2 and A4 for $\bm{B}_{\bar{1}}$ and $\bm{B}_{\bar{2}}$ using the equivalences E1 through E3.)

The modular invariance conditions for the full set of asymmetric twist basis vectors, $\bm{B}_1, \bm{B}_2,\bm{B}_{\bar{1}}, \bm{B}_{\bar{2}}$,  can be compactly stated as
\begin{subequations}
\label{eq:AsymmetricModInv}
\equ{
\label{eq:ModB1B2}
n_{12}^2 = N^2 \mod{4}~, 
\qquad 
m_{34}^2 = M^2 \mod{4}~, 
\qquad 
m_{56}^2 = (N-M)^2 \mod{4}~,
\\[1ex]
\label{eq:ModbarB1barB2}
\bn_{12}^2 = \bN^2 \mod{4}~, 
\qquad 
\bmm_{34}^2 = \bM^2 \mod{4}~, 
\qquad 
\bmm_{56}^2 = (\bN-\bM)^2 \mod{4}~,
\\[1ex]
\label{eq:ModB1barB1}
(2n_{12}-1^2)\cdot \bN_{12} - (2\bn_{12}-1^2)\cdot N_{12} = N^2 - \bN^2 \mod{4}~, 
\\[1ex]
\label{eq:ModB2barB2}
(2m_{34}-1^2)\cdot \bM_{34} - (2\bmm_{34}-1^2)\cdot M_{34} = M^2 - \bM^2 \mod{4}~, 
\\[1ex]
\label{eq|ModB1barB2}
(2n_{12}-1^2)\cdot \bM_{12} - (2\bmm_{34}-1^2)\cdot N_{34} = N^2 - \bM^2 \mod{4}~, 
\\[1ex]
\label{eq:ModB2barB1}
(2m_{34}-1^2)\cdot \bN_{34} - (2\bn_{12}-1^2)\cdot M_{12} = M^2 - \bN^2 \mod{4}~.
}
\end{subequations}
Here we used the shorthands for the sums, like $n_{12}^2 = n_1+n_2$, and inner products, like $(2n_{12}-1^2)\cdot \bN_{12} = (2n_1-1)\bN_1+(2n_2-1)\bN_2$, etc. Furthermore, note that $2n_1-1 = -(-1)^{n_1}$, therefore $2n_{12}-1^2$ is a two--component vector with entries $\pm 1$. The first two equations in \eqref{eq:ModB1B2} and \eqref{eq:ModbarB1barB2} are obtained by squaring these basis vectors. The third equations arise from the inner products of $\bm{B}_1\cdot \bm{B}_2$ and $\bm{B}_{\bar{1}} \cdot \bm{B}_{\bar{2}}$, respectively. The conditions, \eqref{eq:ModB1barB1} through \eqref{eq:ModB2barB1}, correspond to the inner product relations of $\bm{B}_1\cdot \bm{B}_{\bar{1}}$, $\bm{B}_2\cdot \bm{B}_{\bar{2}}$, $\bm{B}_1\cdot \bm{B}_{\bar{2}}$ and $\bm{B}_2\cdot \bm{B}_{\bar{1}}$, respectively. If only some of these asymmetric twist basis vectors are present, the conditions are reduced accordingly. 

\subsubsection*{Extended symmetric and asymmetric basis vectors combined}

From table~\ref{tab:NarainPointGroupClass} it can be inferred that $\bm{B}_{2\bar{2}}$ can be combined with $\bm{B}_1$ or $\bm{B}_{\bar{1}}$ leading to the additional conditions 
\begin{subequations}
\equ{
\label{eq:ModB1_barB1}
n_{12}^2 = N^2 \mod{4}~, 
\qquad 
\bn_{12}^2 = \bN^2 \mod{4}~, 
\\[1ex]
\label{eq:ModB1DotbarB1}
(2n_{12}-1^2)\cdot \bN_{12} - (2\bn_{12}-1^2)\cdot N_{12} = N^2 - \bN^2 \mod{4}~, 
\\[1ex]
\label{eq:ModB22B1_barB1}
m_{56}^2 = m_{34}^2 - (2m_{34}-1^2)\cdot N_{34} \mod{4}~, 
\qquad 
m_{56}^2 = m_{34}^2 - (2m_{34}-1^2)\cdot \overline{N}_{34} \mod{4}~, 
}
\end{subequations}
The equations \eqref{eq:ModB1_barB1} arise from squaring $\bm{B}_1$ and $\bm{B}_{\bar{1}}$ and \eqref{eq:ModB1DotbarB1} from their inner product. The final two equations \eqref{eq:ModB22B1_barB1} are obtained by dotting $\bm{B}_{2\bar{2}}$ with $\bm{B}_1$ and $\bm{B}_{\bar{1}}$. Since these final two conditions both restrict $m_{56}^2$, combined they can be understood as a projection condition: $(2m_{34}-1^2)\cdot (N-\bN)_{34} = 0\mod{4}$. (One could equivalently have started with the basis vectors $\bm{B}_{1\bar{1}}$, $\bm{B}_2$ and $\bm{B}_{\bar{2}}$, but then one has to consider more combinations that are equivalent to those result from this parameterisation.)

\subsection[Order--two T--folds on the {$SO(12)$} lattice]{Order--two T--folds on the \boldsymbol{$SO(12)$} lattice}

The classification of all six--dimensional order--two T--folds on the $SO(12)$ lattice at the free fermionic point have been collected in \cref{tab:NarainRotoTransClass}. This classification gives a complete list of T--fold configurations in terms of inequivalent sets of extended twist basis vectors for each of the admissible order--two Narain point groups. It is complete up to interchange of the subscripts $L$ and $R$. This subsection only states the main results of this classification, the following subsection details of the classification procedure. 

The basic structures of the extended twist basis vectors are listed in \cref{tab:SchemeClass} defining the labels used for them in this work. The labeling is chosen with asymmetric constructions in particular in mind
and signifies differences between pure twists and roto--translations, which are distinguished further by in which directions they also act as shifts. 
For example, the $\Intr_{2L}$ labels i signifies a pure twist, ii a roto--translation in one non--twisted (left--moving) direction and iii a roto--translation in both non--twisted directions. 
And in the orbifolds that have a final symmetry $\Intr_2$ factor, the label A corresponds to a pure twist. Labels B and C correspond to roto--translations with a symetric shift in a single direction, but for label B this shift is in a direction which is also twisted, while for label C this is not the case. Similarly labels D and E are two cases where the shift  in the roto--translations act in two directions. 
Therefore, the five well--known symmetric $\Intr_2^2$ orbifolds on the $SO(12)$ lattice do not get distinct labels here, but are referred to by double labels made out of the labels for the ``first factor $\Intr_2$'' and ``final factor $\Intr_2$'' (like for example a--A for the configuration defined by $\bm{b}_{1\bar{1}}, \bm{b}_{2\bar{2}}$). This table only gives the labeling for the asymmetric $\Intr_{2L}$ and $\Intr_{2L}^2$; asymmetric factors $\Intr_{2R}$ and $\Intr_{2R}^2$ are indicated in the second place. For example,  the label i--i refers to the $\Intr_{2L}\times\Intr_{2R}$ configuration with twist basis vectors $\bm{b}_1, \bm{b}_{\bar{1}}$, the label I--I to the $\Intr_{2L}^2\times\Intr_{2R}^2$ configuration with twist basis vectors $\bm{b}_1,\bm{b}_2, \bm{b}_{\bar{1}}, \bm{b}_{\bar{2}}$, and the label i--i--A the $\Intr_{2L}\times\Intr_{2R}\times \Intr_2$ configuration with twist basis vectors $\bm{b}_1, \bm{b}_{\bar{1}}, \bm{b}_{2\bar{2}}$. These different labels alone sometimes do not distinguish the inequivalent configurations completely. If not, the configurations are further enumerated as $.1$, $.2$, etc. 

The classification \cref{tab:NarainRotoTransClass} is built up as follows. The different Narain point groups are introduced as heading rows. In addition, they indicate the amount of $\cN$--extended supersymmetry the corresponding T--folds all minimally preserve and the untwisted spectrum defined in multiplets of that extended supersymmetry. The supergravity multiplet is always present there. The next rows that follow define the different configurations of T--folds with that point group. They are divided into three columns. The first gives the unique label of the configuration, the next column gives the defining twist basis vectors. The final third column is divided into two parts: the top part labels the massless twisted sectors. If a twisted sector label has a gray colour this signifies that this twisted sector is necessarily projected out by some GGSO projection. The bottom part of the third column gives the massless twisted spectra of the configurations in terms of the appropriate $\cN$--extended supersymmetry representations. If the spectra depend on some detailed choice of GGSO phases, the spectra are given in parametric form in terms of some parameters. In that case the dependence on the GGSO phases of these parameters are given in the header row as well. The details of the spectra computations are explained in the next section.

\begin{table}
\begin{center}
\begin{tabular}{ccc}
    \begin{tabular}{|c||c|}
        \hline
        \multicolumn{2}{|c|}{$\boldsymbol{\Intr_{2L}}^2$} 
        \\ \hline\hline
        I & $\bm{b}_1,~\bm{b}_2$ 
        \\ 
        II & $\bm{b}_1,~\bm{b}_2+\bm{e}_{35\bar{*}}$ 
        \\ 
        III & $\bm{b}_1,~\bm{b}_2+\bm{e}_{3456\bar{*}\bar{*}}$ 
        \\ 
        IV & $\bm{b}_1+\bm{e}_{1\bar{1}},~\bm{b}_2+\bm{e}_{356\bar{*}}$ 
        \\ 
        V & $\bm{b}_1+\bm{e}_{12\bar{1}\bar{2}},~\bm{b}_2+\bm{e}_{3456\bar{*}\bar{*}}$ 
        \\ 
        VI & $\bm{b}_1+\bm{e}_{12\bar{1}\bar{2}},~\bm{b}_2+\bm{e}_{34\bar{*}\bar{*}}$  
        \\[1ex] \hline
    \end{tabular}
    & 
    \begin{tabular}{|c||c|}
        \hline
        \multicolumn{2}{|c|}{$\boldsymbol{\Intr_{2L}$}} 
        \\\hline\hline
        i & $\bm{b}_{1}$ 
        \\
        ii & $\bm{b}_{1}+\bm{e}_{1\bar{*}}$   
        \\
        iii & $\bm{b}_{1}+\bm{e}_{12\bar{*}\bar{*}}$   
        \\ \hline    
        \hline
        \multicolumn{2}{|c|}{First factor $\boldsymbol{\Intr_2}$} 
        \\\hline\hline
        a & $\bm{b}_{1\bar{1}}$ 
        \\
        b & $\bm{b}_{1\bar{1}}+\bm{e}_{1\bar{1}}$   
        \\ \hline 
 
    \end{tabular}    
    & 
    \begin{tabular}{|c||c|}
        \hline
        \multicolumn{2}{|c|}{Final factor $\boldsymbol{\Intr_{2}}$} 
        \\ \hline\hline
        A & $\bm{b}_{2\bar{2}}$ 
        \\
        B & $\bm{b}_{2\bar{2}}+\bm{e}_{5\bar{5}}$   
        \\
        C & $\bm{b}_{2\bar{2}}+\bm{e}_{3\bar{3}}$   
        \\
        D & $\bm{b}_{2\bar{2}}+\bm{e}_{56\bar{5}\bar{6}}$   
        \\
        E & $\bm{b}_{2\bar{2}}+\bm{e}_{35\bar{3}\bar{5}}$  
        \\[1ex] \hline  
        \multicolumn{2}{c}{}\\ 
    \end{tabular}
\end{tabular}
\end{center}
\caption{\label{tab:SchemeClass} The classification labels used to characterise the different order--two T--folds are listed in terms of the twist basis vectors with possible translations. The cases, where it is not yet fixed in which anti--holomorphic directions the translation are pointing, are indicated with $\bar{*}$. The first and final $\Intr_2$ factors refer to constructions with one or two symmetric twist basis vectors. The conventions for $\Intr_{2R}^2$ and $\Intr_{2R}$ are conjugate to those for $\Intr_{2L}^2$ and $\Intr_{2L}$.}
\end{table}

\subsection[Classification procedure of order--two {$SO(12)$} lattice compatible T--folds]{Classification procedure of order--two \boldsymbol{$SO(12)$} lattice compatible T--folds}
\label{sc:ClassificationProcedure}

The main difficulty in the classification is to ensure that each configuration is counted only once: all configurations have to be identified up to the equivalence relations introduced in \cref{sc:EquivalenceRelations}. The parametrisations of \cref{sc:ExtendedTwistBasisVectors} already anticipate many consequences of these relations. Using those a full classification  of a number of relatively simple order--two T--folds, namely those with $\Intr_2$, $\Intr_2^2$, $\Intr_{2L}$ and $\Intr_{2L}^2$ point groups, were obtained. The parametisation constraints A1)--A4) and S1)--S2) aid in identifying these inequivalent configurations. They serve to distinguish the twist vectors labelled by i,\ldots,iii, I,\ldots,VI and A,\ldots,E in \cref{tab:SchemeClass}, which refer to the different twist and roto--translation actions that the basis vectors can generate, and which are used to organise \cref{tab:NarainRotoTransClass} throughout. 

A similar approach can be taken for more complicated point groups, however, it is easy to miss E1--E3 mappings that render configurations equivalent. A computerised implementation was therefore used for all point groups and cross checked against solution sets obtained by hand. 
The computer codes developed enumerate all modular invariant choices of the parameters $n_{12}$, $m_{3456}$, $N$, $M$, and their barred counterparts (without using the restrictions A1)--A4) and S1)--S2)). 
For each choice, the additive set generated by the basis vectors is constructed, and two configurations are identified whenever some mapping with E1--E3 maps one additive set onto the other. All inequivalent configurations are then identified by exhaustive searching so that no equivalence mapping is missed. 
The codes and the resulting lists of configurations for each point group are collected in the
\texttt{Model\_Classification\_Z2N} folder of
\cite{percival2026z2n_typeii_code}.
More details of the classification the inequivalent configurations specific for each point group are provided below.

\subsubsection*{Symmetric $\boldsymbol{\Intr_2}$ and $\boldsymbol{\Intr_2^2}$ orbifolds}

The results for symmetric cases $\Intr_2$ and $\Intr_2^2$ coincide with results known in the literature~\cite{DW,FRTV,Athanasopoulos:2016aws}: the five $\Intr_2^2$ configurations correspond to the geometries $(1$--$1)$ through $(1$--$5)$ in the classification of ref.~\cite{DW}. In detail, there the $SO(12)$ lattice is realised by the vectors $e_1,\ldots,e_6$ and $\sfrac12 e_{135}$ (corresponding to their $\tau$'s in the three complex directions). By a change of basis, $e_1 \ra e_{12}, e_3 \ra e_{34}, e_5 \ra e_{56}$, one arrives at our convention where the $SO(12)$ lattice is spanned by $e_1,\ldots, e_6$ and $\sfrac 12 e_{123456}$. 

An additional point that should be mentioned is that the twist basis vectors \eqref{eq:PureTwistBasisVectorsSym} are not the same as those used in ref.~\cite{Athanasopoulos:2016aws} which followed the NAHE conventions. As a consequence the first $\Intr_2^2$ configuration in \cref{tab:NarainRotoTransClass} with label a--A according to the labeling scheme of \cref{tab:SchemeClass} is the ($1$--$2$) of the DW classification, the second configuration with label a--D is the DW orbifold ($1$--$1$); while the others are matched in the same order.

\subsubsection*{Asymmetric $\boldsymbol{\Intr_{2L}}$ orbifolds}

The options for the asymmetric $\Intr_{2L}$ orbifolds are very limited and distinguished by whether there is no shift, a single shift or a shift in both directions in which the twist element does not act as a twist on the holomorphic side. On the anti--holomorphic side one can in principle pick any directions for the shifts, but using the equivalences one can ensure that the shifts of the roto--translations are $e_{1\bar{1}}$ and $e_{12\bar{1}\bar{2}}$.

\subsubsection*{Asymmetric $\boldsymbol{\Intr_{2L}^2}$ orbifolds}

In light of the parameterisation \eqref{eq:TwistBasisVectors} identifying the different asymmetric $\Intr_{2L}^2$ orbifolds, means that one has to consider all choices for the vectors $n=(n_{12})$, $m=(m_{3456})$ and the six component vectors $N$ and $M$ subject to the equivalence relations E1 and E3. (Equivalence E2 is already taken care of by the parameterisation.) An elegant way of removing redundancies is by considering the lengths of the vectors $N$, $M$, and $[N\!-\!M]$ associated to the non--trivial twist vectors $\bm{B}_1$, $\bm{B}_2$ and $\bm{B}_3$, since lengths are invariant under relabelling of the (anti--)holomorphic coordinates. Moreover, the sum of these lengths is invariant under relabelling of the nontrivial point group elements. Given that the vectors $N$ and $M$ have at most two entries 1 and the rest 0, one arrives at \cref{tab:ClassificationZ2L2models}. Exactly which vector has which length is thus irrelevant, the sum of the lengths distinguishes the cases uniquely. The choice for which vectors have which lengths and the specific parameterisation is subject to the equivalences and the modular invariance conditions~\eqref{eq:ModB1B2}. Thus here first the possible inequivalent vectors $N$ and $M$ are determined and from them possible choices for the vectors $n$ and $m$. 

In order to distinguish the different cases uniquely the parameterisation of the $n=(n_{12})$ and $m=(m_{3456})$ of the last two columns of the ``Parameterisation'' column of \cref{tab:ClassificationZ2L2models} can be used provided that one takes the inner product of $N$ and $M$ into account. For the first three cases the inner products are always zero as one of the vectors is the zero vector. For the final three cases the values of the inner products are essential to distinguish them uniquely. For example, case with the vectors $n$ and $m$ of case VI but with inner product $N\cdot M=2$ instead of $0$, is, in fact, equivalent to case III. 

The enumeration of the cases in \cref{tab:ClassificationZ2L2models} coincides with that in \cref{tab:NarainRotoTransClass}. The results of \cref{tab:ClassificationZ2L2models} will be used for the classification of the asymmetric orbifolds with point groups that involve $\Intr_{2L}^2$ or $\Intr_{2R}^2$ factors.

\renewcommand{\arraystretch}{1.2}
\begin{table}
    \centering
    \begin{tabular}{|c||c|c|c|c|| c|c|c|| c|c|| c|}
        \hline 
         & \multicolumn{3}{c}{\textbf{Lengths}} & \textbf{Sum} & \multicolumn{5}{c||}{\textbf{Parameterisation}} & 
        \\
        \textbf{Label} & $\big|N\big|$ & $\big|M\big|$ & $\big|[N\!-\!M]\big|$ & $\sum$ & $N$ & $M$ & $N\cdot M$ & $n_{12}$ & $m_{3456}$  & $\bm{B}_1\!\cdot\!\bm{B}_2$
        \\ \hline\hline
        \textbf{I} & $0$ & $0$ & $0$ & $0$ & $(0^6)$ & $(0^6)$ & $0$ & $(0^2)$ & $(0^4)$ & $2$ \\ 
        \textbf{II} & $0$ & $1$ & $1$ & $2$ & $(0^6)$ & $(10^5)$ & $0$ & $(0^2)$ & $(1010)$ & $2$ \\ 
        \textbf{III} & $0$ & $\sqrt{2}$ & $\sqrt{2}$ & $2\sqrt{2}$ & $(0^6)$ & $(1^20^4)$ & $0$ & $(0^2)$ & $(1^4)$ & $2$ \\ 
        \textbf{IV} & $1$ & $1$ & $\sqrt{2}$ & $2+\sqrt{2}$ & $(10^5)$ & $(010^4)$ & $0$ & $(10)$ & $(1011)$ & $2$ \\ 
        \textbf{V} & $\sqrt{2}$ & $\sqrt{2}$ & $\sqrt{2}$ & $3\sqrt{2}$ & $(1^20^4)$ & $(1010^3)$ & $1$ & $(1^2)$ & $(1^4)$ & $2$ \\ 
        \textbf{VI} & $\sqrt{2}$ & $\sqrt{2}$ & $2$ & $2+2\sqrt{2}$ & $(1^20^4)$ & $(0^21^20^2)$ & $0$ & $(1^2)$ & $(1^20^2)$ & $4$ 
         \\ \hline 
    \end{tabular}
    \caption{Classification of the asymmetric $\Intr_{2L}^2$ orbifolds on the $SO(12)$ lattice using the equivalence invariance of the sum of lengths.}
    \label{tab:ClassificationZ2L2models}
\end{table}

\subsubsection*{Asymmetric $\boldsymbol{\Intr_{2L}\times \Intr_{2R}}$ orbifolds}

To analyse the remaining cases, the above results can be used, since subgroups of their point groups were considered in the previous cases. The first example are the orbifolds with point group $\Intr_{2L}\times \Intr_{2R}$. The two point group subgroups are parametrised by $n_{12}, \bar{n}_{12} = (0^2), (10), (1^2)$ and vectors $N$ and $\bar{N}$ with the same lengths as $n_{12}$ and $\bar{n}_{12}$, respectively. In these orbifolds the directions 1,2 are distinguished from the other directions 3 through 6, since the twists only act in the latter. The number of inequivalent modular invariant realisations per configurations are listed in \cref{tab:OverviewZ2LZ2Rmodels}; some configurations do not admit any solution; while others multiple that are then simply enumerated. The resulting inequivalent configurations have been tabulated in \Cref{tab:NarainRotoTransClass}.

\subsubsection*{Asymmetric $\boldsymbol{\Intr_{2L}\times \Intr_2}$ orbifolds}

The subgroups of the point group $\Intr_{2L}\times \Intr_2$ are $\Intr_{2L}$ and $\Intr_2$. As suggested in the previous subsection, the basis vectors can be taken to be $\bm{B}_1$ and $\bm{B}_{2\bar{2}}$. The $\Intr_2$ basis vector $\bm{B}_{2\bar{2}}$ are distinguished by $m=\bar{m}=(0^4), (0^210), (100^2), (0^21^2), (1010)$. The $\Intr_{2L}$ basis $\bm{B}_1$ is parametrised by $n=(n_{12})$ and $N$ with $n_{12}=(0^2), (10), (1^2)$. Thus, only $N$ is free but, of course, subject to the modular invariance conditions. One particular consequence is that the length of $N$ is equal to that of $n_{12}$. Its representation is not unique, but because on the holomorphic side the three two--torus directions are distinguished, the redundancies due to the equivalence relations are greatly reduced. Moreover, if $m$ has a non--zero entry in the 3 or 4 directions, but not both, then these two directions are distinguished (and similar for the 5,6 directions). Taking all this into account, one can tabulate all possible choices for $N$ for a given $n$ and $m$. Each choice, for which all modular invariance conditions are satisfied, defines a viable configuration. 

However, T--folds with this point group provide the first example where this procedure still produces configurations that may still be equivalent. For the cases where this happens the configuration higher up the list within the labeling scheme is chosen as the representative. The number of inequivalent modular invariant configurations are listed in \cref{tab:OverviewZ2LZ2models}. Notice that according to this table no configurations *--C are among the inequivalent ones. This does not mean that such configurations do not exist, but only that they are equivalent to other configurations higher up in the list. For example, the configuration ii--C is equivalent to the configuration ii--B. The $10$ inequivalent configurations are tabulated in~\cref{tab:NarainRotoTransClass}.

\subsubsection*{Asymmetric $\boldsymbol{\Intr_{2L}^2\times \Intr_{2R}}$ orbifolds}

The subgroups of the point group $\Intr_{2L}^2\times \Intr_{2R}$ relevant for the classification are $\Intr_{2L}^2$ and $\Intr_{2R}$. For the parametrisation of the $\Intr_{2L}^2$ elements the choices in the last two columns of the ``Parameterisation'' columns of \cref{tab:ClassificationZ2L2models} can be used augmented with vectors $N, M$, and for $\Intr_{2R}$  $\bar{n}_{12},\bar{n}_{12}=(0^2), (10), (1^2)$ and $\bar{N}$. This results in $6\cdot 3=18$ configurations characterised by $n_{12},m_{3456}$ and $\bar{n}_{12}$. For each of them all possible inequivalent choices for $N, M$ and $\bar{N}$ are considered and checked whether they result in modular invariant models according to the relevant conditions from \eqref{eq:AsymmetricModInv}.

\clearpage

\renewcommand{\arraystretch}{1.2}
\begin{table}
    \centering
    \begin{tabular}{|l|| c|c|c|}
        \hline 
        $\boldsymbol{\bar{n}_{12}}$ $\boldsymbol{\Big\backslash}$ $\boldsymbol{n_{12}}$ & \textbf{i:}~~ $\boldsymbol{(00)}$  & \textbf{ii:}~ $\boldsymbol{(01)}$  & \textbf{iii:} $\boldsymbol{(11)}$
        \\ \hline\hline
        \textbf{i:}~~ $\boldsymbol{(00)}$ & $1$ & $1$ & $1$ 
        \\ \hline 
        \textbf{ii:}~ $\boldsymbol{(10)}$ & $1$ & $3$ & $3$ 
        \\ \hline
        \textbf{iii:} $\boldsymbol{(11)}$ & $1$ & $3$ & $3$ 
        \\ \hline 
    \end{tabular}
    \caption{The number of solutions for each of the possible configurations of the asymmetric $\Intr_{2L}\times\Intr_{2R}$ orbifolds parametrised by $n_{12}$ and $\bar{n}_{12}$; $17$ in total.}
    \label{tab:OverviewZ2LZ2Rmodels}
\end{table}

\renewcommand{\arraystretch}{1.2}
\begin{table}
    \centering
    \begin{tabular}{|l|| c|c|c|c|c| }
        \hline 
        $\boldsymbol{{n}_{12}}$ $\boldsymbol{\Big\backslash}$ $\boldsymbol{m_{3456}}$ & \textbf{A:} $\boldsymbol{(0000)}$ & \textbf{B:} $\boldsymbol{(0010)}$ & \textbf{C:} $\boldsymbol{(1000)}$ & \textbf{D:} $\boldsymbol{(0011)}$ & \textbf{E:} $\boldsymbol{(1001)}$
        \\ \hline\hline
        \textbf{i:}~~ $\boldsymbol{(00)}$ & $1$ & $0$ & $0$ & $0$ & $1$ 
        \\ \hline
        \textbf{ii:}~ $\boldsymbol{(10)}$ & $1$ & $1$ & $0$ & $0$ & $1$ 
        \\ \hline
        \textbf{iii:} $\boldsymbol{(11)}$ & $1$ & $1$ & $0$ & $1$ & $2$
        \\ \hline 
    \end{tabular}
    \caption{The number of solutions for each of the possible configurations of the asymmetric $\Intr_{2L}\times\Intr_{2}$ orbifolds parametrised by $m_{3456}$ and ${n}_{12}$; $10$ in total.}
    \label{tab:OverviewZ2LZ2models}
\end{table}

\renewcommand{\arraystretch}{1.2}
\begin{table}
    \centering
    \begin{tabular}{|l|| c|c|c|c|c|c| }
        \hline 
        $\boldsymbol{\bar{n}_{12}}$ $\boldsymbol{\Big\backslash}$ \!\!\tabu{c}{$\boldsymbol{n_{12}}$ \\ $\boldsymbol{m_{3456}}$}\!\!\!  & \textbf{I:}\!\!\tabu{c}{$\boldsymbol{(00)}$ \\ $\boldsymbol{(0000)}$}\!\!\! & \textbf{II:}\!\!\tabu{c}{$\boldsymbol{(00)}$ \\ $\boldsymbol{(1010)}$}\!\!\! & \textbf{III:}\!\!\tabu{c}{$\boldsymbol{(00)}$ \\ $\boldsymbol{(1111)}$}\!\!\! & \textbf{IV:}\!\!\tabu{c}{$\boldsymbol{(10)}$ \\ $\boldsymbol{(1011)}$}\!\!\! & \textbf{V:}\!\!\tabu{c}{$\boldsymbol{(11)}$ \\ $\boldsymbol{(1111)}$}\!\!\! & \textbf{VI:}\!\!\tabu{c}{$\boldsymbol{(11)}$ \\ $\boldsymbol{(1100)}$}\!\!\!
        \\ \hline\hline
        \textbf{i:}~~ $\boldsymbol{(00)}$ & $1$ & $1$ & $1$ & $1$ & $0$ & $0$ 
        \\ \hline
        \textbf{ii:}~ $\boldsymbol{(10)}$ & $0$ & $1$ & $1$ & $2$ & $1$ & $0$ 
        \\ \hline
        \textbf{iii:} $\boldsymbol{(11)}$ & $0$ & $1$ & $1$ & $3$ & $1$ & $2$ 
        \\ \hline 
    \end{tabular}
    \caption{The number of solutions for each of the possible configurations of the asymmetric $\Intr_{2L}^2\times\Intr_{2R}$ orbifolds parametrised by $n_{12},m_{3456}$ and $\bar{n}_{12}$; $17$ in total.}
    \label{tab:OverviewZ2L2Z2Rmodels}
\end{table}

\renewcommand{\arraystretch}{1.2}
\begin{table}
    \centering
    \begin{tabular}{l}

    \begin{tabular}{|l|| c|c|c|| l|| c|c|c|}
        \hline 
        $\boldsymbol{m_{3456}}$ & \multicolumn{3}{l||}{\textbf{A:} $\boldsymbol{(0000)}$} & $\boldsymbol{m_{3456}}$ & \multicolumn{3}{l|}{\textbf{B:} $\boldsymbol{(0010)}$} 
        \\ \hline 
        \!$\boldsymbol{\bar{n}_{12}}$ \!\!$\boldsymbol{\Big\backslash}$\!\! $\boldsymbol{n_{12}}$\! & \textbf{i:}~~ $\boldsymbol{(00)}$  & \textbf{ii:}~ $\boldsymbol{(01)}$  & \textbf{iii:} $\boldsymbol{(11)}$ & 
        \!$\boldsymbol{\bar{n}_{12}}$ \!\!$\boldsymbol{\Big\backslash}$\!\! $\boldsymbol{n_{12}}$\! & \textbf{i:}~~ $\boldsymbol{(00)}$  & \textbf{ii:}~ $\boldsymbol{(01)}$  & \textbf{iii:} $\boldsymbol{(11)}$
        \\ \hline\hline
        \textbf{i:}~~ $\boldsymbol{(00)}$ & $1$ & $1$ & $1$ & \textbf{i:}~~ $\boldsymbol{(00)}$ & $0$ & $0$ & $0$
        \\ \hline 
        \textbf{ii:}~ $\boldsymbol{(10)}$ & $1$ & $3$ & $2$ & \textbf{ii:}~ $\boldsymbol{(10)}$ & $0$ & $1$ & $1$ 
        \\ \hline
        \textbf{iii:} $\boldsymbol{(11)}$ & $1$ & $2$ & $3$ & \textbf{iii:} $\boldsymbol{(11)}$ & $0$ & $1$ & $3$
        \\ \hline
    \end{tabular}
    \\ \\[-2.9ex] 
    \begin{tabular}{|l|| c|c|c|}
        \hline 
        $\boldsymbol{m_{3456}}$ & \multicolumn{3}{l|}{\textbf{E:}\, $\boldsymbol{(1001)}$} 
        \\ \hline 
        \!$\boldsymbol{\bar{n}_{12}}$ \!\!$\boldsymbol{\Big\backslash}$\!\! $\boldsymbol{n_{12}}$\! & \textbf{i:}~~ $\boldsymbol{(00)}$  & \textbf{ii:}~ $\boldsymbol{(01)}$  & \textbf{iii:} $\boldsymbol{(11)}$ 
        \\ \hline\hline
        \textbf{i:}~~ $\boldsymbol{(00)}$ & $1$ & $1$ & $1$
        \\ \cline{1-4}  
        \textbf{ii:}~ $\boldsymbol{(10)}$ & $1$ & $2$ & $2$
        \\ \cline{1-4} 
        \textbf{iii:} $\boldsymbol{(11)}$ & $1$ & $2$ & $2$
        \\ \cline{1-4} 
    \end{tabular}            
    \end{tabular}
    \caption{The number of solutions for each of the possible configurations of the asymmetric $\Intr_{2L}\times\Intr_{2R}\times\Intr_2$ orbifolds parametrised by $n_{12}$, $\bar{n}_{12}$ and $m_{3456}$; $34$ in total.}
    \label{tab:OverviewZ2LZ2RZ2models}
\end{table}

\renewcommand{\arraystretch}{1.2}
\begin{table}
    \centering
    \begin{tabular}{|l|| c|c|c|c|c|c| }
        \hline 
        \!\!\tabu{c}{$\boldsymbol{\bar{n}_{12}}$ \\ $\boldsymbol{\bar{m}_{3456}}$}\!\! $\boldsymbol{\Big\backslash}$ \!\!\tabu{c}{$\boldsymbol{n_{12}}$ \\ $\boldsymbol{m_{3456}}$}\!\!\!\!\!  & \textbf{I:}\!\!\tabu{c}{$\boldsymbol{(00)}$ \\ $\boldsymbol{(0000)}$}\!\!\!\! & \textbf{II:}\!\!\tabu{c}{$\boldsymbol{(00)}$ \\ $\boldsymbol{(1010)}$}\!\!\!\! & \textbf{III:}\!\!\tabu{c}{$\boldsymbol{(00)}$ \\ $\boldsymbol{(1111)}$}\!\!\!\! & \textbf{IV:}\!\!\tabu{c}{$\boldsymbol{(10)}$ \\ $\boldsymbol{(1011)}$}\!\!\!\! & \textbf{V:}\!\!\tabu{c}{$\boldsymbol{(11)}$ \\ $\boldsymbol{(1111)}$}\!\!\!\! & \textbf{VI:}\!\!\tabu{c}{$\boldsymbol{(11)}$ \\ $\boldsymbol{(1100)}$}\!\!\!\!
        \\ \hline\hline
        \textbf{I:}~~~ \!\tabu{c}{$\boldsymbol{(00)}$ $\boldsymbol{(0000)}$}\!\! & 1 & 0 & 0 & 0 & 0 & 0 
        \\ \hline
        \textbf{II:}~~ \!\!\tabu{c}{$\boldsymbol{(00)}$ $\boldsymbol{(1010)}$}\!\! & 0 & 1 & 0 & 1 & 0 & 0 
        \\ \hline
        \textbf{III:}~ \!\!\tabu{c}{$\boldsymbol{(00)}$ $\boldsymbol{(1111)}$}\!\! & 0 & 0 & 1 & 1 & 0 & 0 
        \\ \hline
        \textbf{IV:}~ \!\!\tabu{c}{$\boldsymbol{(10)}$ $\boldsymbol{(1011)}$}\!\! & 0 & 1 & 1 & 3 & 1 & 0 
        \\ \hline
        \textbf{V:}~~ \!\!\tabu{c}{$\boldsymbol{(11)}$ $\boldsymbol{(1111)}$}\!\! & 0 & 0 & 0 & 1 & 1 & 0 
        \\ \hline
        \textbf{VI:}~ \!\!\tabu{c}{$\boldsymbol{(11)}$ $\boldsymbol{(1100)}$}\!\! & 0 & 0 & 0 & 0 & 0 & 3 
        \\ \hline 
    \end{tabular}
    \caption{The number of solutions for each of the possible configurations of the asymmetric $\Intr_{2L}^2\times\Intr_{2R}^2$ orbifolds parametrised by $n_{12},m_{3456}$ and $\bar{n}_{12},\bar{m}_{3456}$; $16$ in total. }
    \label{tab:OverviewZ2L2Z2R2models}
\end{table}

\clearpage 


Just like the previous case, also here some of configurations obtained this way, still turn out to be equivalent to one another, in which case the configuration higher up in the list is taken as the unique representative. The number of inequivalent modular invariant configurations are listed in \cref{tab:OverviewZ2L2Z2Rmodels}; some configuration do not admit any solutions; while others multiple that are then simply enumerated. The resulting inequivalent configurations have been tabulated in \Cref{tab:NarainRotoTransClass}.

\subsubsection*{Asymmetric $\boldsymbol{\Intr_{2L}\times \Intr_{2R} \times \Intr_2}$ orbifolds}

The subgroups of the point group $\Intr_{2L}\times\Intr_{2R}\times \Intr_2$ relevant for the classification are $\Intr_{2L}$, $\Intr_{2R}$ and $\Intr_2$. As suggested in the previous subsection, the basis vectors can be taken to be $\bm{B}_1$, $\bm{B}_{\bar{1}}$ and $\bm{B}_{2\bar{2}}$. The $\Intr_2$ basis vector $\bm{B}_{2\bar{2}}$ are distinguished by $m=\bar{m}=(0^4), (0^210), (100^2), (0^21^2), (1010)$. The basis vectors $\bm{B}_1, \bm{B}_{\bar{1}}$ is parametrised by $n=(n_{12}), \bar{n}=\bar{n}_{12}$ and $N, \bar{N}$ with $n_{12},\bar{n}_{12}=(0^2), (10), (1^2)$. Thus, only $N, \bar{N}$ are subject to the modular invariance conditions, with their lengths equal to that of $n_{12}$ and $\bar{n}_{12}$, respectively. Their representations are not unique, but since the three two--torus directions are distinguished, the redundancies due to the equivalence relations are greatly reduced. Moreover, if $m$ has a non--zero entry in the 3 or 4 directions, but not both, then these two directions are distinguished (and similar for the 5,6 directions).

Again, the so obtained list of configurations need to be checked for possible equivalences which are to be removed. 
The resulting inequivalent configurations are tabulated \Cref{tab:NarainRotoTransClass}. It is important to emphasise that for this and the other two asymmetric orbifolds the two loop modular invariance conditions, third equation in \eqref{eq:12loopModInv}, become restrictive. The number of inequivalent modular invariant realisations per configurations are listed in \cref{tab:OverviewZ2LZ2RZ2models}. \Cref{tab:NarainRotoTransClass} tabulates all the resulting inequivalent choices.

\subsubsection*{Asymmetric $\boldsymbol{\Intr_{2L}^2\times \Intr_{2R}}^2$ orbifolds}

The subgroups of the point group $\Intr_{2L}^2\times \Intr_{2R}^2$ relevant for the classification are $\Intr_{2L}^2$ and $\Intr_{2R}^2$. For the parametrisations of the elements of both  $\Intr_{2L}^2$ and $\Intr_{2R}^2$ can be taken from the last two columns of the ``Parameterisation'' columns of \cref{tab:ClassificationZ2L2models} and can be
 augmented with vectors $N, M$, and $\bar{N}, \bar{M}$. This results in $6\cdot 6=36$ configurations characterised by $n_{12},m_{3456}$ and $\bar{n}_{12}$. For each of them all possible inequivalent choices for $N, M$ and $\bar{N}, \bar{M}$ are considered and checked whether they result in modular invariant models according to the relevant conditions from \eqref{eq:AsymmetricModInv}. The resulting list has be cleaned by removing all equivalent configurations. The number of inequivalent modular invariant realisations per configurations are listed in \cref{tab:OverviewZ2L2Z2R2models}. The resulting inequivalent configurations have been tabulated in \Cref{tab:NarainRotoTransClass}.

\section{Supersymmetric twisted sectors of order--two T--folds}
\label{sc:TwistedSectors}

The twisted spectra of order--two T--folds have some special properties as compared to symmetric orbifolds. In particular, asymmetric twist vectors can give rise to twisted sectors that contain Rarita--Schwinger (spin--3/2) fields. This section is devoted to give a detailed description of the structure of the twisted supersectors. Like for symmetric order--two orbifolds, the twisted sectors of order--two T--folds define six dimensional states that fall into multiplets of six dimensional (extended) supersymmetries. On these sectors various projections act to form four dimensional states. The projections typically depend on certain combinations of GGSO phases. To describe all these cases simultaneously, the projected twisted spectra are given in parametric form in \cref{tab:NarainRotoTransClass} for all configurations of order--two T--folds on the $SO(12)$ lattice at the free fermionic point.

\renewcommand{\arraystretch}{1.5}
\begin{table}[t]
    \centering
    \begin{tabular}{|l|cc||l|cc|}
        \hline
        \multicolumn{1}{|c|}{\textbf{Sector}} & \multicolumn{2}{c||}{\textbf{$\boldsymbol{L/R}$--moving masses}} &
            $\bm{B}_{1\bar{1}}=\bm{B}_1+\bm{B}_{\bar{1}}$ &  $\frac{[n+\bN]_{12}^2}{8}$ & $\frac{[\bn+N]_{12}^2}{8}$
        \\ \cline{4-6}
        \multicolumn{1}{|c|}{$\bm{B}$} & \ \ \ $\boldsymbol{M_L^2(\bm{B})}$\ \ \  &  \ \ \ $\boldsymbol{M_R^2(\bm{B})}$\ \ \ & 
            $\bm{B}_{1\bar{2}}=\bm{B}_1+\bm{B}_{\bar{2}}$ &  $\frac{[n+\bM]_{12}^2}{8}$ &  $\frac{[\bmm+N]_{34}^2}{8}$
        \\ \cline{1-3}\cline{4-6}
        $\bm{B}_1$ & $\frac{n_{12}^2}{8}$ & $ -\frac{1}{2}+\frac{N^2}{8}$  &
            $\bm{B}_{2\bar{1}}=\bm{B}_2+\bm{B}_{\bar{1}}$ &  $\frac{[m+\bN]_{34}^2}{8}$ & $\frac{[\bn+M]_{12}^2}{8}$  \\ \hline
        $\bm{B}_2$ & $\frac{m_{34}^2}{8}$ & $ -\frac{1}{2}+\frac{M^2}{8}$ &
            $\bm{B}_{2\bar{2}}=\bm{B}_2+\bm{B}_{\bar{2}}$ &  $\frac{[m+\bM]_{34}^2}{8}$ & $\frac{[\bmm+M]_{34}^2}{8}$   \\ \hline
        $\bm{B}_3=\bm{B}_1+\bm{B}_2$ & $\frac{m_{56}^2}{8}$ & $ -\frac{1}{2}+\frac{(N-M)^2}{8}$   &
            $\bm{B}_{1\bar{3}}=\bm{B}_1+\bm{B}_{\bar{3}}$ & $\frac{[n+\bN+\bM]_{12}^2}{8}$ & $\frac{[\bmm+N]_{56}^2}{8}$ \\ \hline
        $\bm{B}_{\bar{1}}$ & $ -\frac{1}{2}+\frac{\bN^2}{8}$  & $\frac{\bn_{12}^2}{8}$  &
            $\bm{B}_{2\bar{3}}=\bm{B}_2+\bm{B}_{\bar{3}}$ &  $\frac{[m+\bM+\bN]_{34}^2}{8}$ &  $\frac{[\bmm+M]_{56}^2}{8}$  \\ \hline
        $\bm{B}_{\bar{2}}$ & $ -\frac{1}{2}+\frac{\bM^2}{8}$  & $\frac{\bmm_{34}^2}{8}$ & 
            $\bm{B}_{3\bar{1}}=\bm{B}_3+\bm{B}_{\bar{1}}$ &  $\frac{[m+\bN]_{56}^2}{8}$ & $\frac{[\bn+N+M]_{12}^2}{8}$ \\ \hline
       $\bm{B}_{\bar{3}}=\bm{B}_{\bar{1}}+\bm{B}_{\bar{2}}$ & $ -\frac{1}{2}+\frac{(\bN-\bM)^2}{8}$  & $\frac{\bmm_{56}^2}{8}$ &
           $\bm{B}_{3\bar{2}}=\bm{B}_3+\bm{B}_{\bar{2}}$ &  $\frac{[m+\bM]_{56}^2}{8}$ &  $\frac{[\bmm+N+M]_{34}^2}{8}$ \\ \hline 
        \multicolumn{3}{c|}{} & $\bm{B}_{3\bar{3}}=\bm{B}_3+\bm{B}_{\bar{3}}$ &  $\frac{[m+\bN+\bM]_{56}^2}{8}$ & $\frac{[\bmm+N+M]_{56}^2}{8}$
        \\ \cline{4-6}
    \end{tabular}
    \caption{The left-- and right--moving masses of the vacuum states of the twisted sectors $\ga, \bgb, \ga\bgb$. The first six sectors $\ga$ and $\bgb$ (associated to the vectors $\bm{B}_\ga$ and $\bm{B}_{\bgb}$) can give rise to massless six dimensional (2,1) Rarita-Schwinger multiplets. The sectors $\ga'$, $\bgb'$, $\ga\bgb$, and $\ga\bgb'$ (associated to $\bm{B}_{\ga}'$, $\bm{B}_{\bgb}'$, $\bm{B}_{\ga\bgb}$, and $\bm{B}_{\ga\bgb}'$) can give rise to massless six dimensional (2,0) self--dual tensor or (1,1) vector multiplets depending on whether they reside in type IIB or IIA, respectively.}
    \label{tab:twistedsecs}
\end{table}

\subsection{Parameterisation of twisted sectors}

All possible linear combinations involving the $\Intr_2$ twist basis vectors $\bm{B}_1,\bm{B}_2,\bm{B}_{\bar{1}},\bm{B}_{\bar{2}},\bm{B}_{1\bar{1}},\bm{B}_{2\bar{2}}$ generate the twisted sectors. The twisted sectors associated to vectors $\bm{B}_{\ga}$, $\bm{B}_{\bgb}$, $\bm{B}_{\ga\bgb}$, $\bm{B}_{\ga}'=\bm{B}_{\ga}+\bm{E}$, $\bm{B}_{\bgb}'=\bm{B}_{\bgb}+\bm{E}$, $\bm{B}_{\ga\bgb}'=\bm{B}_{\ga\bgb}+\bm{E}$ are denoted as  $\ga$, $\bgb$, $\ga\bgb$, $\ga'$, $\bgb'$, $\ga\bgb'$, respectively, for short. In \cref{sc:6Dtwisted} these twist vectors are taken as the basis for the twisted supersectors. The twist basis vectors can be combined as follows 
\equ{ \label{eq:ExpansionsTwistVectors}
    \arry{c}{
    \bm{B}_\ga = \ga_1\, \bm{B}_1 + \ga_2\, \bm{B}_2~, 
    \qquad
    \bm{B}_{\bgb} = \bgb_1\, \bm{B}_{\bar{1}} + \bgb_2\, \bm{B}_{\bar{2}}~, 
    \qquad 
    \bm{B}_{\ga\bgb} = \bm{B}_\ga + \bm{B}_\bgb~, 
    \\[1ex]
    \bm{B}_\ga' = \bm{B}_\ga + \bm{E}~, 
    \qquad
    \bm{B}_{\bgb}' = \bm{B}_{\bgb} +  \bm{E}~, 
    \qquad
    \bm{B}_{\ga\bgb}' = \bm{B}_{\ga\bgb} + \bm{E}~,
    }
}
with $\ga_1,\ga_2, \bgb_1,\bgb_2=0,1$ such that $\ga=(\ga_1,\ga_2)$ and $\bgb=(\bgb_1,\bgb_2)$ label $1,2,3$ in binary as $01, 10, 11$, respectively. For symmetric vectors we write $\bm{B}_{\ga\bga} = \ga_1\, \bm{B}_{1\bar{1}}+\ga_2\,\bm{B}_{2\bar{2}}$ such that $\ga_1=\bga_1$ and $\ga_2=\bga_2$. As many properties of symmetric twist vectors are identical to that of $\bm{B}_{\ga\bgb}$, they are not treated separately.

\subsection{Ground state twisted masses}

Many of these twisted sectors do not give rise to massless states. To determine which do, the $L/R$--moving masses $(M_L^2,M_R^2)$ of the vacuum states of all the twisted sectors are listed in Table \ref{tab:twistedsecs}. The masses of the primed sectors $\bm{B}'=\bm{B}+\bm{E}$, obtained by adding the $SO(12)$ symmetry vector $\bm{E}=\one-\Sv-\overline{\Sv}$, are not tabulated separately in this table as they follow from the unprimed ones. Explicitly, adding $\bm{E}$ shifts the masses according to 
\equ{
    M_L^2(\bm{B}'_\ga)=\tfrac14-M_L^2(\bm{B}_\ga)~, \qquad 
    M_R^2(\bm{B}'_\ga)=-\tfrac14-M_R^2(\bm{B}_\ga)~,
} 
with $L\leftrightarrow R$ for the barred sectors $\bgb$, and for the mixed sectors to 
\equ{ 
    M_{L/R}^2(\bm{B}'_{\ga\bgb})=\tfrac14-M_{L/R}^2(\bm{B}_{\ga\bgb})
}
on both sides, where $\ga,\bgb=1,2,3$.

If the findings of A1)--A4) are combined with the modular invariance conditions, it follows that $N, M, \bN$ and $\bM$ have at most two non--zero entries so that a number of modular invariance conditions become strict identities instead of mod conditions. In particular, 
\equ{
    n_{12}^2 = N^2~, 
    \qquad 
    m_{34}^2 = M^2~, 
    \qquad 
    \bn_{12}^2 = \bN^2~, 
    \qquad 
    \bmm_{34}^2 = \bM^2~.
}
Similar relations for $m_{56}$ and $\bmm_{56}$ do not hold, because, for example, $N$ and $M$ could both have two entries equal to 1 that do not overlap, then $(N-M)^2 = 4$ which is consistent with $m_{56}=0$ mod 4. Hence there are two options for $m_{56}=0$: either $N=M=0$ or $N$ and $M$ both have two non--zero entries which do not overlap. The key implications of Table \ref{tab:twistedsecs} 
can be summarised as follows: 
\items{
\item 
If there is shift in a single direction that a twist vector does not act as a twist, 
then the vacuum state of the corresponding sector is never massless. 
\item 
The vacuum state of the sector $\ga$ 
has a tachyonic $R$--moving mass and a zero $L$--moving mass (or vice versa for $\bgb$) 
if there are no shifts in the directions that $\bm{B}_\ga$ or $\bm{B}_{\bgb}$ do not act as a twist. 
\item 
The vacuum state of the sector $\ga'$ (or $\bgb'$) is massless 
if there are shifts in both directions that $\bm{B}_{\ga}$ (or $\bm{B}_{\bgb}$) do not act as a twist. 
\item 
The vacuum states of sectors $\ga\bgb$ are massless 
if there are no shifts in the directions that $\bm{B}_{\ga\bgb}$ do not act as a twist. 
\item 
The vacuum state of sectors $\ga\bgb'$ are massless 
if there are shifts in both directions that $\bm{B}_{\ga\bgb}$ do not act as a twist.  
}
Notice that these findings imply that these sectors $\ga$ and $\ga'$, or $\bgb$ and $\bgb'$, or $\ga\bgb$ and $\ga\bgb'$ can never both be massless simultaneously.

\subsection{Massless six dimensional twisted supermultiplets}
\label{sc:6Dtwisted}

All twisted sectors in order two orbifold models could be considered as six dimensional subsectors of the full theory. Each twisted sector is associated to a twist vector $\Bgb$ in the additive set $\mathscr{B}$. Aside from the mappings \eqref{eq:SusyMappings}, there is a second way that supermultiplets may be formed: if a vector $\Bgb$ in the additive set has four out of the eight fermions overlapping with $\bm{S}$ or $\bar{\bm{S}}$, then 
\equ{ \label{eq:SusyMappings4DII}
\big| \Bgb \big\rangle \leftrightarrow \big| \Bgb + \bm{S}\big\rangle~, 
\qquad 
\big| \Bgb \big\rangle \leftrightarrow \big| \Bgb + \bar{\bm{S}}\big\rangle~. 
}
For asymmetric constructions it is possible that certain sectors make use of different options on the holomorphic and anti--holomorphic sides. Below the resulting supermultiplets are worked out for all the twisted sectors that are massless and it will be shown that they fall in multiplets of $(2,1)$, $(2,0)$, $(1,1)$ and $(1,0)$ supersymmetry. \Cref{tab:TwistedSectorsMultiplets} gives an overview of which sector give rise to which multiplet.

\subsubsection*{$\boldsymbol{(2,1)}$ Rarita--Schwinger multiplets}

The massless twisted sector $\bm{B}_\ga$ forms a $(2,1)$ Rarita--Schwinger multiplet for both the type II theories. The vacuum state $\big|\bm{B}_\ga \big\rangle$ has four $\gch$'s and four $y$ or $w$ fermions and no anti--holomorphic fermions; it is therefore not level matched by itself. The massless fields in target space arise from the following collection of level matched states in the supersector: 
\equ{ \label{eq:RS}
\big|\big|\bm{B}_\ga \big\rangle\!\big\rangle = 
\Big\{~
\bgps^M \big|\bm{B}_\ga \big\rangle~, 
\quad
\bgps^M \big|\bm{B}_\ga +\bm{S}\big\rangle~, 
\quad
\big|\bm{B}_\ga + \bar{\bm{S}}\big\rangle~, 
\quad
\big|\bm{B}_\ga + \bm{S}+\bar{\bm{S}}\big\rangle~\Big\}~.
}
Because of 
\equ{\label{eq:SbSTwistVectors}
    \arry{c}{
    \bm{S}\cdot \bm{B}_{\ga\bgb} = \bm{S}\cdot \bm{B}_{\ga} = 2(1+\gd_{\ga 3})~, 
    \qquad 
    \bar{\bm{S}}\cdot \bm{B}_{\ga\bgb} = \bar{\bm{S}}\cdot \bm{B}_{\bgb} =-2(1+\gd_{\bgb 3})~, 
    \\[1ex]
    \bm{S}\cdot \bm{B}_{\ga\bgb}' = \bm{S}\cdot \bm{B}_{\ga}' = 2(1+\gd_{\ga 3})~, 
    \qquad 
    \bar{\bm{S}}\cdot \bm{B}_{\ga\bgb}' = \bar{\bm{S}}\cdot \bm{B}_{\bgb}' =-2(1+\gd_{\bgb 3})~, 
    }
}
\eqref{eq:CCinterchange} and \eqref{eq:4DGSOtypeII}, it follows that 
\equ{ \label{eq:BSphases}
\CC{\bm{B}_{\ga}}{\bm{S}} = (-1)^{\gd_{\ga 3}}~, 
\qquad 
 \CC{\bm{B}_{\ga}}{\bar{\bm{S}}} = -1~. 
}
This means that these vacuum states are all projected out and at the same time the states in \eqref{eq:RS} are all kept. The string states \eqref{eq:RS} corresponds to $(2,1)$ Rarita--Schwinger multiplet in six dimensions given in \cref{tab:6DMultiplets} as the counting of states in appendix~\ref{app:21RSstates} shows. 

\subsubsection*{$\boldsymbol{(1,0)}$ Hyper multiplets?}

The vector $B_\ga'$ contains four $\gch$'s and four $y$'s or $w$'s and eight $\byy$'s or $\bw$'s in order to produce massless states. As a consequence no $\bar{\bm{S}}$ can be added to this vector and the multiplet structure is reduced to the following collection of states in the supersector: 
\equ{ \label{eq:Hyper}
\big|\big|\bm{B}_{\ga}' \big\rangle\!\big\rangle = 
\Big\{~ 
\big|\bm{B}_{\ga}' \big\rangle~,
\quad 
\big|\bm{B}_{\ga}' + \bm{S}\big\rangle~\Big\}~. 
}
Combined these states would constitute sixteen hyper multiplets of $\cN=(1,0)$ supersymmetry in six dimensions as the counting of states in appendix~\ref{app:10Hstates} shows. 

However, all the basis vectors by themselves preserves at least $\cN=(2,0)$ or $(1,1)$ (as can be inferred from \cref{tab:NarainPointGroupClass} since each of these basis vector defines a $\Intr_{2L}$, $\Intr_{2R}$ or $\Intr_2$ action). But these amounts of supersymmetry do not admit hyper multiplets, see {\em e.g.} \cref{tab:6DMultiplets}. The theory solves this apparent paradox in an elegant way: since $\CC{\bm{B}_\ga'}{\bar{\bm{S}}}=-1$, all these state are projected out as long as the twist basis vector $\bm{B}_\ga$ preserves supersymmetry on the anti--holomorphic side. Indeed, if the GGSO phases are chosen such that the maximal amount of supersymmetry is preserved, then $\bm{B}_\ga$ should preserve $(2,1)$ supersymmetry.

Also the sector $\bm{B}_3$ within models with a $\Intr_{2L}^2$ factor of type VI is massless on the left-- and right--moving side (as for these configurations type $(N-M)^2=4$ see \cref{tab:ClassificationZ2L2models}). Hence this sector would also give rise to $(1,0)$ hyper multiplets. For the same reason, as discussed just above, these states are completely projected out as well.

\subsubsection*{$\boldsymbol{(2,0)}$ Self--dual tensor multiplets}

The twist vector $\bm{B}_{\ga\bgb}$ contains four $\gch$'s, four $y$'s or $w$'s, four $\bgch$'s and four $\byy$'s or $\bw$'s in order that this sector is massless. This supersector then contains the following collection of states
\equ{ \label{eq:TensorVector}
\big|\big|\bm{B}_{\ga\bgb} \big\rangle\!\big\rangle = \Big\{~ 
\big|\bm{B}_{\ga\bgb} \big\rangle~,
\quad 
\big|\bm{B}_{\ga\bgb} + \bm{S}\big\rangle~,
\quad 
\big|\bm{B}_{\ga\bgb} +\bar{\bm{S}}\big\rangle~, 
\quad 
\big|\bm{B}_{\ga\bgb} + \bm{S}+\bar{\bm{S}}\big\rangle~\Big\}~, 
}
in the type IIB theory. All these states are massless, but they are only part of the physical spectrum if they survive the GGSO projections \eqref{eq:GGSOprojections}.  Because of \eqref{eq:CCinterchange},  \eqref{eq:SbSTwistVectors} and \eqref{eq:4DGSOtypeII}, it follows that 
\equ{
\CC{\bm{B}_{\ga\bgb}}{\bm{S}} = (-1)^{\gd_{\ga 3}}~, 
\qquad 
 \CC{\bm{B}_{\ga\bgb}}{\bar{\bm{S}}} = (-1)^{\gd_{\bgb 3}}~. 
}
The string states \eqref{eq:TensorVector} corresponds to eight $(2,0)$ tensor multiplets in six dimensions given in \cref{tab:6DMultiplets} as the counting of states in appendix~\ref{app:TVstates} shows.

\subsubsection*{$\boldsymbol{(1,1)}$ Vector multiplets} 

The string states of the twisted sector $\bm{B}_{\ga\bgb}$ for the type IIA theory are the same as those of the IIB given in \cref{eq:TensorVector} but now result in $(1,1)$ vector mutiplets, see again appendix~\ref{app:TVstates}.

\subsubsection*{Summary}

In \cref{tab:TwistedSectorsMultiplets}, the structure of the massless twisted sectors are summarised. Their multiplets are given as representations of extended supersymmetry in six dimensions and their dimensional reduction to four dimensions. Note, in particular, that the distinction between the six dimensional $(2,0)$ tensor and $(1,1)$ disappears during this reduction and they both give rise to $\mathcal{N}=4$ vector multiplets in four dimensions.

\renewcommand{\arraystretch}{1.5}
\begin{table}[t]
    \centering
    \begin{tabular}{|c| ccl| ccl| ccl|}
        \hline 
        \textbf{Sector} & \multicolumn{6}{c|}{\textbf{Six / four dimensional multiplets}} 
        \\ 
         $\bm{B}$ & \multicolumn{3}{c|}{\textbf{IIB}} & \multicolumn{3}{c|}{\textbf{IIA}}  
        \\ \hline\hline
        $\bm{B}_\ga$ & $2$ &  $(2,1)$ & Rarita--Schwinger & $2$ & $(2,1)$ & Rarita--Schwinger   \\ 
        & $\cong$ & $\mathcal{N}=6$ & Rarita--Schwinger & $\cong$ & $\mathcal{N}=6$ & Rarita--Schwinger 
        \\ \hline
        $\bm{B}_{\bar{\ga}}$ & $2$ & $(2,1)$ & Rarita--Schwinger & $2$ & $(1,2)$  & Rarita--Schwinger     \\
        & $\cong$ & $\mathcal{N}=6$ & Rarita--Schwinger & $\cong$ & $\mathcal{N}=6$ & Rarita--Schwinger 
        \\ \hline
        $\bm{B}_{\ga\bar{\gb}}$ & $8$ & $(2,0)$ & Tensor & $8$ & $(1,1)$ & Vector   \\ 
        & $\cong$ & $\mathcal{N}=4$ & Vector & $\cong$ & $\mathcal{N}=4$ & Vector  
        \\ \hline
        $\bm{B}_{\ga\bar{\gb}}'$ & $8$ & $(2,0)$ & Tensor & $8$ & $(1,1)$ & Vector   \\ 
        & $\cong$ & $\mathcal{N}=4$ & Vector & $\cong$ & $\mathcal{N}=4$ & Vector 
        \\ \hline
    \end{tabular}
    \caption{The six dimensional supersymmetry multiplets of massless twisted sectors for the type IIA/B theories and the dimensional reduction to four dimensions. The sectors $\bm{B}_3$ or $\bm{B}_{\bar{3}}$ are empty in theories of the configuration type VI of the $\Intr_{2L}^2$ or  $\Intr_{2R}^2$ factors, respectively.}
    \label{tab:TwistedSectorsMultiplets}
\end{table}

\subsection{Twist projections of the twisted sectors}
\label{sc:ProjectedTwisted}

The six dimensional multiplets of various levels of extended supersymmetry are projected to four dimensions by the action of the other twist basis vectors. This reduces the amount of supersymmetry and consequently leads to a branching of the multiplets to the representations of the leftover supersymmetries. In addition, each projection associated to a twist basis vector has basically four possible consequences as far as the massless twisted sectors are concerned:
\items{
    \item halves the multiplicity factor of all states; 
    \item selects certain multiplets in the branching;   
    \item projects out all states; 
    \item acts trivially. 
} 
Generically the first possibility occurs because the twist leads to some correlation of the internal spins associated to $y$'s, $w$'s, $\byy$'s or $\bw$'s and the other quantum numbers, hence the halving of the multiplicities. Therefore, only if there is no overlap of the projecting twist basis vector with the state on these internal fermions, the target space--time properties are affected, which results in the selection of particular multiplets. The final two options can happen if different projections act on the internal fermions $y's$ and $w's$ in an identical way: if the GGSO phases are compatible the projectors act in the same fashion so that the second projector acts trivially. If the GGSO phases are opposite, the projections are incompatible and no states survive.

\begin{table}[t]
\centering
\renewcommand{\arraystretch}{1.5}
\begin{tabular}{| c| c| l|}
\hline 
\textbf{Point group} & \textbf{Type} & \textbf{GGSO phases} \\ \hline\hline 
$\Id$ & -- & -- \\ \hline
$\Intr_2$ &  all & -- \\ \hline 
$\Intr_2^2$ & all & 
$\CC{\bm{B}_{1\bar{1}}}{\bm{B}_{2\bar{2}}} = \CC{\bm{B}_{2\bar{2}}}{\bm{B}_{1\bar{1}}}$  \\ \hline 
$\Intr_{2L}$ & all & -- \\ \hline
$\Intr_{2L}\times \Intr_{2R}$ & all & 
$\CC{\bm{B}_{1}}{\bm{B}_{\bar{1}}} = \CC{\bm{B}_{\bar{1}}}{\bm{B}_{1}}$ \\ \hline
$\Intr_{2L}^2$ & I$\cdots$\!V & 
$\CC{\bm{B}_1}{\bm{B}_2} = - \CC{\bm{B}_2}{\bm{B}_1}$  \\ \cline{2-3}
& VI &  $\CC{\bm{B}_1}{\bm{B}_2} = \CC{\bm{B}_2}{\bm{B}_1}$ \\ \hline 
$\Intr_{2L} \times \Intr_2$ & all & 
$\CC{\bm{B}_{1}}{\bm{B}_{2\bar{2}}} = -\CC{\bm{B}_{2\bar{2}}}{\bm{B}_{1}} $  \\ \hline 
$\Intr_{2L}^2\times \Intr_{2R}$ & I$\cdots$\!V--i,ii,iii & 
$\CC{\bm{B}_1}{\bm{B}_2} = - \CC{\bm{B}_2}{\bm{B}_1}$,~  
$\CC{\bm{B}_{1}}{\bm{B}_{\bar{1}}} = \CC{\bm{B}_{\bar{1}}}{\bm{B}_{1}}$,~ 
$\CC{\bm{B}_{2}}{\bm{B}_{\bar{1}}} = \CC{\bm{B}_{\bar{1}}}{\bm{B}_{2}}$ \\ \cline{2-3}
& VI--iii & 
$\CC{\bm{B}_1}{\bm{B}_2} = - \CC{\bm{B}_2}{\bm{B}_1}$,~  
$\CC{\bm{B}_{1}}{\bm{B}_{\bar{1}}} = \CC{\bm{B}_{\bar{1}}}{\bm{B}_{1}}$,~ 
$\CC{\bm{B}_{2}}{\bm{B}_{\bar{1}}} = \CC{\bm{B}_{\bar{1}}}{\bm{B}_{2}}$ \\ \hline 
$\Intr_{2L}\times \Intr_{2R} \times \Intr_2$ & all & 
$\CC{\bm{B}_{1}}{\bm{B}_{\bar{1}}} = \CC{\bm{B}_{\bar{1}}}{\bm{B}_{1}}$,~ 
$\CC{\bm{B}_{1}}{\bm{B}_{2\bar{2}}} = -\CC{\bm{B}_{2\bar{2}}}{\bm{B}_{1}}$,~ 
$\CC{\bm{B}_{\bar{1}}}{\bm{B}_{2\bar{2}}} = -\CC{\bm{B}_{2\bar{2}}}{\bm{B}_{\bar{1}}}$ \\ \hline
$\Intr_{2L}^2\times \Intr_{2R}^2$ & all &  
$\CC{\bm{B}_{1}}{\bm{B}_{\bar{1}}} = \CC{\bm{B}_{\bar{1}}}{\bm{B}_{1}}$,~
$\CC{\bm{B}_{2}}{\bm{B}_{\bar{1}}} = \CC{\bm{B}_{\bar{1}}}{\bm{B}_{2}}$,~ 
$\CC{\bm{B}_{1}}{\bm{B}_{\bar{2}}} = \CC{\bm{B}_{\bar{2}}}{\bm{B}_{1}}$,~ 
$\CC{\bm{B}_{2}}{\bm{B}_{\bar{2}}} = \CC{\bm{B}_{\bar{2}}}{\bm{B}_{2}}$ \\ \cline{2-3}
& I$\cdots$\!V--I$\cdots$\!V & 
$\CC{\bm{B}_1}{\bm{B}_2} = - \CC{\bm{B}_2}{\bm{B}_1}$,~ 
$\CC{\bm{B}_{\bar{1}}}{\bm{B}_{\bar{2}}} = - \CC{\bm{B}_{\bar{2}}}{\bm{B}_{\bar{1}}}$ \\ \cline{2-3}
& VI--VI & 
$\CC{\bm{B}_1}{\bm{B}_2} = \CC{\bm{B}_2}{\bm{B}_1}$,~ 
$\CC{\bm{B}_{\bar{1}}}{\bm{B}_{\bar{2}}} = \CC{\bm{B}_{\bar{2}}}{\bm{B}_{\bar{1}}}$ 
\\ \hline
\end{tabular}
\caption{\label{tab:FixedFreeGGSOphases} Overview of the symmetry properties of the remaining physical GGSO phases per (a)symmetric orbifold.}
\end{table}

\subsubsection*{Free GGSO phases}

As discussed above, the GGSO phases involving the basis vectors $\bm{S}$ and $\bar{\bm{S}}$ are fixed by demanding that the constructions preserve the maximal amount of supersymmetry possible. We also recall that the sign of the phases $\CC{\bm{S}}{\bar{\bm{S}}}$ decides if the T--fold is considered in the type IIA or IIB context. Meanwhile, the phases with $\bm{E}$ fix the internal chiralities and therefore do not affect the spacetime properties of the states. In this work these phases are chosen to be fixed by \eqref{eq:ConventionCC_E}. 

The remaining free phases involve the twist basis vectors only. An overview of the unfixed GGSO phases for each of the asymmetric orbifolds on the $SO(12)$ lattice is given in \cref{tab:FixedFreeGGSOphases}. For most point groups the phases are either always symmetric or anti--symmetric for all types. Only for the orbifolds with point groups that involve a $\Intr_{2L}^2$ or $\Intr_{2R}^2$ factor, the symmetry properties of the phases $\CC{\bm{B}_{\ga}}{\bm{B}_{\gb}}$ and $\CC{\bm{B}_{\bga}}{\bm{B}_{\bgb}}$ depend on the configuration type: 
\equ{
    \CC{\bm{B}_{\gb}}{\bm{B}_{\ga}} = 
    \begin{cases}
        -\CC{\bm{B}_{\ga}}{\bm{B}_{\gb}}~, & \text{I}\ldots\text{V}~, \\
        \phantom{-}\CC{\bm{B}_{\ga}}{\bm{B}_{\gb}}~, & \text{VI}~, 
    \end{cases}
    \qquad
    \CC{\bm{B}_{\bgb}}{\bm{B}_{\bga}} = 
    \begin{cases}
        -\CC{\bm{B}_{\bga}}{\bm{B}_{\bgb}}~, & \text{I}\ldots\text{V}~, \\
        \phantom{-}\CC{\bm{B}_{\bga}}{\bm{B}_{\bgb}}~, & \text{VI}~, 
    \end{cases}    
}
for $\gb\neq\ga$ and $\bgb\neq\bga$. When a GGSO phase is anti--symmetric, it can, in principle, by itself be fixed by a relabeling of the basis vectors involved. But since for most T--folds these basis vectors also appear in other GGSO phases, this would involve many redefinitions of all these phases as well, therefore we keep them as undetermined in our treatment.

\subsubsection*{Projected (2,1) Rarita--Schwinger multiplets}

Consider a massless six dimensional $(2,1)$ Rarita--Schwinger multiplet $||\bm{B}_\ga\rangle\!\rangle$. By further twist projections the remaining states can be classified according to $\mathcal{N}=4$ or $3$ supersymmetry in four dimensions for all models except the $\Intr_{2L}\times\Intr_{2R}$ T--folds. Even though one might expect the remaining states to organise themselves in $\cN=2$ multiplets, the states always pair up to multiplets of $\mathcal{N}=3$ supersymmetry at least. Consequently, in general this multiplet branches as 
\equ{
    \text{6D (2,1) RS} \ra \text{4D }\mathcal{N}=6\text{ RS} \ra  \text{4D }\mathcal{N}=4\text{ RS + V} \ra  \text{4D }\mathcal{N}=3\text{ RS + 3\,V}~. 
}
If $\bm{B}_{\bar{1}}$ and $\bm{B}_{\bar{2}}$ do not overlap with $\bm{B}_\ga$ specific representations are selected, then they can either be Rarita--Schwinger multiplets or vector multiplets. To identify which of the two options is selected, it is sufficient to determine if the spin--$3/2$ states $\bgPs_0|\bm{B}_\ga+\bm{S}\rangle$ are kept under the projections or not by the projections of $\bm{B}_{\bar{1}}$ or $\bm{B}_{\bar{2}}$. If $\bm{B}_{\bar{1}}$ or $\bm{B}_{\bar{2}}$ do overlap with $\bm{B}_\ga$, then there might be more projections that should halve the states: the twisted six dimensional $(2,1)$ Rarita--Schwinger multiplets $||\bm{B}_\ga\rangle\!\rangle$ come with a multiplicity of only two, while $\bm{B}_1$ or $\bm{B}_2$ and $\bm{B}_{\bar{1}}$ and $\bm{B}_{\bar{2}}$ together would naively lead to an eightfold ($1/2^3$) reduction. This is of course not possible, hence after the first halving of the states, the further projections either keep all remaining states or kick them all out. 

The same logic applies to $(2,1)$ or $(1,2)$ Rarita--Schwinger $||\bm{B}_{\bga}\rangle\!\rangle$ subject to the projectors $\bm{B}_{1}$ or $\bm{B}_{2}$ and also for the asymmetric $\Intr_{2L}\times\Intr_{2R}$ orbifolds defining six dimensional T--folds keeping in mind that then the relevant branchings are 
\equ{
    \text{6D (2,1) RS} \ra 
    \text{6D }\begin{cases} 
        (2,0)\text{ RS + T} & \text{IIB}
        \\[1ex]
        (1,1)\text{ RS + V} & \text{IIA}
    \end{cases}
}

 \subsubsection*{Projected (2,0) Tensor or (1,1) Vector multiplets}

The effects of further projections on the twisted (2,0) tensor or (1,1) vector multiplets $||\bm{B}_{\ga\bgb} \rangle\!\rangle$ of the type IIB or IIA theories, can be treated simultaneously. First of all observe that these sectors are $\cN=4$ vector multiplets viewed from a four dimensional perspective. 

For all point groups that preserve $\cN\geq 3$ supersymmetry, there are two options per projecting element: either it kicks out the sector altogether, or the multiplicity is halved. The resulting multiplicity of the states is always integral. (Even for the largest point group, $\Intr_{2L}^2\times\Intr_{2R}^2$, the multiplicities are divided by $2^3=8$ at most, because self--projections always act trivially or projects out the states completely.) 

For point groups that preserve only $\cN=2$, there is another possible effect of the projections induced by the basis vectors: the selection of $\cN=2$ vector or hyper multiplets. The easiest way, to see whether there is a selection and, if so, what is selected, is to determine whether four dimensional vector fields can appear or not. This can be investigated whether the four dimensional helicities, generated by $\gps^\gm$ and $\bgps^\gm$ in the states $|\bm{B}_{\ga\bgb} +\bm{S}+\bar{\bm{S}}\rangle$, are (anti--)correlated or not. If there is no correlation, then equal number of vector and hypers are kept, just the multiplicity is halved. If there is a correlation, then certain combinations of the GGSO phases decide, whether there is a correlation or an anti--correlation. In the case of a correlation both left-- and right--moving helicities are up or both down giving rise to vector fields, while in the case of an anti--correlation the helicities are opposite, then these states give rise to pairs of scalar fields. 

This selection principle is well--established for symmetric $\Intr_2^2$ orbifolds. Only for the a--D configuration the GGSO phase 
\equ{ \non 
    \arry{|c||c|c|}{
        \hline 
    \boldsymbol{\CC{B_{1\bar{1}}}{B_{2\bar{2}}}} & \textbf{IIB} & \textbf{IIA}
    \\[1ex] \hline\hline
    +1 & \text{Vector} & \text{Hyper}
    \\ \hline 
    -1 & \text{Hyper} & \text{Vector}
    \\ \hline 
    }
}
selects whether $\mathcal{N}=2$ vectors or hypers are kept of the $||\bm{B}_{1\bar{1}} \rangle\!\rangle$, $||\bm{B}_{2\bar{2}} \rangle\!\rangle$ and  $||\bm{B}_{3\bar{3}}' \rangle\!\rangle$ multiplets. In the $\Intr_2^2$ part of \cref{tab:NarainRotoTransClass} this selection is indicated using the variable $h=1$ when hypers are kept but no vectors or vise versa for $h=0$.

\subsubsection*{Summary}

In \cref{tab:ProjectionConsequences} the consequences of this analysis is summarised for the T--folds in which non--trivial projections occur. The table specifies the types of multiplets of the various physical (not projected out by $\bm{S}$ or $\bar{\bm{S}}$)  massless twisted sectors with which multiplicities arise. The cases in which some projections selects certain types of multiplets the different options are indicated as well.

These results can be used to arrive at the results presented in the classification table, \cref{tab:NarainRotoTransClass}. In that table the GGSO phases are indicated which select between different types of multiplets for the orbifolds where such selections may occur.

\begin{table}[t]
\centering
    \tabu{|ccc||ccc|}{
        \hline 
        \textbf{Point} & \textbf{4D--SUSY} & \textbf{Configu-} &  \multicolumn{3}{c|}{\textbf{Physical massless twisted sector}} \\ 
        \textbf{group} & $\boldsymbol{\mathcal{N}}$ & \textbf{ration} & ~~~~$\bm{B}_\ga$~~~~ & ~~~~$\bm{B}_\bgb$ & $\bm{B}_{\ga\bgb}, \bm{B}_{\ga\bgb}'$~~~~  
        \\ \hline\hline
        $\Intr_2^2$ & 2 & generic & -- & -- & 4\,(V+H)   \\ 
        & & a--D & -- & -- & 8\,V/\,8\,H  \\ \hline 
        $\Intr_{2L}\times\Intr_{2R}$ & 4 & all & 2\,RS/\,2\,V & 2\,RS/\,2\,V & 8\,V/\,4\,V/\,--  \\ \hline 
        $\Intr_{2L}^2$ & 5 & all & RS & -- & -- \\ \hline 
        $\Intr_{2L}\times\Intr_2$ & 3 & all & RS & -- & 4\,V \\ \hline 
        $\Intr_{2L}^2\times\Intr_{2R}$ & 3 & all & RS+V/\,2\,V & 2\,RS/\,2\,V & 2\,V \\ \hline 
        $\Intr_{2L}\times\Intr_{2R}\times\Intr_2$ & 2 & all & RS/\,H+2\,V+H & RS/\,H+2\,V+H & \tabu{c}{2\,V+2\,H/ \\[-1ex] 4\,V/\,4\,H} \\ \hline 
        $\Intr_{2L}^2\times\Intr_{2R}^2$ & 2 & all & RS/\,H+V & RS/\,H+V & \tabu{c}{2\,V+2\,H/\,V+H/\\[-1ex] 2\,V/\,2\,H} \\ \hline  
    }
    \caption{\label{tab:ProjectionConsequences}
        Overview of the multiplicities of which types of the twisted sectors of the T--folds on the SO(12) lattice after all projections of the twist basis vectors have been applied. Here ``generic'' means that this applies to all configurations of \cref{tab:NarainRotoTransClass} except the ones listed below; if it reads ``all'' then the effects of the projectors is the same for all models of that T--fold type.  
    A cell entry ``--'' indicates that the corresponding twisted sector is not present in that T--fold. A ``/'' signifies that a projection selects between the options listed.}
\end{table}

\subsection[Example: projected states in the pure twist ${\Intr_{2L}^2\times\Intr_{2R}^2}$ T--fold]{Example: projected states in the pure twist $\boldsymbol{\Intr_{2L}^2\times\Intr_{2R}^2}$ T--fold} 

Even though the selection can be determined by considering the spin--3/2 states alone, it is instructive to see how the projections affect all the states in $||\bm{b}_\ga\rangle\!\rangle$ in the pure twist $\Intr_{2L}^2\times\Intr_{2R}^2$ T--fold, {\em e.g.}\ configuration I--I, in the type IIB theory. In \cref{tab:GSOprojectionsSectorb1} all the states of the $(2,1)$ Rarita--Schwinger multiplet are listed and the values of four fermion number operators are indicated so as to facilitate working out the remaining GGSO projections. In order to collect the results compactly, combinations of the following notations have been employed in this table: 
the ten dimensional spin states (of positive chirality) are specified as 
\equ{ \non 
|\bar{\bm{S}}_0 \rangle = \big| \pm(\sfrac12\, \sfrac12\, \sfrac12\, \sfrac12) \big\rangle~,
\quad 
|\bar{\bm{S}}_1 \rangle = \big| \pm(\sfrac12\, \sfrac12\, \sm\sfrac12\, \sm\sfrac12) \big\rangle~,
\quad 
|\bar{\bm{S}}_2 \rangle = \big| \pm(\sfrac12\, \sm\sfrac12\, \sfrac12\, \sm\sfrac12) \big\rangle~,
\quad 
|\bar{\bm{S}}_3 \rangle = \big| \pm(\sfrac12\, \sm\sfrac12\, \sm\sfrac12\, \sfrac12) \big\rangle~,
}
in accordance to the complex notation $\bar{S} = \{ \bar{\gPs}_0,\bar{\gPs}_1,\bar{\gPs}_2,\bar{\gPs}_3 \}$. 
The states on which the GGSO projections of $\bm{S}$ and $\bm{E}$ have already been enforced are denoted by
\equ{
\arry{lcl}{
| \bm{b}_1 \rangle_\gs = \big| \pm ((0\, 0\, \sfrac12\, \sfrac12); \gs(0\, \sfrac12\, \sfrac12))\,\big\rangle~,
& \qquad & 
| \bm{b}_1 +\bm{S}\rangle_\gs = \big| \pm (( \sfrac12\, \sm\sfrac12\,0\, 0); \gs(0\, \sfrac12\, \sfrac12))\,\big\rangle~,
\\[1ex]
| \bm{b}_2 \rangle_\gs = \big| \pm ((0\,  \sfrac12\, 0\,\sfrac12); \gs(\sfrac12\, 0\, \sfrac12))\,\big\rangle~,
& \qquad & 
| \bm{b}_2 + \bm{S}\rangle_\gs = \big| \pm ((\sfrac12\, 0\,\sm\sfrac12\, 0); \gs(\sfrac12\, 0\, \sfrac12))\,\big\rangle~,
\\[1ex] 
| \bm{b}_3 \rangle_\gs = \big| \pm ((0\,  \sfrac12\, \sm\sfrac12\,0); \gs(\sfrac12\, \sfrac12\, 0))\,\big\rangle~,
& \qquad & 
| \bm{b}_3 +\bm{S}\rangle_\gs = \big| \pm ((\sfrac12\, 0\,  0\, \sfrac12); \gs(\sfrac12\, \sfrac12\, 0))\,\big\rangle~,
}
}
where the first four entries indicate external spin associated with $\bm{S}=\{\gPs_{0,1,2,3}\}$  and the latter three entries the internal spin associated with $y^{1,2,3}$, respectively, all in a complex basis, {i.e.}\ $\gPs_0=\psi^2+i\,\psi^3, \gPs_1=\gch^1+i\,\gch^2, \gPs_2=\gch^3+i\,\gch^4,\gPs_3=\gch^5+i\,\gch^6$. The parameter $\gs=\pm$ indicates whether these external and internal spins are correlated or anti--correlated. Note that the external spin of the $\bm{b}_3$ states, $| \bm{b}_3 \rangle_\gs$ and $| \bm{b}_3 +\bm{S}\rangle_\gs$, contain an additional minus sign in accordance to \eqref{eq:BSphases}. 

Table~\ref{tab:GSOprojectionsSectorb1} emphasises a number of striking features: 
\items{
\item 
Each sector $\bm{b}_\ga$ splits up into one Rarita--Schwinger and three vector $\mathcal{N}=3$ multiplets in four dimensions provided that all four twist basis vectors $\bm{b}_{1}$, $\bm{b}_{2}$, $\bm{b}_{\bar{1}}$ and $\bm{b}_{\bar{2}}$ are present. (If fewer twist basis vectors are present, these multiplets can pair up to higher extended supersymmetry multiplets.) 
\item 
By the GGSO projection~\eqref{eq:GGSOprojections} associated to $\bm{b}_{\bar{1}}$ and $\bm{b}_{\bar{2}}$, the GGSO phases $\CC{\bm{b}_\ga}{\bm{b}_{\bar{1}}}$ and $\CC{\bm{b}_\ga}{\bm{b}_{\bar{2}}}$ select whether the $\mathcal{N}=3$ Rarita--Schwinger multiplet is kept (when both equal $+1$) or one of the $\mathcal{N}=3$ vector multiplets. 
\item 
For the sectors $\bm{b}_1$ and $\bm{b}_2$ one GGSO projection enforced by $\bm{b}_1$ and $\bm{b}_2$is trivial; while the other requires a correlation or anti--correlation between external and internal spins associated to the complexified fermions $\chi$'s and $y$'s, respectively, parameterised by $\gs=\pm 1$. For many practical purposes this just halves the number of states. 
\item 
The GGSO projections enforced by $\bm{b}_1$ and $\bm{b}_2$ on the sector $\bm{b}_3$ are compatible, because the fermion numbers associated to these basis vectors equal the (anti--)correlation variables $\sm\gs$ and $\gs$, respectively. This is ultimately due to the opposite sign of $\CC{\bm{b}_\ga}{\bm{S}}$ for $\ga=3$, see \eqref{eq:BSphases}. 
}
These findings confirm the general statements made above. Note that here only the selection option of four options discussed for the projectors $\bm{b}_{\bar{1}}$ and $\bm{b}_{\bar{2}}$ occurs since $\bm{b}_{\bar{1}}$ and $\bm{b}_{\bar{2}}$ do not overlap with $\bm{b}_\ga$.

\begin{table}
\centering
\begin{tabular}{|c|c| cccc | cc|}
\hline
\textbf{Sector} $\bm{b}_\ga$ \textbf{states} & $\boldsymbol{\mathcal{N}=3}$ & 
$\bm{b}_{1}$ & $\bm{b}_{2}$ & $\bm{b}_{\bar{1}}$ &$\bm{b}_{\bar{2}}$ & 
$\bm{b}_{1}$ & $\bm{b}_{2}$ 
\\ 
 $\bar{\gPs}_a^t |\bm{B}\rangle$ with $\bm{B}=\bm{b}_\ga + s \bm{S} + (1-t) \bar{\bm{S}}_a$ 
& \textbf{multiplets} & 
\multicolumn{4}{c||}{$\boldsymbol{(-)^{\bm{b}_*\cdot F}}$} & 
\multicolumn{2}{c|}{$\boldsymbol{  \gd_{\bm{B}} \CC{\bm{B}}{\bm{b}_*}  }$} 
\\ \hline\hline 
$\bgPs_0 | \bm{b}_1 +\bm{S} \rangle_\gs, \bgPs_0 | \bm{b}_1 \rangle_\gs; | \bm{b}_1 +\bm{S} + \bar{\bm{S}}_0 \rangle_\gs, |\bm{b}_1 + \bar{\bm{S}}_0\rangle_\gs$  & \!Rarita--Schwinger\! & 
$+$ & $\gs$ & $+$ & $+$ & $+$ & $+$
\\\hline  
$| \bm{b}_1 +\bm{S} + \bar{\bm{S}}_1 \rangle_\gs, |\bm{b}_1 + \bar{\bm{S}}_1\rangle_\gs; \bgPs_1 | \bm{b}_1 \rangle_\gs, \bgPs_1 | \bm{b}_1 +\bm{S} \rangle_\gs$  & Vector & 
$+$ & $\gs$ & $+$ & $-$ & $+$ & $+$
\\\hline 
$| \bm{b}_1 +\bm{S} + \bar{\bm{S}}_2 \rangle_\gs, |\bm{b}_1 + \bar{\bm{S}}_2\rangle_\gs; \bgPs_2 | \bm{b}_1 \rangle_\gs, \bgPs_2 | \bm{b}_1 +\bm{S} \rangle_\gs$ & Vector & 
$+$ & $\gs$ & $-$ & $+$ & $+$ & $+$
\\\hline 
$| \bm{b}_1 +\bm{S} + \bar{\bm{S}}_3 \rangle_\gs, |\bm{b}_1 + \bar{\bm{S}}_3\rangle_\gs; 
\bgPs_3 | \bm{b}_1 \rangle_\gs, \bgPs_3 | \bm{b}_1 +\bm{S} \rangle_\gs$
& Vector & 
$+$ & $\gs$ & $-$ & $-$ & $+$ & $+$
\\\hline\hline 

$\bgPs_0 | \bm{b}_2 +\bm{S} \rangle_\gs, \bgPs_0 | \bm{b}_2 \rangle_\gs; | \bm{b}_2 +\bm{S} + \bar{\bm{S}}_0 \rangle_\gs, |\bm{b}_2 + \bar{\bm{S}}_0\rangle_\gs$  & \!Rarita--Schwinger\! & 
$\gs$ & $+$ & $+$ & $+$ & $-$ & $+$
\\\hline  
$| \bm{b}_2 +\bm{S} + \bar{\bm{S}}_1 \rangle_\gs, |\bm{b}_2 + \bar{\bm{S}}_1\rangle_\gs; \bgPs_1 | \bm{b}_2 \rangle_\gs, \bgPs_1 | \bm{b}_2 +\bm{S} \rangle_\gs$  & Vector & 
$\gs$ & $+$ & $+$ & $-$ & $-$ & $+$
\\\hline 
$| \bm{b}_2 +\bm{S} + \bar{\bm{S}}_2 \rangle_\gs, |\bm{b}_2 + \bar{\bm{S}}_2\rangle_\gs; \bgPs_2 | \bm{b}_2 \rangle_\gs, \bgPs_2 | \bm{b}_2 +\bm{S} \rangle_\gs$ & Vector & 
$\gs$ & $+$ & $-$ & $+$ & $-$ & $+$
\\\hline 
$| \bm{b}_2 +\bm{S} + \bar{\bm{S}}_3 \rangle_\gs, |\bm{b}_2 + \bar{\bm{S}}_3\rangle_\gs; 
\bgPs_3 | \bm{b}_2 \rangle_\gs, \bgPs_3 | \bm{b}_2 +\bm{S} \rangle_\gs$
& Vector & 
$\gs$ & $+$ & $-$ & $-$ & $-$ & $+$
\\\hline\hline 
$\bgPs_0 | \bm{b}_3 +\bm{S} \rangle_\gs, \bgPs_0 | \bm{b}_3 \rangle_\gs; | \bm{b}_3 +\bm{S} + \bar{\bm{S}}_0 \rangle_\gs, |\bm{b}_3 + \bar{\bm{S}}_0\rangle_\gs$  & \!Rarita--Schwinger\! & 
$\sm\gs$ & $\gs$ & $+$ & $+$ & $-$ & $+$
\\\hline  
$| \bm{b}_3 +\bm{S} + \bar{\bm{S}}_1 \rangle_\gs, |\bm{b}_3 + \bar{\bm{S}}_1\rangle_\gs; \bgPs_1 | \bm{b}_3 \rangle_\gs, \bgPs_1 | \bm{b}_3 +\bm{S} \rangle_\gs$  & Vector & 
$\sm\gs$ & $\gs$ & $+$ & $-$ & $-$ & $+$
\\\hline 
$| \bm{b}_3 +\bm{S} + \bar{\bm{S}}_2 \rangle_\gs, |\bm{b}_3 + \bar{\bm{S}}_2\rangle_\gs; \bgPs_2 | \bm{b}_3 \rangle_\gs, \bgPs_2 | \bm{b}_3 +\bm{S} \rangle_\gs$ & Vector & 
$\sm\gs$ & $\gs$ & $-$ & $+$ & $-$ & $+$
\\\hline 
$| \bm{b}_3 +\bm{S} + \bar{\bm{S}}_3 \rangle_\gs, |\bm{b}_3 + \bar{\bm{S}}_3\rangle_\gs; 
\bgPs_3 | \bm{b}_3 \rangle_\gs, \bgPs_3 | \bm{b}_3 +\bm{S} \rangle_\gs$
& Vector & 
$\sm\gs$ & $\gs$ & $-$ & $-$ & $-$ & $+$
\\ \hline 
\end{tabular}
\caption{\label{tab:GSOprojectionsSectorb1}
    All the states of the sectors $\bm{b}_{1,2,3}$ classified as $\mathcal{N}=3$ multiplets, where $s,t = 0,1$ and $a=0,\ldots,3$. The four fermion numbers associated to the twist basis vectors $\bm{b}_*=\bm{b}_1,\bm{b}_2,\bm{b}_{\bar{1}},\bm{b}_{\bar{2}}$ are given as well as two GGSO phases associated to $\bm{b}_*=\bm{b}_{1,2}$ so that the effects of the GGSO projections~\eqref{eq:GGSOprojections} can be worked out effectively. }
\end{table}

\subsection{Details of the spectra of the (a)symmetric order--two orbifolds}

This subsection discusses the resulting spectra of all the possible order--two T--folds on the $SO(12)$ lattice at the free fermionic point. The full details of the discussed spectra can be found in \cref{tab:NarainRotoTransClass}. If the spectra depend on certain GGSO phases, this table states the number of states parametrically. Below these parameterisations are elaborated on; in the table they are repeated for completeness. To crosscheck the results of \cref{tab:NarainRotoTransClass}, the massless spectra were also determined independently with the codes in the \texttt{Spectrum\_Analysis} folder of ref.~\cite{percival2026z2n_typeii_code}, which compute the spectrum of each configuration directly from its basis vectors and GGSO phases.

\subsubsection*{Symmetric $\boldsymbol{\Intr_2}$ and $\boldsymbol{\Intr_2^2}$ orbifolds}

The spectra of the symmetric $\Intr_2$ and $\Intr_2^2$ on the $SO(12)$ lattice are well--known and are reproduced. They were used as elementary crosschecks of our procedures. The spectra of configuration a--D depends via
\equ{ \label{eq:Parameter_h_Z22}
    h=\sfrac 12\Big(1+\CC{\bar{\bm{S}}}{\bar{\bm{S}}}\CC{\bm{B}_{1\bar{1}}}{\bm{B}_{2\bar{2}}}\Big)
}
on the GGSO (or discrete torsion) phase $\CC{\bm{B}_{1\bar{1}}}{\bm{B}_{2\bar{2}}}$. If the $h=1$ the twisted spectrum of this configuration contains 24 twisted hyper multiplets, if $h=0$ it contains 24 twisted vector multiplets.

\subsubsection*{Asymmetric $\boldsymbol{\Intr_{2L}}$ orbifolds}

These models have (at least) $\cN=6$ supersymmetry and hence the spectrum can only contain the supergravity multiplet and Rarita--Schwinger multiplets. Only configuration i has a non--empty twisted sector which contains two Rarita--Schwinger multiplets, see \cref{tab:NarainRotoTransClass}.

\subsubsection*{Asymmetric $\boldsymbol{\Intr_{2L}\times \Intr_{2R}}$ orbifolds}

The $\Intr_{2L}\times\Intr_{2R}$ orbifolds possess $\cN=4$ supersymmetry and hence the spectrum may contain Rarita--Schwinger and vector multiplets besides the supergravity multiplet. According to \cref{tab:FixedFreeGGSOphases}, $\CC{\bm{B}_1}{\bm{B}_{\bar{1}}}$ is the only physical GGSO phase for these orbifolds. In configurations where the twisted sectors $1$ or $\bar{1}$ are massless, this phase determines whether the model gives rise to Rarita--Schwinger ($r=1$) or vector multiplets ($r=0$), where 
\equ{ \label{eq:RSmultiplet}
    r = \sfrac12 \Big(1+ \CC{\bm{B}_1}{\bm{B}_{\bar{1}}} \Big)
    \quad\text{or}\quad 
    \CC{\bm{B}_1}{\bm{B}_{\bar{1}}} = -(-1)^r~. 
}
This relation follows directly from GGSO projection conditions. The resulting spectra from the sectors $1$ or $\bar{1}$ can be compactly parameterised as: $2r$(RS--V)+4\,V. Only five configurations may lead to models that may contain Rarita--Schwinger multiplets in this way. 

Curiously, configuration ii--ii.2 depends on this parameter even though it never contains Rarita--Schwinger multiplets. Instead, if $r=0$ all eight vector multiplets in the sector $1\bar{1}'$ are kept, while if $r=1$ all states in this sector are projected out. Full details of the possible spectra of these orbifolds can be found in \cref{tab:NarainRotoTransClass}.

\subsubsection*{Asymmetric $\boldsymbol{\Intr_{2L}^2}$ orbifolds}

The $\Intr_{2L}^2$ orbifolds possess $\cN=5$ supersymmetry and hence the spectrum can only contain the supergravity multiplet and a number of Rarita--Schwinger multiplets. Only the first three configurations I, II and III have non--empty twisted sector(s) consisting of Rarita--Schwinger multiplets. In particular, the twisted sector 3 of configuration VI is fully projected out irrespectively of the phase $\CC{\bm{B}_{1}}{\bm{B}_{2}}$. Full details of the possible spectra of these orbifolds can be found in \cref{tab:NarainRotoTransClass}.

\subsubsection*{Asymmetric $\boldsymbol{\Intr_{2L}\times \Intr_2}$ orbifolds}

The $\Intr_{2L}\times\Intr_{2}$ orbifolds possess $\cN=3$ supersymmetry and hence the spectrum may contain Rarita--Schwinger and vector multiplets besides the supergravity multiplet. According to \cref{tab:FixedFreeGGSOphases}, all GGSO phases for this class of orbifolds are fixed or physically irrelevant. Only the configurations i--A and i--E give rise to a single Rarita--Schwinger multiplet and a number of vector multiplets in their twisted sectors; all other models at most some number of twisted vector multiplets. Eight configurations have empty twisted spectra. Full details of the possible spectra of these orbifolds can be found in \cref{tab:NarainRotoTransClass}.

\subsubsection*{Asymmetric $\boldsymbol{\Intr_{2L}^2\times \Intr_{2R}}$ orbifolds}

The $\Intr_{2L}^2\times\Intr_{2R}$ orbifolds possess $\cN=3$ supersymmetry and hence the spectrum may contain Rarita--Schwinger and vector multiplets besides the supergravity multiplet. Full details of the possible spectra of these orbifolds can be found in \cref{tab:NarainRotoTransClass}; below some specific features are highlighted. 

The sectors $1,2,3$ and $\bar{1}$ may give rise to Rarita--Schwinger multiplets. As a generalisation of \eqref{eq:RSmultiplet} define the parameters 
\equ{ \label{eq:RSmultiplets}
    r_1 = \sfrac12 \Big(1+ \CC{\bm{B}_1}{\bm{B}_{\bar{1}}} \Big)~, 
    \quad 
    r_2 = \sfrac12 \Big(1+ \CC{\bm{B}_2}{\bm{B}_{\bar{1}}} \Big)~, 
    \quad
    r_3 =\sfrac12 \Big(1+ \CC{\bm{B}_1}{\bm{B}_{\bar{1}}} \CC{\bm{B}_2}{\bm{B}_{\bar{1}}} \Big)~, 
    \quad 
    r_{\bar{1}} = r_1r_2~. 
}
The last relation states that in the supersector $\bar{1}$ Rarita--Schwinger multiplets only appear provided that both GGSO phases are $+1$, see \cref{tab:GSOprojectionsSectorb1}. Since $r_3$ depend on the combination of both GGSO phases, they can be expressed in $r_1$ and $r_2$ as:
\equ{
    r_3 = 2r_1r_2 -r_1-r_2+1 = \sfrac12 \Big(1+ (-1)^{r_1+r_2} \Big)~.
}
The latter representation most clearly shows that $r_3=1$ only if $r_1$ and $r_2$ are equal.
The spectra of supersectors $1,2,3$ can be expressed as $r_\ga$(RS--V)+2\,V and for supersector $\bar{1}$ as $2r_{\bar{1}}$(RS--V)+2\,V. In particular, for configuration I--i the total number Rarita--Schwinger multiplets can, therefore, be expressed as 
\equ{ \label{eq:Z2L2Z2R_RS_relation}
    (r_1+r_2+r_3+2r_{\bar{1}})\text{(RS--V)}+8\,\text{V} = (4r_1r_2+1)\text{(RS--V)}+8\,\text{V}
}
see \cref{tab:NarainRotoTransClass}. In other words, configuration I--i of the $\Intr_{2L}^2\times\Intr_{2R}$ orbifold only gives rise to five Rarita--Schwinger multiplets if $r_1=r_2=1$; otherwise to just a single one. 

The supersectors $\ga\bgb$ and $\ga\bgb'$ generically give rise to two vector multiplets, {\em i.e.}\ half of the original states are projected out. However, sometimes some of these sectors are kept completely or fully projected out. This happens in configurations III--i, IV--i, IV--ii.1, IV--ii.2, II-iii and III-iii. For example, for configuration III-iii the sector $1\bar{1}'$ is projected out if $r_1=1$ and kept if $r_1=0$. Combined this sector gives rise to $4(1-r_1)$ vector multiplets. Thus this is precisely anti--correlated to whether the supersector $1$ gives rise to a Rarita--Schwinger multiplet.

\subsubsection*{Asymmetric $\boldsymbol{\Intr_{2L}\times \Intr_{2R} \times \Intr_2}$ orbifolds}

The $\Intr_{2L}\times\Intr_{2R}\times\Intr_2$ orbifolds possess $\cN=2$ supersymmetry and hence the spectrum may contain Rarita--Schwinger, vector and hyper multiplets besides the supergravity multiplet.
According to \cref{tab:FixedFreeGGSOphases} there are three relevant GGSO phases $\CC{B_1}{B_{\bar{1}}}$, $\CC{B_1}{B_{2\bar{2}}}$ and $\CC{B_{\bar{1}}}{B_{2\bar{2}}}$.

Like for the $\Intr_{2L}\times\Intr_{2R}$ orbifolds, the parameter 
\equ{ \label{eq:Z2LZ2RZ2_RS_relation}
    r = \sfrac12 \Big(1+ \CC{\bm{B}_1}{\bm{B}_{\bar{1}}} \Big)~, 
}
determines whether Rarita--Schwinger multiplets arise from the sectors $1$ and $\bar{1}$, if present ({\em i.e.} for the nine configurations i--i--A, i--ii--A, ii--i--A, i--iii--A, iii--i--A, i--ii--E, ii--i--E, i--iii--E, iii--i--E). In addition, for five specific configurations (ii--ii--A.2, i--iii--A, iii--i--A, i--iii--E, iii--i--E) this parameter also decides whether certain supersectors are projected out altogether. 

The twisted sectors $\ga\bgb$ and $\ga\bgb'$ may give rise to $\cN=2$ vector or hyper multiplets or both. Provided that there is a selection of vector or hyper multiplets within the supersectors $1\bar{1}$, $2\bar{2}$, $3\bar{3}$ or one of their primed versions, the parameter 
\begin{subequations}
\label{eq:Z2LZ2RZ2_h}
\equ{ \label{eq:Z2LZ2RZ2_h0_relation}
    h= \sfrac12 \Big(1+ \CC{\bar{\bm{S}}}{\bar{\bm{S}}} \CC{\bm{B}_1}{\bm{B}_{2\bar{2}}} \CC{\bm{B}_{\bar{1}}}{\bm{B}_{2\bar{2}}} \Big) 
} 
equals $1$ if this sector gives rise to hypers and $0$ if this sector gives rise to vectors. Similarly, provided that there is a selection of vector or hyper multiplets within the supersector $2\bar{3}$, $3\bar{2}$ or one of their primed versions, the parameter 
\equ{ \label{eq:Z2LZ2RZ2_h1_relation}
    h'= \sfrac12 \Big(1 - \CC{\bar{\bm{S}}}{\bar{\bm{S}}} \CC{\bm{B}_1}{\bm{B}_{\bar{1}}} \CC{\bm{B}_1}{\bm{B}_{2\bar{2}}} \CC{\bm{B}_{\bar{1}}}{\bm{B}_{2\bar{2}}} \Big) 
} 
\end{subequations}
equals $1$ if this sector gives rise to hypers and $0$ if this sector gives rise to vectors. The GGSO phase $\CC{\bm{S}}{\bar{\bm{S}}}$ appears in equations~\eqref{eq:Z2LZ2RZ2_h} as it determines the correlation between target space chiralities arising from the left-- and right--moving fermionic zero modes. 

Notice that the parameters $r, h, h'$ only depend on two combinations of the three free GGSO phases, since both $h$ and $h'$ depend on the combined phases 
\equ{ 
    \CC{\bm{B}_{1\bar{1}}}{\bm{B}_{2\bar{2}}} = \CC{\bm{B}_1}{\bm{B}_{2\bar{2}}} \CC{\bm{B}_{\bar{1}}}{\bm{B}_{2\bar{2}}}~.   
}
This parameterisation is used to display the spectra of this orbifold in \cref{tab:NarainRotoTransClass}. Consequently, one can show that $h' = r+h-2rh$ or equivalently $h'=r+h\ \text{mod}\ 2$ since $r,h,h' = 0,1$.

\subsubsection*{Asymmetric $\boldsymbol{\Intr_{2L}^2\times \Intr_{2R}}^2$ orbifolds}

As before, the parameters $r_\ga$ and $r_\bgb$, defined as  
\equ{ \label{eq:Parameters_r}
    r_\ga=\sfrac14\Big(1+\sum\limits_\bgb\CC{\bm{B}_\ga}{\bm{B}_\bgb}\Big)~,
    \qquad 
    r_\bgb=\sfrac14\Big(1+\sum\limits_\ga\CC{\bm{B}_\ga}{\bm{B}_\bgb}\Big)~, 
    \qquad 
    r=\sum\limits_\ga r_\ga = \sum\limits_{\bgb} r_{\bgb}~, 
}
equal $1$ if the supersector $\ga,\bgb$ gives rise to Rarita--Schwinger multiplets. 
For certain configurations the parameter 
\equ{ \label{eq:Parameters_p}
    p_{\ga\bgb} = \sfrac12\Big( 1 - \CC{\bm{B}_\ga}{\bm{B}_\bgb} \Big)
}
selects whether the sector $\ga\bgb'$ is fully kept or completely projected out. In addition, it is convenient to introduce the parameters 
\equ{
    h_{\ga\bgb} = \sfrac12\Big( 1 +  \CC{\bar{\bm{S}}}{\bar{\bm{S}}} \CC{\bm{B}_{\gb\bga}}{\bm{B}_{\ga\bgb}} \Big)~, 
    \qquad 
    h_{\ga\bgb}' = \sfrac12\Big( 1 -  \CC{\bar{\bm{S}}}{\bar{\bm{S}}} \CC{\bm{B}_\ga}{\bm{B}_\bgb} \CC{\bm{B}_{\gb\bga}}{\bm{B}_{\ga\bgb}} \Big)~, 
}
where $\ga,\bgb=1,2$ and $\gb=3-\ga$ and $\bga=3-\bgb$; 
or explicitly 
\begin{subequations}
    \label{eq:Parameters_h}
   \equ{
    h_{2\bar{2}} = h_{1\bar{1}} = \sfrac12\Big( 1 +  \CC{\bar{\bm{S}}}{\bar{\bm{S}}} \CC{\bm{B}_{1\bar{1}}}{\bm{B}_{2\bar{2}}} \Big)~, 
    \qquad 
    h_{1\bar{2}} = h_{2\bar{1}} = \sfrac12\Big( 1 +  \CC{\bar{\bm{S}}}{\bar{\bm{S}}} \CC{\bm{B}_{2\bar{1}}}{\bm{B}_{1\bar{2}}} \Big)~, 
    \\[2ex]
    h_{1\bar{1}}'  = \sfrac12\Big( 1 -  \CC{\bar{\bm{S}}}{\bar{\bm{S}}} \CC{\bm{B}_{1}}{\bm{B}_{\bar{1}}} \CC{\bm{B}_{2\bar{2}}}{\bm{B}_{1\bar{1}}} \Big)~, 
    \qquad 
    h_{1\bar{2}}'  = \sfrac12\Big( 1 -  \CC{\bar{\bm{S}}}{\bar{\bm{S}}} \CC{\bm{B}_{1}}{\bm{B}_{\bar{2}}} \CC{\bm{B}_{2\bar{1}}}{\bm{B}_{1\bar{2}}} \Big)~, 
    \\[2ex]
    h_{2\bar{2}}'  = \sfrac12\Big( 1 - \CC{\bar{\bm{S}}}{\bar{\bm{S}}} \CC{\bm{B}_{2}}{\bm{B}_{\bar{2}}} \CC{\bm{B}_{1\bar{1}}}{\bm{B}_{2\bar{2}}} \Big)~, 
    \qquad 
    h_{2\bar{1}}'  = \sfrac12\Big( 1 -  \CC{\bar{\bm{S}}}{\bar{\bm{S}}} \CC{\bm{B}_{2}}{\bm{B}_{\bar{1}}} \CC{\bm{B}_{1\bar{2}}}{\bm{B}_{2\bar{1}}} \Big)~, 
} 
\end{subequations}
They can be expressed in terms of the fundamental GGSO phase by using the expansion 
\equ{ \label{eq:ExpansionBab}
    \CC{\bm{B}_{\gb\bga}}{\bm{B}_{\ga\bgb}} = \CC{\bm{B}_{\gb}}{\bm{B}_{\ga}} \CC{\bm{B}_{\bga}}{\bm{B}_{\ga}} \CC{\bm{B}_{\gb}}{\bm{B}_{\bgb}} \CC{\bm{B}_{\bga}}{\bm{B}_{\bgb}}~. 
}
and their symmetry properties. 
Since if $p_{1\bar{3}}$ (or $p_{3\bar{1}}$) equals $1$ then $\CC{B_1}{B_{\bar{1}}}=-\CC{B_1}{B_{\bar{2}}}$ (or $\CC{B_1}{B_{\bar{1}}}=-\CC{B_2}{B_{\bar{1}}}$), 
one can show that 
\equ{
    p_{1\bar{3}}\, h_{2\bar{2}}' = p_{1\bar{3}}\, h_{2\bar{1}}'~, 
    \qquad
    p_{3\bar{1}}\, h_{2\bar{2}}' = p_{3\bar{1}}\, h_{1\bar{2}}'~. 
}
These relations are used to simplify the expressions of the spectra somewhat in \cref{tab:NarainRotoTransClass}.

\clearpage

\renewcommand{\arraystretch}{1.2}
\setlength{\LTcapwidth}{\linewidth}
\setlength{\LTcapleft}{0pt}
\begin{longtable}[c]{| c| l| l|}
\caption{\label{tab:NarainRotoTransClass} This classification table gives the inequivalent twist basis vectors for all inequivalent configurations of all order--two T--folds on the $SO(12)$ lattice at the free fermionic point for each of the order--two Narain point groups. The massless twisted sectors are indicated. When they are indicated in gray this signifies that these twisted sectors are massless but projected out by some GGSO projection. The spectra are given in terms of super multiplets with respect to the extended supersymmetry indicated using the following notations for the four dimensional super multiplets: SG~=~SuperGravity; RS~=~Rarita--Schwinger; V~=~Vector and H~=~Hyper multiplets. The spectra often depend on some detailed choice of GGSO phases. For each configuration where this is the case, its spectra are given in parametric form. The parameters used for this are defined in terms of the relevant GGSO phases where the next point group is introduced. 
}
\\
\hline 
\multicolumn{3}{|c|}{\textbf{Point group} \hfill \textbf{4D Extended SUSY : Untwisted sector multiplets }} \\ \hline 
\textbf{Label} & \textbf{Twist basis vectors} & \textbf{Twisted sectors / multiplets} 
\\ \hline\hline 
\endfirsthead
\hline 
\multicolumn{3}{|c|}{{\em \tablename\ \thetable{} -- continued from previous page}} \\
\hline 
\multicolumn{3}{|c|}{\textbf{Point group} \hfill \textbf{4D Extended SUSY : Untwisted sector multiplets}} \\ \hline 
\textbf{Label} & \textbf{Twist basis vectors} & \textbf{Twisted sectors / multiplets}  \\ \hline\hline 
\endhead
\hline \multicolumn{3}{|r|}{{\em continued on next page...}} \\ \hline 
\endfoot
\hline 
\endlastfoot
\multicolumn{3}{|c|}{$\Id$ \hfill $\mathcal{N}=8$ : SG}  \\ \hline 
-- & &  \\ 
 &   &  
\\ \hline\hline 
%
%
\multicolumn{3}{|c|}{$\Intr_2$ \hfill $\mathcal{N}=4$ : SG + 6\,V} \\ \hline 
a  & $\bm{b}_{1\bar{1}}$ & ${1\bar{1}}$   \\ 
   &  & 8\,V  \\ \hline 
b & $\bm{b}_{1\bar{1}}+\bm{e}_{1\bar{1}}$ &  \\ 
  &  & 
\\ \hline\hline 
%
%
\multicolumn{3}{|c|}{$\Intr_2^2$ \hfill $\mathcal{N}=2$ : SG+3\,V+4\,H} \\ 
\multicolumn{3}{|r|}{$h=\sfrac 12\Big(1+\CC{\bar{\bm{S}}}{\bar{\bm{S}}}\CC{\bm{B}_{1\bar{1}}}{\bm{B}_{2\bar{2}}}\Big)=0,1$}
\\ \hline 
a--A & $\bm{b}_{1\bar{1}},~\bm{b}_{2\bar{2}}$ & ${1\bar{1}}; {2\bar{2}}; {3\bar{3}}$  \\ 
(1--2)  &  & 12\,V+12\,H\\ \hline
a--D & $\bm{b}_{1\bar{1}},~\bm{b}_{2\bar{2}}+\bm{e}_{56\bar{5}\bar{6}}$ & ${1\bar{1}}; {2\bar{2}}; {3\bar{3}}'$  \\
(1--1)  &  & 
24$h$\,(H--V)+24\,V

\\[0.5ex] \hline
a--B & $\bm{b}_{1\bar{1}},~\bm{b}_{2\bar{2}}+\bm{e}_{5\bar{5}}$ &${1\bar{1}}; {2\bar{2}}$   \\ 
(1--3)  &  & 8\,V+8\,H \\ \hline
a--E & $\bm{b}_{1\bar{1}},~\bm{b}_{2\bar{2}}+\bm{e}_{35\bar{3}\bar{5}}$ & ${1\bar{1}}$  \\ 
(1--4)  &  & 4\,V+4\,H\\ \hline 
b--E & $\bm{b}_{1\bar{1}}+\bm{e}_{1\bar{1}},~\bm{b}_{2\bar{2}}+\bm{e}_{35\bar{3}\bar{5}}$ &  \\
(1--5)  &  &    
\\  \hline\hline
%

%
%
\multicolumn{3}{|c|}{$\Intr_{2L}$ \hfill $\mathcal{N}=6$ : SG} \\ \hline 
i  & $\bm{b}_1$  & ${1}$   \\ 
   &  &  2\,RS \\ \hline 
ii & $\bm{b}_1+\bm{e}_{1\bar{1}}$ &  \\ 
   &  &   \\ \hline 
iii & $\bm{b}_1+\bm{e}_{12\bar{1}\bar{2}}$ &  \textcolor{gray}{${1}'$}   \\ 
   &  & 
\\ \hline \hline 
%

%
\multicolumn{3}{|c|}{$\Intr_{2L}\times \Intr_{2R}$ \hfill $\mathcal{N}=4$ : SG+2\,V}  \\ 
\multicolumn{3}{|r|}{$r=\sfrac12\Big(1+\CC{\bm{B}_{1}}{\bm{B}_{\bar{1}}}\Big)=0,1$}
\\ \hline 
i--i & $\bm{b}_1,$ & ${1}; {\bar{1}}; {1\bar{1}}$  \\
  & $\bm{b}_{\bar{1}}$ & 
4$r$\,(RS--2V)+12\,V   \\[0.5ex] \hline
ii--i & $\bm{b}_1+\bm{e}_{1\bar{1}},$ & ${\bar{1}}$  \\
  &$\bm{b}_{\bar{1}}$ &  
2$r$\,(RS--2V)+4\,V  \\[0.5ex] \hline
i--ii & $\bm{b}_1,$ & ${1}$ \\ 
  &$\bm{b}_{\bar{1}}+\bm{e}_{1\bar{1}}$ &  
2$r$\,(RS--2V)+4\,V \\[0.5ex] \hline
iii--i & $\bm{b}_1+\bm{e}_{12\bar{1}\bar{2}},$ & ${\bar{1}}; {1\bar{1}}'$\textcolor{gray}{$; {1}'$} \\
  & $\bm{b}_{\bar{1}}$ &    
2$r$\,(RS--6V)+12\,V \\[0.5ex] \hline
i--iii & $\bm{b}_1,$ & $1; {1\bar{1}}'$\textcolor{gray}{$; {\bar{1}}'$}  \\
  & $\bm{\bar{b}}_1+\bm{e}_{12\bar{1}\bar{2}}$ &  
2$r$\,(RS-6V)+12\,V  \\[0.5ex] \hline
ii--ii.1 & $\bm{b}_1+\bm{e}_{1\bar{1}},$ & ${1\bar{1}}$  \\
  & $\bm{b}_{\bar{1}}+\bm{e}_{1\bar{1}}$ & 4\,V  \\ \hline
ii--ii.2 & $\bm{b}_1+\bm{e}_{1\bar{2}},$ & ${1\bar{1}}'$  \\
  & $\bm{b}_{\bar{1}}+\bm{e}_{2\bar{1}}$ & 8(1--$r$)V  \\ \hline
ii--ii.3 & $\bm{b}_1+\bm{e}_{1\bar{3}},$ &  \\
  & $\bm{b}_{\bar{1}}+\bm{e}_{3\bar{1}}$ &  \\ \hline
iii--ii.1 & $\bm{b}_1+\bm{e}_{12\bar{1}\bar{2}},$ & \textcolor{gray}{${1}'$} \\
  &$\bm{\bar{b}}_1+\bm{e}_{1\bar{1}}$ &   \\ \hline
ii--iii.1 & $\bm{b}_1+\bm{e}_{1\bar{1}},$ & \textcolor{gray}{$\bar{1}'$} \\
   & $\bm{\bar{b}}_1+\bm{e}_{12\bar{1}\bar{2}}$ &    \\ \hline
iii--ii.2 & $\bm{b}_1+\bm{e}_{12\bar{2}\bar{3}},$ & ${1\bar{1}}'$\textcolor{gray}{$; {1}'$} \\
   &$\bm{b}_{\bar{1}}+\bm{e}_{3\bar{1}}$ & 4\,V  \\ \hline
ii--iii.2 & $\bm{b}_1+\bm{e}_{1\bar{3}},$ & ${1\bar{1}}'$\textcolor{gray}{$; \bar{1}'$}  \\ 
   &$\bm{b}_{\bar{1}}+\bm{e}_{23\bar{1}\bar{2}}$ & 4\,V \\ \hline
iii-ii.3 & $\bm{b}_1+\bm{e}_{12\bar{3}\bar{4}},$ & \textcolor{gray}{${1}'$} \\
   & $\bm{\bar{b}}_1+\bm{e}_{1\bar{1}}$ & \\ \hline
ii--iii.3 & $\bm{b}_1+\bm{e}_{1\bar{1}},$ &  \textcolor{gray}{${\bar{1}}'$} \\
   & $\bm{b}_{\bar{1}}+\bm{e}_{34\bar{1}\bar{2}}$ &   \\ \hline
iii--iii.1 & $\bm{b}_1+\bm{e}_{12\bar{1}\bar{2}},$ &  ${1\bar{1}}$\textcolor{gray}{$; {1}'; {\bar{1}}'$}  \\ 
   & $\bm{\bar{b}}_1+\bm{e}_{12\bar{1}\bar{2}}$ & 4\,V  \\ \hline
iii--iii.2 & $\bm{b}_1+\bm{e}_{12\bar{1}\bar{3}},$ & \textcolor{gray}{${1}'; {\bar{1}}'$}  \\
   & $\bm{b}_{\bar{1}}+\bm{e}_{13\bar{1}\bar{2}}$ &   \\ \hline
iii--iii.3 & $\bm{b}_1+\bm{e}_{12\bar{3}\bar{4}},$ &   ${1\bar{1}}'$\textcolor{gray}{$; {1}'; {\bar{1}}'$} \\ 
   & $\bm{\bar{b}}_1+\bm{e}_{34\bar{1}\bar{2}}$ & 4\,V
\\ \hline\hline 
%

%
\multicolumn{3}{|c|}{$\Intr_{2L}^2$ \hfill $\mathcal{N}=5$ : SG}  \\ \hline 
I & $\bm{b}_1,~\bm{b}_2$ & $1; 2; 3$  \\
  & $$ & 3\,RS \\ \hline
II & $\bm{b}_1,~\bm{b}_2+\bm{e}_{35\bar{1}}$ & $1$  \\ 
  & $$ & RS  \\ \hline
III & $\bm{b}_1,~\bm{b}_2+\bm{e}_{3456\bar{1}\bar{2}}$ & $1$\textcolor{gray}{$; 2'; 3'$} \\
  & $$ & RS \\ \hline
IV & $\bm{b}_1+\bm{e}_{1\bar{1}},~\bm{b}_2+\bm{e}_{356\bar{2}}$ & \textcolor{gray}{$3'$} \\ 
  & $$ &   \\ \hline
V & $\bm{b}_1+\bm{e}_{12\bar{1}\bar{2}},~\bm{b}_2+\bm{e}_{3456\bar{1}\bar{3}}$ & \textcolor{gray}{$1'; 2'; 3'$} \\ 
  & $$ &  \\ \hline
\pagebreak
VI & $\bm{b}_1+\bm{e}_{12\bar{1}\bar{2}},~\bm{b}_2+\bm{e}_{34\bar{3}\bar{4}}$ & \textcolor{gray}{$3$; $1'; 2'$}   \\  
  & $$ &  
\\ \hline\hline
%
%

%
%
\multicolumn{3}{|c|}{$\Intr_{2L} \times \Intr_2$ \hfill $\mathcal{N}=3$ : SG + 3\,V} \\ \hline
i--A  & $\bm{b}_1,$ & ${1}; {2\bar{2}}; {3\bar{2}}$   \\ 
   & $\bm{b}_{2\bar{2}}$ & RS+11\,V \\ \hline 
ii--A  & $\bm{b}_1+\bm{e}_{1\bar{1}},$ & ${2\bar{2}}; {3\bar{2}}$  \\
   & $\bm{b}_{2\bar{2}}$ & 8\,V  \\ \hline
iii--A  & $\bm{b}_1+\bm{e}_{12\bar{1}\bar{2}},$ & ${2\bar{2}}; {3\bar{2}}$\textcolor{gray}{$; {1}'$}  \\ 
   & $\bm{b}_{2\bar{2}}$ & 8\,V  \\ \hline
ii--B  & $\bm{b}_1+\bm{e}_{1\bar{3}},$ & ${2\bar{2}}$  \\ 
   & $\bm{b}_{2\bar{2}}+\bm{e}_{5\bar{5}}$ & 4\,V  \\ \hline 
iii--B  & $\bm{b}_1+\bm{e}_{12\bar{1}\bar{3}},$ & ${2\bar{2}}$\textcolor{gray}{$; {1}'$}  \\
   & $\bm{b}_{2\bar{2}}+\bm{e}_{5\bar{5}}$ & 4\,V  \\  \hline
iii--D & $\bm{b}_1+\bm{e}_{12\bar{3}\bar{4}},$ & ${2\bar{2}}; {3\bar{2}}'$\textcolor{gray}{$; {1}'$}  \\ 
   & $\bm{b}_{2\bar{2}}+\bm{e}_{56\bar{5}\bar{6}}$ & 8\,V   \\ \hline
i--E & $\bm{b}_1,$ & ${1}$  \\  
   & $\bm{b}_{2\bar{2}}+\bm{e}_{35\bar{3}\bar{5}}$ & RS+3\,V  \\ \hline
ii--E & $\bm{b}_1+\bm{e}_{1\bar{1}},$ &  \\
   & $\bm{b}_{2\bar{2}}+\bm{e}_{35\bar{3}\bar{5}}$ &  \\ \hline
iii--E.1 & $\bm{b}_1+\bm{e}_{12\bar{1}\bar{2}},$ & \textcolor{gray}{${1}'$} \\
   & $\bm{b}_{2\bar{2}}+\bm{e}_{35\bar{3}\bar{5}}$ &   \\ \hline
%
iii--E.2 & $\bm{b}_1+\bm{e}_{12\bar{3}\bar{4}},$ & \textcolor{gray}{${1}'$} \\ 
   & $\bm{b}_{2\bar{2}}+\bm{e}_{35\bar{3}\bar{5}}$ &   
\\ \hline\hline
%

%
%
\multicolumn{3}{|c|}{$\Intr_{2L}^2\times \Intr_{2R}$ \hfill $\mathcal{N}=3$ : SG + V} \\ 
\multicolumn{3}{|r|}{$r_{\bar{1}}=r_1r_2$~, $r_\ga=\sfrac12\Big(1+\CC{\bm{B}_\ga}{\bm{B}_{\bar{1}}}\Big)=0,1$~, $\ga=1,2,3$}\\ \hline
I--i  & $\bm{b}_1,~\bm{b}_2,$ 
   & ${1}; {2}; {3}; {\bar{1}}; {1\bar{1}}; {2\bar{1}}; {3\bar{1}}$  \\
   & $\bm{b}_{\bar{1}}$ & $(r_1+r_2+r_3+2r_{\bar{1}})$\,(RS-V)+14\,V \\ \hline
II--i  & $\bm{b}_1,~\bm{b}_2+\bm{e}_{35\bar{1}},$ 
   & ${1}; {\bar{1}}; {1\bar{1}}$  \\ 
   &$\bm{b}_{\bar{1}}$ & $(r_1+2r_{\bar{1}})$\,(RS-V)+6\,V \\ \hline
III--i  & $\bm{b}_1,~\bm{b}_2+\bm{e}_{3456\bar{1}\bar{2}},$ 
   & ${1}; {\bar{1}}; {1\bar{1}}; {2\bar{1}}'; {3\bar{1}}'$\textcolor{gray}{$; {2}'; {3}'$} \\ 
   & $\bm{b}_{\bar{1}}$ 
   &  $(r_1+2r_{\bar{1}})$\,(RS-V)+($7-2r_2-2r_3$)\,2V \\ \hline
IV--i  & $\bm{b}_1+\bm{e}_{1\bar{1}},~\bm{b}_2+\bm{e}_{356\bar{2}},$ 
   & ${\bar{1}}; {3\bar{1}}'$\textcolor{gray}{$; {3}'$}  \\ 
   & $\bm{b}_{\bar{1}}$ 
   &  $2r_{\bar{1}}$\,(RS--V)+($6-4r_3$)\,V \\ \hline
%
II--ii  & $\bm{b}_1,~\bm{b}_2+\bm{e}_{35\bar{3}},$ 
   & ${1}$  \\
   & $\bm{b}_{\bar{1}}+\bm{e}_{1\bar{1}}$ & $r_1$\,(RS-V)+2\,V \\ \hline
III--ii  & $\bm{b}_1,~\bm{b}_2+\bm{e}_{3456\bar{2}\bar{4}},$ 
   & ${1}; {2\bar{1}}'; {3\bar{1}}'$\textcolor{gray}{$; {2}'; {3}'$} \\ 
   & $\bm{b}_{\bar{1}}+\bm{e}_{1\bar{1}}$ 
   &  $r_1$\,(RS-V)+6\,V   \\ \hline
IV--ii.1  & $\bm{b}_1+\bm{e}_{1\bar{2}},~\bm{b}_2+\bm{e}_{356\bar{3}},$ & ${1\bar{1}}'; {3\bar{1}}'$\textcolor{gray}{$; {3}'$} \\
   & $\bm{b}_{\bar{1}}+\bm{e}_{2\bar{1}}$ 
   &  ($3-2r_1$)\,2V \\ \hline
%
IV--ii.2 & $\bm{b}_1+\bm{e}_{1\bar{3}},~\bm{b}_2+\bm{e}_{356\bar{4}},$ & $\textcolor{gray}{{3}'}$ \\ 
   & $\bm{b}_{\bar{1}}+\bm{e}_{5\bar{1}}$ 
   &  \\ \hline 
V--ii & $\bm{b}_1+\bm{e}_{12\bar{2}\bar{3}},~\bm{b}_2+\bm{e}_{3456\bar{2}\bar{4}},$ &  ${1\bar{1}}'; {2\bar{1}}'$\textcolor{gray}{$; {1}'; {2}'; {3}'$} \\ 
   & $\bm{b}_{\bar{1}}+\bm{e}_{5\bar{1}}$ 
   & 4\,V  \\ \hline
%
%
%
II--iii & $\bm{b}_1,~\bm{b}_2+\bm{e}_{35\bar{1}},$ 
   & ${1}; {1\bar{1}}'$\textcolor{gray}{$; {\bar{1}}'$} \\ 
   & $\bm{b}_{\bar{1}}+\bm{e}_{12\bar{1}\bar{2}}$ 
   &  $r_1$\,(RS--5V)+6\,V \\ \hline
III--iii & $\bm{b}_1,~\bm{b}_2+\bm{e}_{3456\bar{3}\bar{4}},$ 
   & ${1}; {1\bar{1}}'; {2\bar{1}}'; {3\bar{1}}'$\textcolor{gray}{$; {2}'; {3}'; {\bar{1}}'$}  \\ 
   & $\bm{b}_{\bar{1}}+\bm{e}_{12\bar{1}\bar{2}}$ 
   & $r_1$\,(RS--5V)+10\,V  \\ \hline
IV--iii.1 & $\bm{b}_1+\bm{e}_{1\bar{1}},~\bm{b}_2+\bm{e}_{356\bar{2}},$ & ${3\bar{1}}$\textcolor{gray}{$; {3}'; {\bar{1}}'$} \\
   & $\bm{b}_{\bar{1}}+\bm{e}_{56\bar{1}\bar{2}}$ 
   & 2\,V  \\ \hline
IV--iii.2 & $\bm{b}_1+\bm{e}_{1\bar{1}},~\bm{b}_2+\bm{e}_{356\bar{3}},$ & ${2\bar{1}}'$\textcolor{gray}{$; {3}'; {\bar{1}}'$} \\
   & $\bm{b}_{\bar{1}}+\bm{e}_{45\bar{1}\bar{2}}$ 
   & 2\,V \\ \hline
%
%
%
IV--iii.3 & $\bm{b}_1+\bm{e}_{1\bar{3}},~\bm{b}_2+\bm{e}_{356\bar{4}},$ & ${1\bar{1}}'; {2\bar{1}}'; {3\bar{1}}'$\textcolor{gray}{$; {3}'; {\bar{1}}'$} \\ 
   & $\bm{b}_{\bar{1}}+\bm{e}_{24\bar{1}\bar{2}}$
   & 6\,V \\ \hline
V--iii & $\bm{b}_1+\bm{e}_{12\bar{1}\bar{3}},~\bm{b}_2+\bm{e}_{3456\bar{3}\bar{4}},$ & ${2\bar{1}}'$\textcolor{gray}{$; {1}'; {2}'; {3}'; {\bar{1}}'$} \\
   & $\bm{b}_{\bar{1}}+\bm{e}_{15\bar{1}\bar{2}}$ 
   & 2\,V  \\ \hline
%
VI--iii.1 & $\bm{b}_1+\bm{e}_{12\bar{1}\bar{2}},~\bm{b}_2+\bm{e}_{34\bar{3}\bar{4}},$ 
   & ${1\bar{1}}; {3\bar{1}}; {2\bar{1}}'$\textcolor{gray}{; {3}; ${1}'; {2}'; {\bar{1}}'$}  \\ 
   & $\bm{b}_{\bar{1}}+\bm{e}_{12\bar{1}\bar{2}}$ 
   &  6\,V \\ \hline
VI--iii.2 & $\bm{b}_1+\bm{e}_{12\bar{1}\bar{3}},~\bm{b}_2+\bm{e}_{34\bar{2}\bar{4}},$ 
   & $3\bar{1}$\textcolor{gray}{$; 3; {1}'; {2}'; {\bar{1}}'$}  \\ 
   & $\bm{b}_{\bar{1}}+\bm{e}_{13\bar{1}\bar{2}}$ 
   & 2\,V   
%
%
\\ \hline\hline
%
%
\multicolumn{3}{|c|}{$\Intr_{2L}\times \Intr_{2R} \times \Intr_2$ \hfill $\mathcal{N}=2$ : SG+V+2\,H} \\
\multicolumn{3}{|r|}{$r=\sfrac12\Big(1+\CC{\bm{B}_{1}}{\bm{B}_{\bar{1}}}\Big),~ 
h=\sfrac 12\Big(1+\CC{\bar{\bm{S}}}{\bar{\bm{S}}}\CC{\bm{B}_{1\bar{1}}}{\bm{B}_{2\bar{2}}}\Big)~
h'=\sfrac 12\Big(1-\CC{\bm{B}_{1}}{\bm{B}_{\bar{1}}}\CC{\bar{\bm{S}}}{\bar{\bm{S}}}\CC{\bm{B}_{1\bar{1}}}{\bm{B}_{2\bar{2}}}\Big)=0,1$}
\\ \hline 
i--i--A  & $\bm{b}_1~, \bm{b}_{\bar{1}},$ 
   & $1; \bar{1}; 1\bar{1}; 2\bar{2}; 3\bar{2}; 2\bar{3}; 3\bar{3}$  \\
   & $\bm{b}_{2\bar{2}}$ & $2r$\,(RS--H)+14\,(V+H)  \\ \hline
i--ii--A  & $\bm{b}_1~, \bm{b}_{\bar{1}}+\bm{e}_{1\bar{1}},$ 
   & $1; 2\bar{2}; 3\bar{2}; 2\bar{3}; 3\bar{3}$  \\ 
   & $\bm{b}_{2\bar{2}}$ & $r$\,(RS--H)+10\,(V+H)  \\ \hline
ii--i--A  & $\bm{b}_1+\bm{e}_{1\bar{1}}~, \bm{b}_{\bar{1}},$ 
   & $\bar{1}; 2\bar{2}; 3\bar{2}; 2\bar{3}; 3\bar{3}$  \\
   & $\bm{b}_{2\bar{2}}$ & $r$\,(RS--H)+10\,(V+H)  \\ \hline
ii--ii--A.1  & $\bm{b}_1+\bm{e}_{1\bar{1}}~, \bm{b}_{\bar{1}}+\bm{e}_{1\bar{1}},$ 
   & $1\bar{1}; 2\bar{2}; 3\bar{2}; 2\bar{3}; 3\bar{3}$  \\ 
   & $\bm{b}_{2\bar{2}}$ & 10\,(V+H)  \\ \hline
ii--ii--A.2  & $\bm{b}_1+\bm{e}_{1\bar{2}}~, \bm{b}_{\bar{1}}+\bm{e}_{2\bar{1}},$ 
   & $2\bar{2}; 3\bar{2}; 2\bar{3}; 3\bar{3}; 1\bar{1}'$ \\ 
   & $\bm{b}_{2\bar{2}}$ & 
4(($4-2r$)$h$+2$h'$)\,(H--V)+($24-8r$)\,V  \\ \hline
ii--ii--A.3  & $\bm{b}_1+\bm{e}_{1\bar{5}}~, \bm{b}_{\bar{1}}+\bm{e}_{5\bar{1}},$ 
   & $2\bar{2}; 3\bar{2}; 2\bar{3}$  \\
   & $\bm{b}_{2\bar{2}}$ & 6\,(V+H)  \\ \hline
i--iii--A  & $\bm{b}_1~, \bm{b}_{\bar{1}}+\bm{e}_{12\bar{1}\bar{2}},$ 
   & $1; 2\bar{2}; 3\bar{2}; 2\bar{3}; 3\bar{3}; 1\bar{1}'$\textcolor{gray}{$; \bar{1}'$}  \\ 
   & $\bm{b}_{2\bar{2}}$ 
   &  $r$\,(RS--H)+4(($4-2r$)$h$+2$h'$)\,(H--V) +2\,H+($26-8r$)\,V \\ \hline
iii--i--A & $\bm{b}_1+\bm{e}_{12\bar{1}\bar{2}}~, \bm{b}_{\bar{1}},$ 
   & $\bar{1}; 2\bar{2}; 3\bar{2}; 2\bar{3}; 3\bar{3}; 1\bar{1}'$\textcolor{gray}{$; 1'$}   \\ 
   & $\bm{b}_{2\bar{2}}$ 
   & $r$\,(RS--H)+4(($4-2r$)$h$+2$h'$)\,(H--V) +2\,H+($26-8r$)\,V   \\ \hline
ii--iii--A.1  & $\bm{b}_1+\bm{e}_{1\bar{1}}~, \bm{b}_{\bar{1}}+\bm{e}_{12\bar{1}\bar{2}},$ 
   & $2\bar{2}; 3\bar{2}; 2\bar{3}; 3\bar{3}$\textcolor{gray}{$; \bar{1}'$}  \\ 
   & $\bm{b}_{2\bar{2}}$ 
   &  8\,(V+H)  \\ \hline
iii--ii--A.1 & $\bm{b}_1+\bm{e}_{12\bar{1}\bar{2}}~, \bm{b}_{\bar{1}}+\bm{e}_{1\bar{1}},$ 
   & $2\bar{2}; 3\bar{2}; 2\bar{3}; 3\bar{3}$\textcolor{gray}{$; 1'$}   \\
   & $\bm{b}_{2\bar{2}}$ 
   &  8\,(V+H)  \\ \hline
ii--iii--A.2 & $\bm{b}_1+\bm{e}_{1\bar{5}}~, \bm{b}_{\bar{1}}+\bm{e}_{25\bar{1}\bar{2}},$ 
   & $2\bar{2}; 3\bar{2}; 2\bar{3}; 1\bar{1}'$\textcolor{gray}{$; \bar{1}'$}  \\ 
   & $\bm{b}_{2\bar{2}}$ 
   & 12$h'$\,(H--V)+2\,H+14\,V   \\ \hline
iii--ii--A.2 & $\bm{b}_1+\bm{e}_{12\bar{2}\bar{5}}~, \bm{b}_{\bar{1}}+\bm{e}_{5\bar{1}},$ 
   & $2\bar{2}; 3\bar{2}; 2\bar{3}; 1\bar{1}'$\textcolor{gray}{$; \bar{1}'$}  \\ 
   & $\bm{b}_{2\bar{2}}$ 
   & 12$h'$\,(H--V)+2\,H+14\,V   \\ \hline
iii--iii--A.1 & $\bm{b}_1+\bm{e}_{12\bar{1}\bar{2}}~, \bm{b}_{\bar{1}}+\bm{e}_{12\bar{1}\bar{2}},$ 
   & $1\bar{1}; 2\bar{2}; 3\bar{2}; 2\bar{3}; 3\bar{3}$\textcolor{gray}{$; 1'; \bar{1}'$}   \\ 
   & $\bm{b}_{2\bar{2}}$ 
   &  10\,(V+H)  \\ \hline
iii--iii--A.2 & $\bm{b}_1+\bm{e}_{12\bar{1}\bar{5}}~, \bm{b}_{\bar{1}}+\bm{e}_{15\bar{1}\bar{2}},$ 
   & $2\bar{2}; 3\bar{2}; 2\bar{3}$\textcolor{gray}{$; 1'; \bar{1}'$}   \\ 
   & $\bm{b}_{2\bar{2}}$ 
   & 6\,(V+H)   \\ \hline
iii-iii--A.3 & $\bm{b}_1+\bm{e}_{12\bar{5}\bar{6}}~, \bm{b}_{\bar{1}}+\bm{e}_{56\bar{1}\bar{2}},$ 
   & $2\bar{2}; 3\bar{2}; 2\bar{3}; 3\bar{3}'; 1\bar{1}'$\textcolor{gray}{$; 1'; \bar{1}'$}   \\ 
   & $\bm{b}_{2\bar{2}}$ 
   &  12$h'$\,(H--V)+4\,H+16\,V  \\ \hline
ii--ii--B & $\bm{b}_1+\bm{e}_{1\bar{3}}~, \bm{b}_{\bar{1}}+\bm{e}_{3\bar{1}},$ 
   & $2\bar{2}$  \\  
   & $\bm{b}_{2\bar{2}}+\bm{e}_{5\bar{5}}$ & 2\,(V+H)  \\ \hline
ii--iii--B & $\bm{b}_1+\bm{e}_{1\bar{3}}~, \bm{b}_{\bar{1}}+\bm{e}_{23\bar{1}\bar{2}},$ 
   & $2\bar{2}; 1\bar{1}'$\textcolor{gray}{$; \bar{1}'$}   \\ 
   & $\bm{b}_{2\bar{2}}+\bm{e}_{5\bar{5}}$ 
   & 4\,(V+H)   \\ \hline
iii--ii--B & $\bm{b}_1+\bm{e}_{12\bar{2}\bar{3}}~, \bm{b}_{\bar{1}}+\bm{e}_{3\bar{1}},$ 
   & $2\bar{2}; 1\bar{1}'$\textcolor{gray}{$; 1'$ }   \\ 
   & $\bm{b}_{2\bar{2}}+\bm{e}_{5\bar{5}}$ 
   & 4\,(V+H)  \\ \hline
iii--iii--B.1 & $\bm{b}_1+\bm{e}_{12\bar{1}\bar{3}}~, \bm{b}_{\bar{1}}+\bm{e}_{13\bar{1}\bar{2}},$ 
   & $2\bar{2}$\textcolor{gray}{$; 1'; \bar{1}'$}   \\ 
   & $\bm{b}_{2\bar{2}}+\bm{e}_{5\bar{5}}$ 
   &  2\,(V+H)   \\ \hline
iii--iii--B.2 & $\bm{b}_1+\bm{e}_{12\bar{3}\bar{5}}~, \bm{b}_{\bar{1}}+\bm{e}_{35\bar{1}\bar{2}},$ 
   & $2\bar{2}; 3\bar{3}; 1\bar{1}'$\textcolor{gray}{$; 1'; \bar{1}'$}   \\ 
   & $\bm{b}_{2\bar{2}}+\bm{e}_{5\bar{5}}$ 
   & 12$h$\,(H--V)+12\,V  \\ \hline
iii--iii--B.3 & $\bm{b}_1+\bm{e}_{12\bar{3}\bar{6}}~, \bm{b}_{\bar{1}}+\bm{e}_{36\bar{1}\bar{2}},$ 
   & $2\bar{2}; 1\bar{1}'; 3\bar{3}'$\textcolor{gray}{$; 1'; \bar{1}'$}   \\ 
   & $\bm{b}_{2\bar{2}}+\bm{e}_{5\bar{5}}$ 
   &  6\,(V+H)  \\ \hline
i--i--E & $\bm{b}_1~, \bm{b}_{\bar{1}},$ 
   & $1; \bar{1}; 1\bar{1}$  \\ 
   & $\bm{b}_{2\bar{2}}+\bm{e}_{35\bar{3}\bar{5}}$ & $2r$\,(RS--H)+6\,(V+H)  \\ \hline
i--ii--E & $\bm{b}_1~, \bm{b}_{\bar{1}}+\bm{e}_{1\bar{1}},$ 
   & $1$  \\
   & $\bm{b}_{2\bar{2}}+\bm{e}_{35\bar{3}\bar{5}}$ &  $r$\,(RS--H)+2\,(V+H) \\ \hline
%
ii--i--E & $\bm{b}_1+\bm{e}_{1\bar{1}}~, \bm{b}_{\bar{1}},$ 
   & $\bar{1}$  \\ 
   & $\bm{b}_{2\bar{2}}+\bm{e}_{35\bar{3}\bar{5}}$ & $r$\,(RS--H)+2\,(V+H)  \\ \hline
ii--ii--E.1 & $\bm{b}_1+\bm{e}_{1\bar{1}}~, \bm{b}_{\bar{1}}+\bm{e}_{1\bar{1}},$ & $1\bar{1}$ \\ 
   & $\bm{b}_{2\bar{2}}+\bm{e}_{35\bar{3}\bar{5}}$ & 2\,V+2\,H \\ \hline
ii--ii--E.2 & $\bm{b}_1+\bm{e}_{1\bar{2}}~, \bm{b}_{\bar{1}}+\bm{e}_{2\bar{1}},$ & $1\bar{1}'$ \\ 
   & $\bm{b}_{2\bar{2}}+\bm{e}_{35\bar{3}\bar{5}}$ & 4(1--$r$)\,(V+H)  \\ \hline
i--iii--E & $\bm{b}_1~, \bm{b}_{\bar{1}}+\bm{e}_{12\bar{1}\bar{2}},$ 
   & $1; 1\bar{1}'$\textcolor{gray}{$; \bar{1}'$}   \\ 
   & $\bm{b}_{2\bar{2}}+\bm{e}_{35\bar{3}\bar{5}}$ 
   &  $r$\,(RS--4V--5H)+6\,(V+H)  \\ \hline
iii--i--E & $\bm{b}_1+\bm{e}_{12\bar{1}\bar{2}}~, \bm{b}_{\bar{1}},$ 
   & $\bar{1}; 1\bar{1}'$\textcolor{gray}{;  $1'$}  \\
   & $\bm{b}_{2\bar{2}}+\bm{e}_{35\bar{3}\bar{5}}$ 
   & $r$\,(RS--4V--5H)+6\,(V+H)  \\ \hline
ii--iii--E.1 & $\bm{b}_1+\bm{e}_{1\bar{1}}~, \bm{b}_{\bar{1}}+\bm{e}_{12\bar{1}\bar{2}},$ & \textcolor{gray}{$\bar{1}'$}  \\ 
   & $\bm{b}_{2\bar{2}}+\bm{e}_{35\bar{3}\bar{5}}$ 
   &    \\ \hline
iii--ii--E.1 & $\bm{b}_1+\bm{e}_{12\bar{1}\bar{2}}~, \bm{b}_{\bar{1}}+\bm{e}_{1\bar{1}},$ & \textcolor{gray}{$1'$}  \\ 
   & $\bm{b}_{2\bar{2}}+\bm{e}_{35\bar{3}\bar{5}}$ 
   &    \\ \hline
ii--iii--E.2 & $\bm{b}_1+\bm{e}_{1\bar{1}}~, \bm{b}_{\bar{1}}+\bm{e}_{34\bar{1}\bar{2}},$ & \textcolor{gray}{$\bar{1}'$}  \\ 
   & $\bm{b}_{2\bar{2}}+\bm{e}_{35\bar{3}\bar{5}}$ 
   &    \\ \hline
iii--ii--E.2 & $\bm{b}_1+\bm{e}_{12\bar{3}\bar{4}}~, \bm{b}_{\bar{1}}+\bm{e}_{1\bar{1}},$ &\textcolor{gray}{$1'$}  \\ 
   & $\bm{b}_{2\bar{2}}+\bm{e}_{35\bar{3}\bar{5}}$ 
   &    \\ \hline
iii--iii--E.1 & $\bm{b}_1+\bm{e}_{12\bar{1}\bar{2}}~, \bm{b}_{\bar{1}}+\bm{e}_{12\bar{1}\bar{2}},$ & $1\bar{1}$\textcolor{gray}{$; 1'; \bar{1}'$} \\ 
   & $\bm{b}_{2\bar{2}}+\bm{e}_{35\bar{3}\bar{5}}$ 
   &  2\,(V+H)  \\ \hline
iii--iii--E.2 & $\bm{b}_1+\bm{e}_{12\bar{3}\bar{4}}~, \bm{b}_{\bar{1}}+\bm{e}_{34\bar{1}\bar{2}},$ & $1\bar{1}'$\textcolor{gray}{$; 1'; \bar{1}'$} \\
   & $\bm{b}_{2\bar{2}}+\bm{e}_{35\bar{3}\bar{5}}$ 
   &  2\,(V+H) 
%
%
\\ \hline\hline
%
%
\multicolumn{3}{|c|}{$\Intr_{2L}^2\times \Intr_{2R}^2$ \hfill $\mathcal{N}=2$ : SG + H} \\ 
\multicolumn{3}{|r|}{$r_\ga=\sfrac14\Big(1+\sum\limits_\bgb\CC{\bm{B}_\ga}{\bm{B}_\bgb}\Big)$, $r_\bgb=\sfrac14\Big(1+\sum\limits_\ga\CC{\bm{B}_\ga}{\bm{B}_\bgb}\Big)$, $r=\sum\limits_\ga r_\ga = \sum\limits_{\bgb} r_{\bgb}$} \\ 
\multicolumn{3}{|r|}{$p{}_{\ga\bgb} = \sfrac12\Big( 1 - \CC{\bm{B}_\ga}{\bm{B}_\bgb} \Big)$, 
$h_{\ga\bgb} = \sfrac12\Big( 1 +  \CC{\bar{\bm{S}}}{\bar{\bm{S}}} \CC{\bm{B}_{\gb\bga}}{\bm{B}_{\ga\bgb}} \Big)$, 
$h_{\ga\bgb}' = \sfrac12\Big( 1 -  \CC{\bar{\bm{S}}}{\bar{\bm{S}}} \CC{\bm{B}_\ga}{\bm{B}_\bgb} \CC{\bm{B}_{\gb\bga}}{\bm{B}_{\ga\bgb}} \Big)$
}

\\ \hline
I--I  & $\bm{b}_1,~\bm{b}_2,$ 
   & $1; 2; 3; \bar{1}; \bar{2}; \bar{3}; 1\bar{1}; 1\bar{2}; 2\bar{1}; 2\bar{2}; 3\bar{1}; 3\bar{2}; 1\bar{3}; 2\bar{3}; 3\bar{3}$  \\ 
   & $\bm{b}_{\bar{1}},~\bm{b}_{\bar{2}}$ 
   & 2$r$\,(RS--H)+15\,(V+H) \\ \hline
II--II  & $\bm{b}_1,~\bm{b}_2+\bm{e}_{35\bar{1}},$  
   &  $1; \bar{1}; 1\bar{1}$  \\ 
   & $\bm{b}_{\bar{1}},~\bm{b}_{\bar{2}}+\bm{e}_{1\bar{3}\bar{5}}$ 
   & ($r_1$+$r_{\bar{1}}$)\,(RS--H)+3\,(V+H) \\ \hline
II--IV  & $\bm{b}_1,~\bm{b}_2+\bm{e}_{35\bar{5}},$  
   &  $1; 1\bar{3}'$\textcolor{gray}{$; \bar{3}'$} \\ 
   & $\bm{b}_{\bar{1}}+\bm{e}_{1\bar{1}},~\bm{b}_{\bar{2}}+\bm{e}_{2\bar{3}\bar{5}\bar{6}}$ 
   & $r_1$\,(RS--H)+($1+2p{}_{1\bar{3}}$)\,(V+H) \\ \hline
IV--II  & $\bm{b}_1+\bm{e}_{1\bar{1}},~\bm{b}_2+\bm{e}_{356\bar{2}},$  
   &  $\bar{1}; 3\bar{1}'$\textcolor{gray}{$; 3'$}  \\ 
   & $\bm{b}_{\bar{1}},~\bm{b}_{\bar{2}}+\bm{e}_{5\bar{3}\bar{5}}$ 
   & $r_{\bar{1}}$\,(RS--H)+($1+2p_{3\bar{1}}$)\,(V+H) \\ \hline
III--III  & $\bm{b}_1,~\bm{b}_2+\bm{e}_{3456\bar{1}\bar{2}},$ 
   &  $1; \bar{1}; 1\bar{1}; 1\bar{2}'; 2\bar{1}'; 2\bar{2}'; 3\bar{1}'; 3\bar{2}'; 1\bar{3}'; 2\bar{3}'; 3\bar{3}';$  \textcolor{gray}{$2'; 3'; \bar{2}'; \bar{3}'$} \\ 
   & $\bm{b}_{\bar{1}},~\bm{b}_{\bar{2}}+\bm{e}_{12\bar{3}\bar{4}\bar{5}\bar{6}}$ 
   &  ($r_1+r_{\bar{1}}$)\,(RS--H)+2[$2(p_{2\bar{1}}h_{2\bar{1}}+p_{1\bar{2}}h_{1\bar{2}}+p_{3\bar{1}}h_{1\bar{2}}'+p_{1\bar{3}}h_{2\bar{1}}')+h_{2\bar{2}}'$ \\ 
   & & $+h_{2\bar{1}}+h_{2\bar{1}}'+h_{2\bar{1}}'$]\,(H--V) 
   +[$11+4(p_{1\bar{2}}+p_{2\bar{1}}+p_{3\bar{1}}+p_{1\bar{3}})$]\,V+3\,H
   \\ \hline
III--IV  & $\bm{b}_1,~\bm{b}_2+\bm{e}_{3456\bar{2}\bar{4}},$ 
   &  $1; 2\bar{1}'; 2\bar{2}'; 3\bar{1}'; 3\bar{2}'; 1\bar{3}'; 2\bar{3}'; 3\bar{3}'$\textcolor{gray}{$; 2'; 3'; \bar{3}'$}  \\ 
   & $\bm{b}_{\bar{1}}+\bm{e}_{1\bar{1}},~\bm{b}_{\bar{2}}+\bm{e}_{2\bar{3}\bar{5}\bar{6}}$ 
   & $r_1$\,(RS--H)+4($h_{2\bar{1}}'+h_{2\bar{2}}'+p_{1\bar{3}}h_{2\bar{2}}'$)\,(H--V)  
   +($11+4p_{1\bar{3}}$)\,V + 3\,H \\ \hline
IV--III  & $\bm{b}_{1}+\bm{e}_{1\bar{1}},~\bm{b}_{2}+\bm{e}_{356\bar{2}},$ 
   &  $\bar{1}; 1\bar{2}'; 2\bar{2}'; 3\bar{1}'; 3\bar{2}'; 1\bar{3}'; 2\bar{3}'; 3\bar{3}'$\textcolor{gray}{$; 3'; \bar{2}'; \bar{3}'$}  \\ 
   & $\bm{b}_{\bar{1}},~\bm{b}_{\bar{2}}+\bm{e}_{24\bar{3}\bar{4}\bar{5}\bar{6}}$ 
   &  $r_{\bar{1}}$\,(RS--H)+4($h_{1\bar{2}}'+h_{2\bar{2}}'+p_{3\bar{1}}h_{2\bar{2}}'$)\,(H--V)  
   +($11+4p_{3\bar{1}}$)\,V + 3\,H \\ \hline
IV--IV.1  & $\bm{b}_1+\bm{e}_{1\bar{2}},~\bm{b}_2+\bm{e}_{356\bar{4}},$  
   &  $1\bar{1}'; 2\bar{2}'; 3\bar{1}'; 3\bar{2}'; 1\bar{3}'; 2\bar{3}'; 3\bar{3}'$\textcolor{gray}{$; 3'; \bar{3}'$}  \\ 
   & $\bm{b}_{\bar{1}}+\bm{e}_{2\bar{1}},~\bm{b}_{\bar{2}}+\bm{e}_{4\bar{3}\bar{5}\bar{6}}$ 
   &  2[$2h_{1\bar{1}}'+2h_{2\bar{2}}'+(1+2p_{1\bar{1}}+2p_{2\bar{2}})h_{1\bar{1}}$]\,(H--V)+($10+4p_{1\bar{1}}+4p_{2\bar{2}}$)\,V \\ \hline
IV--IV.2  & $\bm{b}_1+\bm{e}_{1\bar{2}},~\bm{b}_2+\bm{e}_{356\bar{5}},$  
   &  $1\bar{1}'; 3\bar{1}'; 1\bar{3}'$\textcolor{gray}{$; 3'; \bar{3}'$}  \\ 
   & $\bm{b}_{\bar{1}}+\bm{e}_{2\bar{1}},~\bm{b}_{\bar{2}}+\bm{e}_{5\bar{3}\bar{5}\bar{6}}$ 
   & 2($1+p_{1\bar{1}}$)\,(V+H) \\ \hline
IV--IV.3  & $\bm{b}_1+\bm{e}_{1\bar{5}},~\bm{b}_2+\bm{e}_{356\bar{6}},$  
   &  $3\bar{3}$\textcolor{gray}{$; 3'; \bar{3}'$}  \\ 
   & $\bm{b}_{\bar{1}}+\bm{e}_{5\bar{1}},~\bm{b}_{\bar{2}}+\bm{e}_{6\bar{3}\bar{5}\bar{6}}$ 
   & V+H \\ \hline
%
IV--V  & $\bm{b}_1+\bm{e}_{1\bar{1}},~\bm{b}_2+\bm{e}_{356\bar{3}},$  &  $2\bar{3}'; 3\bar{3}'; 1\bar{2}'; 2\bar{1}';1\bar{3}'$\textcolor{gray}{$; 3'; \bar{1}'; \bar{2}'; \bar{3}'$} \\ 
   & $\bm{b}_{\bar{1}}+\bm{e}_{45\bar{1}\bar{2}},~\bm{b}_{\bar{2}}+\bm{e}_{25\bar{3}\bar{4}\bar{5}\bar{6}}$ 
   &  $6h_{2\bar{1}}$\,(H--V)+8\,V+2\,H \\ \hline
V--IV  & $\bm{b}_1+\bm{e}_{12\bar{4}\bar{5}},~\bm{b}_2+\bm{e}_{3456\bar{2}\bar{5}},$  & $1\bar{2}'; 2\bar{1}';3\bar{1}'; 3\bar{2}'; 3\bar{3}'$\textcolor{gray}{$; 1'; 2'; 3'; \bar{3}'$} \\ 
   & $\bm{b}_{\bar{1}}+\bm{e}_{1\bar{1}},~\bm{b}_{\bar{2}}+\bm{e}_{3\bar{3}\bar{5}\bar{6}}$ 
   & $6h_{1\bar{2}}$\,(H--V)+8\,V+2\,H  \\ \hline
V--V & $\bm{b}_1+\bm{e}_{12\bar{1}\bar{3}},~\bm{b}_2+\bm{e}_{3456\bar{1}\bar{5}},$  & $2\bar{2}'; 3\bar{1}'; 1\bar{3}'$\textcolor{gray}{$; 1'; 2'; 3'; \bar{1}'; \bar{2}'; \bar{3}'$} \\ 
   & $\bm{b}_{\bar{1}}+\bm{e}_{13\bar{1}\bar{2}},~\bm{b}_{\bar{2}}+\bm{e}_{15\bar{3}\bar{4}\bar{5}\bar{6}}$ 
   & $6h_{2\bar{2}}'$\,(H--V)+6\,V  \\ \hline
VI--VI.1 & $\bm{b}_1+\bm{e}_{12\bar{1}\bar{2}},~\bm{b}_2+\bm{e}_{34\bar{3}\bar{4}},$ 
   &  $1\bar{1}; 2\bar{2}; 3\bar{1}; 3\bar{2}; 1\bar{3}; 2\bar{3}; 3\bar{3}; 1\bar{2}'; 2\bar{1}';$  \textcolor{gray}{$3; \bar{3}; 1'; 2'; \bar{1}'; \bar{2}'$}  \\  
   & $\bm{b}_{\bar{1}}+\bm{e}_{12\bar{1}\bar{2}},~\bm{b}_{\bar{2}}+\bm{e}_{34\bar{3}\bar{4}}$ 
   & $6(h_{1\bar{2}}'+h_{2\bar{1}}')$\,(H--V)+15\,V+3\,H \\ \hline
%
VI--VI.2 & $\bm{b}_1+\bm{e}_{12\bar{1}\bar{2}},~\bm{b}_2+\bm{e}_{34\bar{3}\bar{5}},$ 
   &  $1\bar{1};  3\bar{1}; 1\bar{3}; 1\bar{2}'; 2\bar{1}'$\textcolor{gray}{$; 3; \bar{3}; 1'; 2'; \bar{1}'; \bar{2}'$}  \\  
   & $\bm{b}_{\bar{1}}+\bm{e}_{12\bar{1}\bar{2}},~\bm{b}_{\bar{2}}+\bm{e}_{35\bar{3}\bar{4}}$ 
   &  5\,(V+H) \\ \hline
VI--VI.3 & $\bm{b}_1+\bm{e}_{12\bar{1}\bar{3}},~\bm{b}_2+\bm{e}_{34\bar{2}\bar{5}},$ 
   &  $3\bar{1}; 1\bar{3}; 2\bar{2}'$\textcolor{gray}{$; 3; \bar{3}; 1'; 2'; \bar{1}'; \bar{2}'$}   \\  
38   & $\bm{b}_{\bar{1}}+\bm{e}_{13\bar{1}\bar{2}},~\bm{b}_{\bar{2}}+\bm{e}_{25\bar{3}\bar{4}}$ 
   & $6h_{2\bar{2}}'$\,(H--V)+6\,V  \\ \hline
\end{longtable}


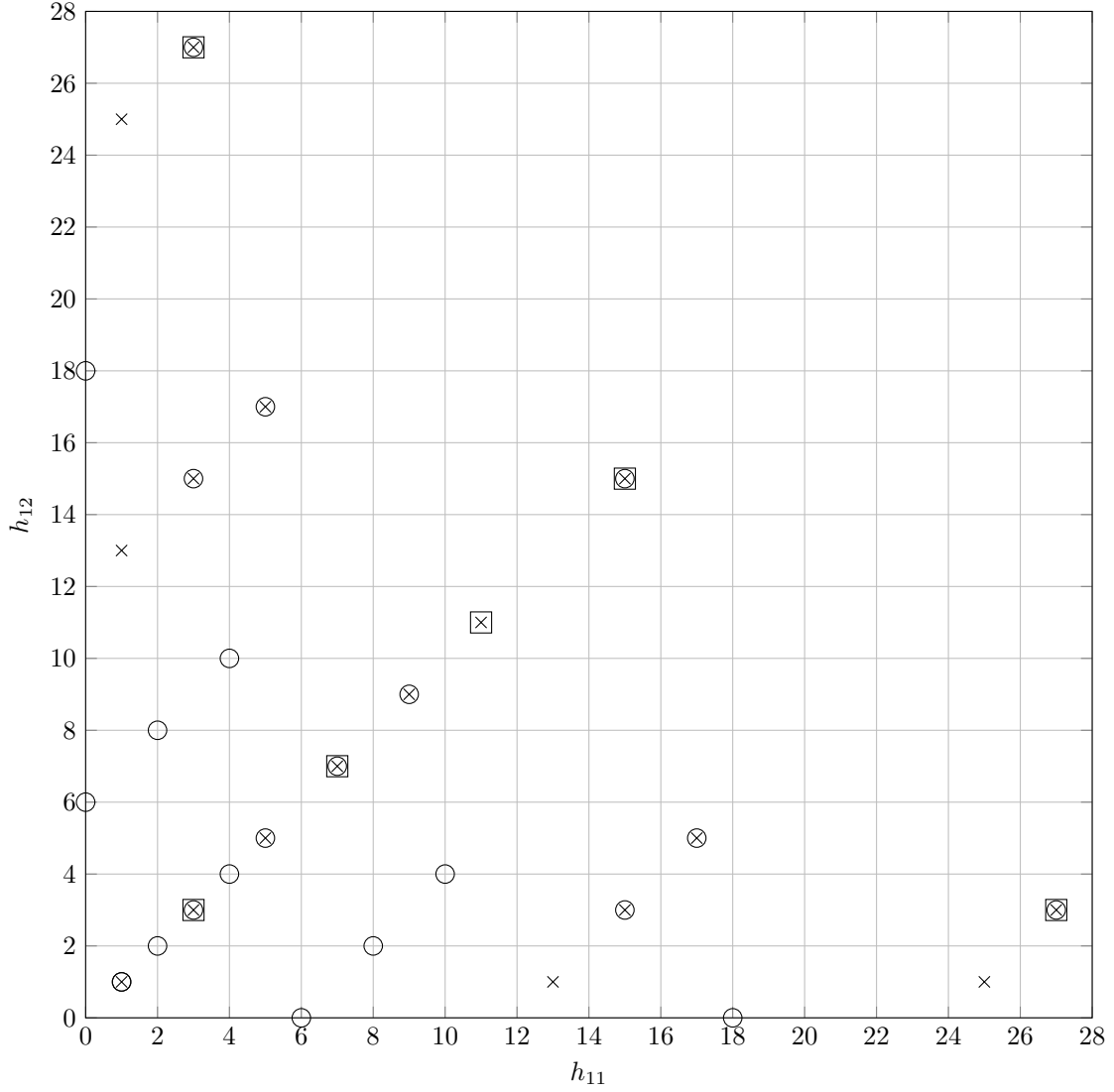
\begin{figure}[t]
\centering
\begin{tikzpicture}
\begin{axis}[width=15cm,height=15cm,
xmin=0,xmax=28,xlabel=$h_{11}$,
ymin=0,ymax=28,ylabel=$h_{12}$,
grid=major,
scatter/classes={%
    z22={mark=square,draw=black,scale=2},    
    zLzRz2={mark=x,draw=black,scale=1.5},         
    zL2zR2={mark=o,draw=black,scale=1.75}}]        
%
\addplot[scatter,only marks,%
    scatter src=explicit symbolic]%
table[meta=label] {
x y label
15  15  z22 
27  3   z22 
3   27  z22
11  11  z22
7   7   z22  
3   3   z22 
15  15  zLzRz2
11  11  zLzRz2
25  1   zLzRz2
1   25  zLzRz2
7   7   zLzRz2
27  3   zLzRz2
3   27  zLzRz2
9   9   zLzRz2
15  3   zLzRz2
3   15  zLzRz2
5   17  zLzRz2
17  5   zLzRz2
3   3   zLzRz2
5   5   zLzRz2
13  1   zLzRz2
1   13  zLzRz2
1   1   zLzRz2
15  15  zL2zR2
3   3   zL2zR2
1   1   zL2zR2
17  5   zL2zR2
5   17  zL2zR2
3   27  zL2zR2
27  3   zL2zR2
15  3   zL2zR2
3   15  zL2zR2
7   7   zL2zR2
10  4   zL2zR2
4   10  zL2zR2
8   2   zL2zR2
2   8   zL2zR2
0   18  zL2zR2
18  0   zL2zR2
2   2   zL2zR2
4   4   zL2zR2
1   1   zL2zR2
0   6   zL2zR2
6   0   zL2zR2
9   9   zL2zR2
5   5   zL2zR2
};
\end{axis}
\end{tikzpicture}
\caption{\label{fig:h11h12ScatterPlot} The possible effective Hodge numbers $(h_{11},h_{12})$ for the $\cN=2$ T--folds, $\square: \Intr_2^2$\,; $\cross: \Intr_{2L}\times\Intr_{2R}\times\Intr_2$ and $\text{\raisebox{0.5ex}{\scalebox{0.75}{$\bigcirc$}}}: \Intr_{2L}^2\times\Intr_{2R}^2$, are displayed  }
\end{figure}

\section{Applications and extensions}
\label{sc:ApplicationsExtensions}

\subsection{Search for configurations with low effective Hodge numbers}
\label{sc:LowHodgeNumbers}

In light of moduli stabilisation, models with small number of scalars are of interest as then there are fewer moduli that need to be stabilised by some (non--perturbative) mechanism. There is renewed interest in this question see {\em e.g.}~\cite{Baykara:2023plc,Aldazabal:2025zht}. 

For smooth Calabi--Yau compactifications of type IIB strings the Hodge numbers $(h_{11},h_{12})$ are related to  the number of hyper and vector multiplets in the resulting $\cN=2$ effective field theories in four dimensions: 
\equ{ \label{eq:EffHodgeNumbers}
    h_{11} = N_H-1~, 
    \qquad
    h_{12} = N_V~. 
}
(For type IIA the Hodge numbers are interchanged.) As hyper multiplets contain 4 and vector multiplets 2 scalars, the total number of scalars equals 
\equ{
 N_\text{scalars} = 4\, N_H + 2\, N_V = 4\,(h_{11}+1) +2\, h_{12}~, 
}
and hence, constructions with low Hodge numbers attract special attention. For T--folds the notion of Hodge numbers is somewhat unclear, as they are not manifolds. But as suggested by {\em e.g.}~\cite{Anastasopoulos:2009kj,Pradisi:2010ee,Bianchi:2009mu} in analogy to Calabi--Yau compactifications, the number of hyper and vector multiplets can be used to define the ``effective Hodge numbers'' via \eqref{eq:EffHodgeNumbers}. 

In our classification of order--two T--folds, only the orbifolds $\Intr_2^2$, $\Intr_{2L}\times\Intr_{2R}\times\Intr_{2}$ and $\Intr_{2L}^2\times\Intr_{2R}^2$ possess $\cN=2$ supersymmetry. For those orbifolds \cref{fig:h11h12ScatterPlot} plots the effective Hodge number combinations that can be deduced from our classification \cref{tab:NarainRotoTransClass}. From this figure it can be read off that T--folds with the smallest number of scalars are those with effective Hodge numbers $(1,1)$ and they can be produced from both $\Intr_{2L}\times\Intr_{2R}\times\Intr_{2}$ and $\Intr_{2L}^2\times\Intr_{2R}^2$. Let us look at both cases in detail: 

There is a single configuration of the $\Intr_{2L}^2\times\Intr_{2R}^2$ T--fold with effective Hodge numbers $(1,1)$, namely configuration IV--IV.3 according to \cref{tab:NarainRotoTransClass}. This configuration was also identified by \cite{Anastasopoulos:2009kj}. Note that beside the universal hyper multiplet in the untwisted sector (containing the dilaton, axion, and two RR--scalars) the additional hyper and vector reside in the twisted sector $3\bar{3}$. 

In addition to this, there are a number of $\Intr_{2L}\times\Intr_{2R}\times\Intr_{2}$ T--fold configurations with effective Hodge numbers $(1,1)$. In fact, for all these cases the twisted massless spectrum is empty, the two hyper and single vector multiplets all reside in the untwisted sector, see \cref{tab:NarainRotoTransClass}. For the configurations ii--iii--E.1,2 and their left/right conjugates iii--ii--E.1,2 this is true irrespectively of the GGSO phases between $\bm{B}_1,\bm{B}_{\bar{1}}$ and between $\bm{B}_{1\bar{1}}, \bm{B}_{2\bar{2}}$. In addition, for the configuration ii--ii--E.2 the twisted sector is empty if $\CC{\bm{B}_1}{\bm{B}_{\bar{1}}}=+1$. 

Ref.~\cite{Dolivet:2007sz} provides a construction of a free fermionic model free of hyper multiplets. It focuses on two models: Their $\cN=2+1=3$ model corresponds to the $\Intr_{2L}^2\times\Intr_{2R}$ T--fold configuration IV--ii.2 in our classification \cref{tab:NarainRotoTransClass}. (To see this, first write their model in our conventions: it is a $\Intr_{2L}\times\Intr_{2R}^2$ orbifold with basis vectors $\bm{B}_1=\bm{b}_1+\bm{e}_{1\bar{5}}$, $\bm{B}_{\bar{1}}=\bm{b}_{\bar{1}}+\bm{e}_{5\bar{1}}$ and $\bm{B}_{\bar{2}}=\bm{b}_{\bar{2}}+\bm{e}_{6\bar{3}\bar{5}\bar{6}}$. Next, fully interchange the left-- and right--moving sectors to obtain the $\Intr_{2L}^2\times\Intr_{2R}$ variant of their model. Then by using equivalence E3 to interchange the anti--holomorphic coordinate labels $\bar{3}\leftrightarrow\bar{5}$ and $\bar{4}\leftrightarrow\bar{6}$, the basis vectors become those of the $\Intr_{2L}^2\times\Intr_{2R}$ T--fold configuration IV--ii.2.) Their second model corresponds to a supersymmetry breaking shift orbifold of the first model; which is not considered in our paper. It is the second model which the authors of~\cite{Dolivet:2007sz} claim is free of hyper multiplets.

\begin{table}
    \centering
    \begin{tabular}{|c||l | l| l|}
    \hline 
    \!\!\!\textbf{$\boldsymbol{(h_{11},h_{21})}$}\!\!\! & \textbf{Twist basis vectors} & \textbf{Label} & \textbf{Equivalent using} 
    \\ \hline\hline 
        $(1,1)$ & $\arry{l}{\bm{b}_1+\bm{e}_{1\bar{5}}, \bm{b}_2+\bm{e}_{3\bar{1}\bar{2}\bar{3}\bar{4}\bar{5}}, \\[-0.5ex]  \bm{b}_{\bar{1}}+\bm{e}_{5\bar{1}}, \bm{b}_{\bar{2}}+\bm{e}_{12345\bar{3}}}$ & IV--IV.3 & 
        \tabu{p{6.5cm}}{E1:~add $\bm{E}$ to $\bm{B}_{2},\bm{B}_{\bar{2}}$; 
        E2:~$(y,\byy\!\leftrightarrow\!w,\bw)^{12}$;\\[-0.5ex]   
        E3:~$3\!\leftrightarrow\! 4, \bar{3}\!\leftrightarrow\! \bar{4}$} 
        \\ \hline 
        $(2,2)$ & $\arry{l}{\bm{b}_1+\bm{e}_{1\bar{2}}, \bm{b}_2+\bm{e}_{3\bar{1}\bar{2}\bar{3}\bar{4}\bar{5}}, \\[-0.5ex]  \bm{b}_{\bar{1}}+\bm{e}_{2\bar{1}}, \bm{b}_{\bar{2}}+\bm{e}_{12345\bar{3}}}$ & IV--IV.2 &  
        \tabu{p{6.5cm}}{E1:~add\,$\bm{E}$ to $\bm{B}_{2},\bm{B}_{\bar{2}}$; 
        E2:~$(y\!\leftrightarrow\!w)^{12}$,\\[-0.5ex]  $(\byy\!\leftrightarrow\!\bw)^{12}$;  
        E3:~$35\!\leftrightarrow\! 46, \bar{3}\bar{5}\!\leftrightarrow\! \bar{4}\bar{6}$}
        \\ \hline 
        $(3,3)$ & $\arry{l}{\bm{b}_1+\bm{e}_{12\bar{1}\bar{2}\bar{3}\bar{4}\bar{5}\bar{6}}, \bm{b}_2+\bm{e}_{236\bar{1}}, \\[-0.5ex]  \bm{b}_{\bar{1}}+\bm{e}_{123456\bar{1}\bar{2}}, \bm{b}_{\bar{2}}+\bm{e}_{1\bar{2}\bar{3}\bar{6}}}$ & II--II  & 
        \tabu{p{6.5cm}}{E1:~add $\bm{E}$ to $\bm{B_{1}},\bm{B}_{\bar{2}}$;\\[-0.5ex]  
        E2:~$(y\!\leftrightarrow\!w)^{2\ldots 6}$, $(\byy\!\leftrightarrow\!\bw)^{2\ldots 6}$ }
        \\  \hline 
        $(4,4)$ & $\arry{l}{\bm{b}_1+\bm{e}_{1\bar{5}}, \bm{b}_2+\bm{e}_{3\bar{1}\bar{2}\bar{4}\bar{5}\bar{6}}, \\[-0.5ex] \bm{b}_{\bar{1}}+\bm{e}_{5\bar{1}}, \bm{b}_{\bar{2}}+\bm{e}_{12456\bar{3}}}$ & IV--IV.2 & 
        \tabu{p{6.5cm}}{E1:~add $\bm{E}$ to $\bm{B}_{2},\bm{B}_{\bar{2}}$; 
        E2:~$(y,\byy\!\leftrightarrow\!w,\bw)^{1256}$;\\[-0.5ex]  
        E3:~$1\bar{1}\!\rightarrow\! 3\bar{3}\!\rightarrow\! 2\bar{2}\!\rightarrow\! 4\bar{4}\!\rightarrow$; 
        E1:~$\bm{B}_{1,\bar{1}}\!\leftrightarrow\!\bm{B}_{2,\bar{2}}$} 
        \\ \hline 
        $(5,5)$ & $\arry{l}{\bm{b}_1+\bm{e}_{126\bar{1}\bar{2}}, \bm{b}_2+\bm{e}_{346\bar{3}\bar{5}}, \\[-0.5ex]  \bm{b}_{\bar{1}}+\bm{e}_{12\bar{1}\bar{2}\bar{6}}, \bm{b}_{\bar{2}}+\bm{e}_{35\bar{3}\bar{4}\bar{6}}}$ & VI--VI.2  & 
        \tabu{p{6.5cm}}{E2:~$(y \!\leftrightarrow\! w)^6$, $(\byy \!\leftrightarrow\! \bw)^6$}  
        \\ \hline 
        $(9,9)$ & $\arry{l}{\bm{b}_1+\bm{e}_{12\bar{1}\bar{2}}, \bm{b}_2+\bm{e}_{34\bar{5}\bar{6}},\\[-0.5ex]  \bm{b}_{\bar{1}}+\bm{e}_{12\bar{1}\bar{2}}, \bm{b}_{\bar{2}}+\bm{e}_{56\bar{3}\bar{4}}}$ & VI--VI.1 & 
        \tabu{p{6.5cm}}{E1:~$\bm{B}_{\bar{2}} \!\rightarrow\! \bm{B}_{\bar{2}}+\bm{B}_{\bar{1}}+\bm{E}$; 
        E3:~$\bar{3}\bar{4}\!\leftrightarrow\!\bar{5}\bar{6}$} 
        \\ \hline
        $(0,6)$ & $\arry{l}{\bm{b}_1+\bm{e}_{12\bar{1}\bar{5}}, \bm{b}_2+\bm{e}_{34\bar{3}\bar{6}}, \\[-0.5ex]  \bm{b}_{\bar{1}}+\bm{e}_{15\bar{1}\bar{2}}, \bm{b}_{\bar{2}}+\bm{e}_{36\bar{3}\bar{4}}}$ & VI--VI.3 & 
        \tabu{p{6.5cm}}{E1:~$\bm{B}_{2,\bar{2}} \!\rightarrow\! \bm{B}_{2,\bar{2}}+\bm{B}_{1,\bar{1}}+\bm{E}$; \\[-0.5ex] 
        E3:~$3\bar{3}\!\rightarrow\! 6\bar{6}\!\rightarrow\! 4\bar{4}\!\rightarrow\! 5\bar{5}\!\rightarrow$} 
        \\ \hline 
        $(2,8)$ & $\arry{l}{\bm{b}_1+\bm{e}_{1\bar{4}}, \bm{b}_2+\bm{e}_{356\bar{2}}, \\[-0.5ex]  \bm{b}_{\bar{1}}+\bm{e}_{4\bar{1}}, \bm{b}_{\bar{2}}+\bm{e}_{2\bar{3}\bar{5}\bar{6}}}$
         & IV--IV.1 & 
        \tabu{p{6.5cm}}{E3:~$\bar{1}\bar{2}\!\leftrightarrow\!\bar{3}\bar{4}$; E1:~$\bm{B}_{\bar{1}}\!\leftrightarrow\!\bm{B}_{\bar{2}}$; \\[-0.5ex] 
        E2:~$(\byy\!\leftrightarrow\!\bw)^{56}$} 
        \\ \hline 
        $(4,10)$ & $\arry{l}{\bm{b}_1+\bm{e}_{1\bar{5}}, \bm{b}_2+\bm{e}_{346\bar{2}\bar{5}}, \\[-0.5ex] \bm{b}_{\bar{1}}+\bm{e}_{5\bar{1}}, \bm{b}_{\bar{2}}+\bm{e}_{25\bar{3}\bar{4}\bar{6}}}$
         & IV--IV.1 & 
        \tabu{p{6.5cm}}{E1:~$\bm{B}_{2,\bar{2}} \!\rightarrow\! \bm{B}_{2,\bar{2}}+\bm{B}_{1,\bar{1}}$; \\[-0.5ex] E3:~$34\!\leftrightarrow\! 65$, $\bar{1}\!\rightarrow\!\bar{3}\!\rightarrow\!\bar{6}\!\rightarrow$, $\bar{2}\!\rightarrow\!\bar{4}\!\rightarrow\!\bar{5}\!\rightarrow$;\\[-0.5ex]  
        E2:~$(y \!\leftrightarrow\! w)^1$,  $(\byy\!\leftrightarrow\! \bw)^{\bar{3}\bar{5}\bar{6}}$; 
        E1:~$\bm{B}_{\bar{1}}\!\leftrightarrow\!\bm{B}_{\bar{2}}$
        }  
        \\ \hline 
        $(0,18)$ & $\arry{l}{\bm{b}_1+\bm{e}_{1\bar{1}\bar{2}\bar{4}\bar{5}\bar{6}}, \bm{b}_2+\bm{e}_{356\bar{2}\bar{3}\bar{4}\bar{5}\bar{6}}, \\[-0.5ex] \bm{b}_{\bar{1}}+\bm{e}_{12456\bar{1}}, \bm{b}_{\bar{2}}+\bm{e}_{23456\bar{3}\bar{5}\bar{6}}}$
         & IV--IV.1 & 
        \tabu{p{6.5cm}}{E1:~add $\bm{E}$ to $\bm{B}_1,\bm{B}_2,\bm{B}_{\bar{1}},\bm{B}_{\bar{2}}$;\\[-0.5ex]  
        E2:~$(y\!\leftrightarrow\! w)^{1\ldots 6}$, $(\byy\!\leftrightarrow\!\bw)^{1\ldots 4}$;\\[-0.5ex] 
        E3:~$13\!\leftrightarrow\! 24$, $\bar{1}\bar{2}\!\leftrightarrow\!\bar{4}\bar{3}$;  
        E1:~$\bm{B}_{\bar{1}}\!\leftrightarrow\!\bm{B}_{\bar{2}}$} 
        \\ \hline 
        $(6,0)$ & $\arry{l}{\bm{b}_1+\bm{e}_{12\bar{1}\bar{3}}, \bm{b}_2+\bm{e}_{34\bar{2}\bar{5}}, \\[-0.5ex]  \bm{b}_{\bar{1}}+\bm{e}_{13\bar{1}\bar{2}}, \bm{b}_{\bar{2}}+\bm{e}_{25\bar{3}\bar{4}}}$ & VI--VI.3 & 
        \tabu{p{6.5cm}}{Identical}         
        \\ \hline 
        $(8,2)$ & $\arry{l}{\bm{b}_1+\bm{e}_{1\bar{2}}, \bm{b}_2+\bm{e}_{356\bar{4}}, \\[-0.5ex] \bm{b}_{\bar{1}}+\bm{e}_{2\bar{1}}, \bm{b}_{\bar{2}}+\bm{e}_{4\bar{3}\bar{5}\bar{6}}}$
         & IV--IV.1  & 
        \tabu{p{6.5cm}}{Identical} 
        \\ \hline 
        $(10,4)$ & $\arry{l}{\bm{b}_1+\bm{e}_{12\bar{4}\bar{5}}, \bm{b}_2+\bm{e}_{36\bar{5}}, \\[-0.5ex]  \bm{b}_{\bar{1}}+\bm{e}_{45\bar{1}\bar{2}}, \bm{b}_{\bar{2}}+\bm{e}_{5\bar{3}\bar{6}}}$
         & IV--IV.1 & 
        \tabu{p{6.5cm}}{E1:~$\bm{B}_{1,\bar{1}}\!\rightarrow\! \bm{B}_{1,\bar{1}}+\bm{B}_{2,\bar{2}}$; \\[-0.5ex]
        E2:~$(y\!\leftrightarrow\! w)^{36}$, $(\byy\!\leftrightarrow\!\bw)^{\bar{1}\bar{2}\bar{3}\bar{6}}$; \\[-0.5ex]
        E3:~$12\!\rightarrow\! 65 \!\rightarrow\! 34 \!\rightarrow$, $\bar{1}\bar{2} \!\rightarrow\! \bar{6}\bar{5}$;  
        E1:~$\bm{B}_{1}\!\leftrightarrow\!\bm{B}_{2}$}
        \\ \hline 
        $(18,0)$ & $\arry{l}{\bm{b}_1+\bm{e}_{1\bar{2}\bar{3}\bar{4}\bar{5}\bar{6}}, \bm{b}_2+\bm{e}_{356\bar{1}\bar{2}\bar{4}\bar{5}\bar{6}}, \\[-0.5ex]  \bm{b}_{\bar{1}}+\bm{e}_{23456\bar{1}}, \bm{b}_{\bar{2}}+\bm{e}_{12456\bar{3}\bar{5}\bar{6}}}$
         & IV--IV.1 & 
        \tabu{p{6.5cm}}{E1:~add $\bm{E}$ to $\bm{B}_1,\bm{B}_2,\bm{B}_{\bar{1}},\bm{B}_{\bar{2}}$;  \\[-0.5ex] 
        E2:~$(y\!\leftrightarrow\! w)^{1\ldots 6}$,  $(\byy\!\leftrightarrow\!\bw)^{1\ldots 6}$; \\[-0.5ex] 
        E3:~$13\!\leftrightarrow\! 24$, $\bar{1}\bar{3}\!\leftrightarrow\! \bar{2}\bar{4}$} 
        \\ \hline 
        $(3,15)$ & $\arry{l}{\bm{b}_1+\bm{e}_{\bar{3}\bar{4}\bar{5}\bar{6}}, \bm{b}_2+\bm{e}_{\bar{1}\bar{2}\bar{5}\bar{6}}, \\[-0.5ex]  \bm{b}_{\bar{1}}+\bm{e}_{3456}, \bm{b}_{\bar{2}}+\bm{e}_{1256}}$ 
         & VI--VI.1 & 
        \tabu{p{6.5cm}}{E1:~add $\bm{E}$ to $\bm{B}_1,\bm{B}_2,\bm{B}_{\bar{1}},\bm{B}_{\bar{2}}$;\\[-0.5ex]  
        E2:~$(y\!\leftrightarrow\! w)^{1\ldots 6}$, $(\byy\!\leftrightarrow\! \bw)^{\bar{1}\ldots\bar{6}}$} 
        \\ \hline 
        $(17,5)$ & $\arry{l}{\bm{b}_1+\bm{e}_{126\bar{1}\bar{2}\bar{3}\bar{4}\bar{5}\bar{6}}, \bm{b}_2+\bm{e}_{5\bar{3}\bar{4}\bar{5}\bar{6}}, \\[-0.5ex]  \bm{b}_{\bar{1}}+\bm{e}_{123456\bar{1}\bar{2}\bar{6}}, \bm{b}_{\bar{2}}+\bm{e}_{3456\bar{5}}}$
         & III--III  & 
        \tabu{p{6.5cm}}{E1:~add $\bm{E}$ to $\bm{B}_1,\bm{B}_2,\bm{B}_{\bar{1}},\bm{B}_{\bar{2}}$;\\[-0.5ex]  
        E2:~$(y \!\leftrightarrow\! w)^{1\ldots 5}$, $(\byy \!\leftrightarrow\! \bw)^{\bar{1}\ldots \bar{5}}$} 
        \\ \hline 
        $(15,3)$ & $\arry{l}{\bm{b}_1+\bm{e}_{12\bar{1}\bar{2}}, \bm{b}_2+\bm{e}_{34\bar{3}\bar{4}}, \\[-0.5ex] \bm{b}_{\bar{1}}+\bm{e}_{12\bar{1}\bar{2}}, \bm{b}_{\bar{2}}+\bm{e}_{34\bar{3}\bar{4}}}$
         & VI--VI.1 & 
        \tabu{p{6.5cm}}{Identical}       
        \\ \hline 
        $(5,17)$ & $\arry{l}{\bm{b}_1+\bm{e}_{12\bar{3}\bar{4}}, \bm{b}_2+\bm{e}_{34\bar{1}\bar{2}\bar{3}\bar{4}\bar{5}\bar{6}}, \\[-0.5ex] \bm{b}_{\bar{1}}+\bm{e}_{34\bar{1}\bar{2}}, \bm{b}_{\bar{2}}+\bm{e}_{123456\bar{3}\bar{4}}}$
         & III--III & 
        \tabu{p{6.5cm}}{E1:~add $\bm{E}$ to $\bm{B}_2,\bm{B}_{\bar{2}}$; \\[-0.5ex] 
        E2:~$(y\!\leftrightarrow\!w)^{1256}$, $(\byy\!\leftrightarrow\!\bw)^{\bar{1}\bar{2}\bar{5}\bar{6}}$; \\[-0.5ex] 
        E3:~$12\!\leftrightarrow\! 34$, $\bar{1}\bar{2}\!\leftrightarrow\! \bar{3}\bar{4}$; 
        E1:~$\bm{B}_{1,\bar{1}}\!\leftrightarrow\!\bm{B}_{2,\bar{2}}$
        } 
        \\ \hline 
     \end{tabular}
    \caption{\label{tab:ComparisonAnastasopoulos}
    The models scan~\cite{Anastasopoulos:2009kj,Pradisi:2010ee,Bianchi:2009mu} are mapped to the configuration of our classification using the indicated equivalences. The order of their models is according to the lines of Table 1 of ref.~\cite{Pradisi:2010ee}.}
\end{table}

\subsection[Comparison with a previous scan of ${\Intr_{2L}^2\times\Intr_{2R}^2}$ T--folds]{Comparison with a previous scan of $\boldsymbol{\Intr_{2L}^2\times\Intr_{2R}^2}$ T--folds}

The authors of the papers \cite{Anastasopoulos:2009kj,Pradisi:2010ee,Bianchi:2009mu} perform a systematic search scan of $\Intr_{2L}^2\times\Intr_{2R}^2$ asymmetric orbifold configurations over all possible roto--translations with a fixed choice of the discrete torsion phases. As this is the most complete investigation of this type of T--folds, it is interesting to compare their results with our classifications\footnote{We are indebted to Massimo Bianchi, Jose Morales and Gianfranco Pradisi for an extended email exchange on which the comparison described in this section is based.}.  

The authors found 18 distinct models listed in Table 1 of their publication \cite{Anastasopoulos:2009kj,Pradisi:2010ee,Bianchi:2009mu}. Our classification of the $\Intr_{2L}^2\times\Intr_{2R}^2$ T--folds contains 16 distinct configurations. 

For a more detailed comparison the input data of their models is collected in \cref{tab:ComparisonAnastasopoulos} using the notation of our paper. The table indicates to which configurations in our classification their models are equivalent to using which equivalences E1--E3. Consequently, their 18 models correspond to only eight inequivalent $\Intr_{2L}^2\times\Intr_{2R}^2$ T--fold configurations. All their models are left/right symmetric in the sense that the twist basis vectors all form left/right symmetric pairs. According to \cref{tab:OverviewZ2L2Z2R2models} there are 10 such configurations in our classification. Apparently none of their models correspond to configurations  I--I or V--V. 

The apparent differences between their search and our classification do not signal a contradiction: Their objective was for a certain fixed GGSO choice to find all possible different effective Hodge numbers, while we performed a classification of all configurations up to the equivalences E1--E3 irrespectively of any choice of GGSO phases. It is well--known that different choices of basis vectors may lead to models that are equivalent modulo differences in GGSO phases. Only when spectra are computed can these choices lead to different spectra and therefore further distinctions of models. 

In addition, they found some interesting results that all their models fall in three finite series of values of the effective Hodge numbers and that the effective Euler number, $\gch = 2|h_{11}-h_{12}|$, is always a multiple of 12. We could confirm that all the suggested effective Hodge numbers can be realised from the equivalent configurations that we identified for a certain choice of GGSO phases, except for their models with numbers $(6,12)$ or $(12,6)$. Instead, we can get for the configuration IV--IV.1 the effective Hodge numbers\footnote{We use these Hodge numbers to refer these models of theirs in \cref{tab:ComparisonAnastasopoulos}.} $(0,18)$ or $(18,0)$ apart from options $(2,8)$, $(8,2)$, $(4,10)$ and $(10,4)$ (which they realise with other models equivalent to this configuration) depending on the choice of GGSO phases. In fact, with our parametric representation of the spectra all GGSO choices for all inequivalent configurations can be considered. None of them leads to effective Hodge numbers $(6,12)$ or $(12,6)$. However, we find that additional effective Hodge numbers can be produced, namely: $(15,15)$, $(3,12)$, $(12,3)$, $(15,0)$, $(0,15)$. Note that the effective Euler numbers associated the latter four effective Hodge numbers are multiples of 6 not of 12.

\subsection{Mirror symmetries on T--folds} 
\label{sc:MirrorSymmetries}

\subsubsection*{Mirror symmetry on symmetric \boldsymbol{$\Intr_2^2$} orbifolds}

Mirror symmetry on symmetric orbifolds can be identified in an easy fashion: one switches between the IIA and IIB by changing the sign of phase $\CC{\bm{S}}{\bm{S}}$ and one looks for a compensating change in some GGSO phase. On the SO(12) lattice there is only a single $\Intr_2^2$ orbifold configuration, dubbed a--D in our classification, in which the numbers of vector and hyper multiplets depend on a GGSO phase. This dependence is encoded in the parameter $h$, defined in \eqref{eq:Parameter_h_Z22}, as can be seen from \cref{tab:NarainRotoTransClass}. From this definition it is not difficult to see that a sign change of $\CC{\bm{S}}{\bm{S}}$ can be compensated by the mapping: 
\equ{
    \text{MS:}~\CC{\bm{B}_{1\bar{1}}}{\bm{B}_{2\bar{2}}} \rightarrow -\CC{\bm{B}_{1\bar{1}}}{\bm{B}_{2\bar{2}}}~. 
}
This would then be the mirror symmetry mapping and it of course can act on all $\Intr_2^2$ orbifolds, but only on the configuration a--D it has the effect on the spectrum that 24 hyper multiplets are exchanged for 24 vector multiplets or vice versa. This is of course a well--known result~\cite{Vafa:1994}.

The same procedure can be applied to other asymmetric orbifolds as well. To see how it acts, it is convenient to focus on T--folds with $\cN=2$ supersymmetry so that there are configurations which possess spectra that have distinct numbers of hyper and vector multiplets. There are two such asymmetric orbifolds $\Intr_{2L}\times\Intr_{2R}\times\Intr_2$ and $\Intr_{2L}^2\times\Intr_{2R}^2$. Let us first consider the former: 

\subsubsection*{Mirror symmetries on asymmetric \boldsymbol{$\Intr_{2L}\times\Intr_{2R}\times\Intr_2$} orbifolds}

For the $\Intr_{2L}\times\Intr_{2R}\times\Intr_2$ there are two parameters $h$ and $h'$ which parameterise if there are hypers (when they are $1$) or vectors (when they are $0$) in the spectra of certain configurations 
(ii--ii--A.2, 
i--iii--A, 
iii--i--A, 
ii--iii--A.2, 
iii--ii--A.2, 
iii--iii-A.3, 
and iii--iii--B.2 
to be exact, according to \cref{tab:NarainRotoTransClass}). 
In equations \eqref{eq:Z2LZ2RZ2_h} the explicit expressions of these parameters are given which can be expanded in the fundamental GGSO phases by \eqref{eq:ExpansionBab}. The sign flip of the phase $\CC{\bm{S}}{\bm{S}}$ can now be compensated by two mappings:
\equ{
    \text{MS$_L$:}~\CC{\bm{B}_{1}}{\bm{B}_{2\bar{2}}} \rightarrow -\CC{\bm{B}_{1}}{\bm{B}_{2\bar{2}}}~, 
    \quad\text{or}\quad  
    \text{MS$_R$:}~\CC{\bm{B}_{\bar{1}}}{\bm{B}_{2\bar{2}}} \rightarrow -\CC{\bm{B}_{\bar{1}}}{\bm{B}_{2\bar{2}}}~. 
}
Hence, apparently there are two mirror mappings: one involving $\bm{B}_1$ on the left--moving side and one involving $\bm{B}_{\bar{1}}$ on the right--moving side, which could be dubbed left-- and right--mirror symmetry maps. Both have the same effect on the spectra of the configurations listed above that involve $h$ and $h'$.

\subsubsection*{Mirror symmetries on asymmetric \boldsymbol{$\Intr_{2L}^2\times\Intr_{2R}^2$} orbifolds}

For the $\Intr_{2L}^2\times\Intr_{2R}^2$ there are six parameters \eqref{eq:Parameters_h} which parameterise if there are hypers  or vectors in the spectra of a large number of configurations, see \cref{tab:NarainRotoTransClass}. The sign flip of the phase $\CC{\bm{S}}{\bm{S}}$ can be compensated by two mappings:
\equ{
    \text{MS$_L$:}~\CC{\bm{B}_{1}}{\bm{B}_{2}} \rightarrow -\CC{\bm{B}_{1}}{\bm{B}_{2}}~, 
    \quad\text{or}\quad  
    \text{MS$_R$:}~\CC{\bm{B}_{\bar{1}}}{\bm{B}_{\bar{2}}} \rightarrow -\CC{\bm{B}_{\bar{1}}}{\bm{B}_{\bar{2}}}~. 
}
So again left-- and right--mirror symmetry maps can be identified. 

As was noted in the discussion about \cref{tab:FixedFreeGGSOphases}, these phases are anti--symmetric for the configurations I$\cdots$\,V--I$\cdots$\,V and hence can be fixed by relabeling of these twist basis vectors. Hence, for these configurations the left-- or right--mirror symmetry maps may be realised by the interchange of the basis vectors $\bm{B}_1$ and $\bm{B}_2$ or $\bm{B}_{\bar{1}}$ and $\bm{B}_{\bar{2}}$, respectively.

\subsubsection*{Mirror symmetries on other asymmetric orbifolds}

On the asymmetric orbifolds $\Intr_{2L}^2$, $\Intr_{2R}^2$, $\Intr_{2L}\times \Intr_2$, $\Intr_{2R}\times \Intr_2$, $\Intr_{2L}^2\times \Intr_{2R}$, $\Intr_{2L}\times \Intr_{2R}^2$ one of the four left-- or right--mirror symmetry maps can act. But since their spectra are $\cN=3$ or higher there are no hyper multiplets anymore which could display any effect of these mappings at the massless level of the theory.

\subsection{Orientifolds of order--two T--folds}
\label{sc:Orientifolds}

The study of asymmetric order--two orbifolds in this paper has focussed on type II models with only closed strings. It is well--known that by gauging the worldsheet parity operator $\gO: \gs \ra 2\pi - \gs$ such models can be extended to include open strings that end on D--branes. 
Orientifolds of both the type IIB and type IIA are possible. In type IIB theories the basic worldsheet parity $\gO$ can be used as the starting point of type I orientifold constructions as the spin--structures on the left-- and right--moving sides are identical. (IIA constructions may be obtained from IIB ones by an odd number of T--dualities.)

Basically such an orientifold procedure follows the following steps~\cite{polchinski_96,Angelantonj:2002ct}: 
\enums{ 
    \item Define a (potentially dressed) worldsheet parity $\gO'$ that is a proper symmetry of the theory. 
    \item Determine the O(rientifold)--planes that result from modding out this dressed worldsheet parity.
    \item Determine the essential properties of D-branes needed to obtain a consistent theory. 
    \item Construct the required D--branes explicitly. 
}
A dressed worldsheet parity $\gO'$ is the basic worldsheet parity $\gO$ accompanied by some additional transformations. If $\gO$ is a symmetry of the type II construction, accompanying transformations are not mandatory, and may involve freely acting involutions of the original type II string background. If $\gO$ is not a symmetry of the type II construction, then the accompanying transformations are essential to ensure that the dressed worldsheet parity is a symmetry. The method of open descendants \cite{type0string1,Angelantonj:2002ct} can be used to determine the types of O--planes arising when the $\gO'$ is gauged and what the essential properties of the required D--branes are to cancel all RR--tadpoles. The boundary state method \cite{Green:1996um} can be used to construct invariant D--branes explicitly. The purpose of this subsection is not to execute the full orientifolding program, but to see to what extent our classification results can be of use for this. 

If a T--fold admits an orientifold projection, often it may admit a number of orientifold projections by combining the orientifold projection with additional involutions which the configuration may admit. 
In this work we do not determine which (if any) involutions the T--fold configurations admit, hence our analysis only indicates if an orientifold projection is possible; not how many distinct options there are for this.

\subsubsection*{Dressed worldsheet parities in the fermionic formulation}

On the complete set of left-- and right--moving fermions $f$ and $\bar{f}$ a dressed worldsheet parity acts as 
\equ{
    \gO'(f) = T\, \bar{f}~, 
    \qquad
    \gO'(\bar{f}) = T^{-1}\, f~,
}
with some matrix $T\in SO(20;\Real)$ in general, since there are 20 fermions on either side and the worldsheet parity is an order--two symmetry of the free worldsheet action. (Obviously, the basic worldsheet parity $\gO$ has $T=\Id$.)

With slight abuse of notation interpret the $\Bga_L$ and $\Bga_R$ as diagonal matrices with the entries of vector $\Bga$ that defines the fermionic boundary conditions \eqref{eq:feEarmichanges}. The dressed worldsheet parity action on the boundary conditions can be represented as 
\equ{ \label{eq:WSPBoundaryMap}
    \gO'(\Bga_L) = T\, \Bga_R\, T^{-1}~, 
    \qquad 
    \gO'(\Bga_R) = T^{-1}\, \Bga_L\, T~. 
}
Since only real fermions are considered, the entries of the basis vectors could be chosen to be 0 or 1 modulo 2 only, this transformation should admit the interpretation as defining boundary conditions of the fermions, {\em i.e.}\ $\gO'(\Bga_L)$ and $\gO'(\Bga_R)$ have to be diagonal matrices with 0 or 1 entries modulo 2. This means that the dressing matrix $T$ may correspond to (asymmetric) freely acting involutions, permutations or reflections of some of the left--moving or some of the right--moving fermions.  In terms of basic twist and shift vectors used in this work this reads 
\equ{
    \gO'(\bm{b}_\ga) = T\,\bm{b}_{\bga}\,T^{-1}~, 
    \qquad
    \gO'(\bm{e}_i) = T\,\bm{e}_{\bi}\,T^{-1}~,
    \qquad 
    \gO'(\bm{b}_{\bgb}) = T^{-1}\,\bm{b}_{\gb}\,T~,
    \qquad 
    \gO'(\bm{e}_{\bj}) = T^{-1}\,\bm{e}_{j}\,T~,
}
and linearly extended for vectors which are sums of these. Even if the dressing matrix is trivial, the effect of the worldsheet parity is to interchange the holomorphic and anti--holomorphic labels of the basis vectors. For a type II fermionic model to admit an orientifold projection its additive set of the worldsheet parity transformed basis vectors should be identical to the original additive set: $\gO'(\BgX) = \BgX$.

\subsubsection*{T--folds with left/right asymmetric point groups}

T--folds with left/right asymmetric point groups do not admit any orientifold projection. For example, asymmetric  $\Intr_{2L}$ orbifolds have twist basis vector $\bm{B}=(\bm{B}_L, \bm{B}_R)$ with $\tr(\bm{B}_L)=8+2t$ and $\tr(\bm{B}_R)=2t$ counting the number of unit entries of the twist basis vector on the left-- and right--moving sides, respectively. Here $t=0,1,3$ labels the three types, i, ii, iii, of $\Intr_{2L}$ T--folds (see \cref{tab:NarainRotoTransClass}). The traces of the worldsheet parity transformed matrices $\tr \gO'(\bm{B}_L) =2t$ and $\tr \gO'(\bm{B}_R) = 8+2t$ are opposite no matter which dressing matrix $T$ is used. This signals that the worldsheet parity transformed point group is always $\Intr_{2R}$. Similar arguments can be given for the other left/right asymmetric point groups, like $\Intr_{2L}^2$ and $\Intr_{2L}^2\times\Intr_{2R}$. Thus, only T--folds with left/right symmetric point groups may admit orientifold projections.

\subsubsection*{Left/right symmetric order--two T--folds}

 Consequently, symmetric orbifolds always admit orientifold projections as the boundary conditions on the left and right moving sides are identical. Thus, in particular, the symmetric order--two orbifolds, with point groups $\Intr_2$ and $\Intr_2^2$, do. For asymmetric orbifolds this is no longer automatically the case. 

Thus, orientifold constructions from the elementary worldsheet parity $\gO$ can start from left/right--symmetric type IIB configurations~\cite{Anastasopoulos:2009kj,Bianchi:2009mu}. If the basis vectors appear in left/right symmetric pairs, orientifold projections with the basic worldsheet parity $\gO$ are definitely possible. Going through the list of all order--two (a)symmetric configurations on the $SO(12)$ lattice at the free fermionic point, the left/right--symmetric T--folds can be easily identified; the list of possible point groups and configurations is given in \cref{tab:LeftRightSymTfolds}.

\begin{table}[t]
    \centering
    \begin{tabular}{|l||c|l|}
    \hline
    \textbf{Point group} & $\boldsymbol{\#}$ & \textbf{Left/right symmetric T--fold configurations} 
    \\ \hline\hline 
    $\Intr_2$ & 2 & all \\ \hline 
    $\Intr_2^2$ & 5 & all \\ \hline
    $\Intr_{2L}\times\Intr_{2R}$ & 7 & i--i; ii--ii.1-3; iii--iii.1-3 \\ \hline 
    $\Intr_{2L}\times\Intr_{2R}\times\Intr_2$ & 16 & i--i--A; ii--ii--A.1-3; iii--iii--A.1-3; ii--ii--B; iii--iii--B.1-3;  
    \\     & & 
    i--i--E; ii--ii--E.1,2; iii--iii--E.1,2 \\ \hline 
    $\Intr_{2L}^2\times\Intr_{2R}^2$ & 10 & I--I; II--II; III--III; IV--IV.1-3; V--V; VI--VI.1-3 
    \\ \hline 
    \end{tabular}
    \caption{\label{tab:LeftRightSymTfolds} This table lists the left/right symmetric order--two T--fold configurations on the $SO(12)$ lattice. Per left/right symmetric point group the number of such configurations is indicated. These configurations admit an orientifold projection involving the basic worldsheet parity $\gO$ to a type I orientifold from the IIB theory.}
\end{table}

\subsubsection*{Matching up with some existing asymmetric orientifolds in the literature}

Orientifolds of asymmetric orbifolds have been considered in the literature in the past. For example, in a series of papers \cite{Blumenhagen:1998uf,blumenhagen_00_ao} D-brane models are constructed on certain asymmetric orbifolds in six dimensions without and with target space supersymmetry breaking. Their construction corresponds to certain $\Intr_{2L}\times\Intr_{2R}$ orbifolds. 

The authors of refs.~\cite{Anastasopoulos:2009kj,Pradisi:2010ee,Bianchi:2009mu} perform the open descendants of the minimal $\Intr_{2L}^2\times\Intr_{2R}^2$ T--fold configuration IV--IV.3 with effective Hodge numbers $(1,1)$ and find that no tadpoles need to be canceled so that no D--branes are needed. 
In addition, ref.~\cite{Bianchi:2009mu} speculates on properties of D--branes on T--folds with left/right asymmetric point groups. However as discussed above, we dispute whether one can define an orientifold in such a case in the first place.

\subsection{Enhanced supergravities}
\label{sc:EnhancedSUGRAs}

As discussed in \cref{sc:6Dtwisted}, the asymmetric twist basis vectors may give rise to twisted sectors that contain Rarita--Schwinger multiplets containing as their highest components massless spin--$3/2$ Rarita--Schwinger fields. It is an interesting question what the status of these multiplets is. One interpretation is that they are multiplets containing gravitini (as these are massless spin--$3/2$ fields). For this interpretation to hold up, these Rarita--Schwinger fields should be associated to the graviton, {\em i.e.}\ form some representation of the (local) extended supersymmetry algebra. In other words, these Rarita--Schwinger multiplets should be part of the supergravity multiplet of some higher extended supersymmetries, see {\em e.g.}\ ref.~\cite{Strathdee:1986jr}. If not, they should probably be interpreted in their own right. 

In ref.~\cite{Dolivet:2007sz}, such twisted sectors with Rarita--Schwinger multiplets with presumed gravitini were discussed. In this reference it was argued that the ``twisting mechanism'', as the authors dubbed it, first kicks out some of the original gravitini from the supergravity multiplet, but then the twisted sectors again introduce new additional gravitini such that extended supergravity multiplets are formed\footnote{Similar supersymmetry enhancements were discussed in refs.~\cite{Imamura:1992np,Sasada:1994iv,aafs,Brunner:1999jq,Gkountoumis:2023fym, Dabholkar:1998kv, FIDR,  Bianchi:2022wku}. 
}

A technical aside might be that the original supergravity of the II theories lives everywhere, while the twisted sectors only at the fixed loci of the orbifold action. So the new gravitini might not have identical properties of the original projected out ones. In particular, their interactions might be distinct. However, since the geometrical compactification interpretation of asymmetric orbifolds is somewhat complicated, one might argue that one should only consider the models as four dimensional theories. 

If the statement that the spin--$3/2$ fields are gravitini, thus associated with the graviton, is true, then for all the models where twisted Rarita--Schwinger multiplets may arise, they always have to fit in higher extended supergravities. In \cref{tab:EnhancedSUGRAs} the configurations are collected in which at least one twisted Rarita--Schwinger multiplet arises. The spectra are first given as representations of the original extended supersymmetry and then, in the final column, the largest consistent enhancement is indicated.

From the classification results obtained in this paper one can conclude that for all configurations in which twisted Rarita--Schwinger multiplets arise, they can be absorbed in enhanced supergravity multiplets. In particular, all pure asymmetric twist models (no roto--translations associated to the asymmetric twists) with the maximal number of Rarita--Schwinger multiplets, enhance to $\cN=8$ supergravity. When there are less than the maximal number of Rarita--Schwinger multiplets ({\em i.e.}\ configurations in which some asymmetric twist vectors involve roto--translations), then the models always enhance such that all these Rarita--Schwinger multiplets are absorbed in the supergravity multiplet. This can be confirmed by considering \cref{tab:EnhancedSUGRAs}. In this table the spectra with the additional Rarita--Schwinger multiplets are given as representations of the amount of supersymmetry preserved by the collection of basis vectors and the multiplets of the maximal enhanced supersymmetry absorbing all Rarita--Schwinger multiplets in the higher extended supergravity.
It is instructive to consider a number of such cases in detail (using the results of \cref{tab:NarainRotoTransClass}):

\begin{table}
    \centering
    \scalebox{0.925}{
    \begin{tabular}{|lll||l |ll| l|}
    \hline
    \textbf{Point}& $\boldsymbol{\cN}$ & \textbf{Configuration} & \textbf{Original} & \multicolumn{2}{l|}{$\boldsymbol{\cN}$\textbf{--enhanced}} & \textbf{Twisted SUSY}
    \\
    \textbf{group} &   &\textbf{} & \textbf{multiplets} & & \textbf{multiplets} & \textbf{generators}
    \\ \hline\hline
    $\Intr_{2L}$ & 6 & i & SG+2\,RS & 8 & SG & $\Sv + \bm{B}_1$
    \\ \hline
    $\Intr_{2L}\times\Intr_{2R}$ & 4 & i--i & SG+4\,RS+6\,V & 8 & SG & $\Sv + \bm{B}_1, \bSv + \bm{B}_{\bar{1}}$
    \\ \cline{3-7}   
    & & i--ii; ii--i &  SG+2\,RS+2\,V & 6 & SG & $\Sv + \bm{B}_1; \bSv + \bm{B}_{\bar{1}}$
    \\ \cline{3-7} 
    & & i--iii; iii--i &  SG+2\,RS+2\,V & 6 & SG & $\Sv + \bm{B}_1; \bSv + \bm{B}_{\bar{1}}$
    \\ \hline
    $\Intr_{2L}^2$ & 5 & I & SG+3\,RS & 8 & SG & $\Sv + \bm{B}_\ga$
    \\ \cline{3-7} 
    & & II; III &  SG+RS & 6 & SG & $\Sv + \bm{B}_1$
    \\ \hline
    $\Intr_{2L}\times\Intr_2$ & 3 & i--A & SG+RS+12\,V & 4 & SG+12\,V & $\Sv + \bm{B}_1$
    \\ \cline{3-7}  
    & & i--E & SG+RS+4\,V & 4 & SG+4\,V & $\Sv + \bm{B}_1$
    \\ \hline 
    $\Intr_{2L}^2\times\Intr_{2R}$ & 3 & I--i, $r=1$ &   SG+RS+14\,V & 4 & SG+14\,V  & $\Sv + \bm{B}_3$
    \\ 
    & & \phantom{I--i; }$r=5$ & SG+5\,RS+10\,V & 8 & SG & $\Sv + \bm{B}_\ga, \bSv + \bm{B}_{\bar{1}}$
    \\ \cline{3-7} 
    & & II--i, $r=1$ & SG+ RS+6\,V & 4 & SG+6\,V  & $\Sv + \bm{B}_1$
    \\ 
    & & \phantom{II--i, }$r=3$ &  SG+3\,RS+4\,V & 6 & SG & $\Sv + \bm{B}_1, \bSv + \bm{B}_{\bar{1}}$
    \\ \cline{3-7} 
    & & III--i, $r=1$ & SG+RS+10\,V & 4 & SG+10\,V  & $\Sv + \bm{B}_1$
    \\ 
    & & \phantom{III--i, }$r=3$ & SG+2\,RS+9\,V & 6 & SG & $\Sv + \bm{B}_1, \bSv + \bm{B}_{\bar{1}}$
    \\ \cline{3-7} 
    & & IV--i & SG+RS+(6-$4r_3$)\,V & 4 & SG+(6-$4r_3$)\,V  & $\Sv + \bm{B}_1$
    \\ \cline{3-7} 
    & & II--ii & SG+RS+2\,V & 4 & SG+2\,V  & $\Sv + \bm{B}_1$
    \\ \cline{3-7} 
    & & III--ii & SG+RS+6\,V & 4 & SG+6\,V  & $\Sv + \bm{B}_1$
    \\ \hline 
    $\Intr_{2L}\times\Intr_{2R}\times\Intr_2$ & 2 & i--i--A &  SG+2\,RS+15\,V+14\,H & 4 & SG+14\,V & $\Sv + \bm{B}_1, \bSv + \bm{B}_{\bar{1}}$
    \\ \cline{3-7}
    & & i--ii--A; ii--i--A & SG+RS+11\,V+11\,H & 3 & SG+11\,V & $\Sv + \bm{B}_1; \bSv + \bm{B}_{\bar{1}}$
    \\ \cline{3-7}
    & & i--iii--A; iii--i--A & SG+RS+11\,V+ 11\,H & 3 & SG+11\,V & $\Sv + \bm{B}_1; \bSv + \bm{B}_{\bar{1}}$
    \\ \cline{3-7}
    & & i--i--E & SG+2\,RS+7\,V+6\,H & 4 & SG+6\,V & $\Sv + \bm{B}_1, \bSv + \bm{B}_{\bar{1}}$
    \\ \cline{3-7}
    & & i--ii--E; ii--i--E & SG+RS+3\,V+3\,H & 3 & SG+3\,V  & $\Sv + \bm{B}_1; \bSv + \bm{B}_{\bar{1}}$
    \\ \cline{3-7}
    & & i--iii--E; iii--i--E & SG+RS+3\,V+3\,H & 3 & SG+3\,V & $\Sv + \bm{B}_1; \bSv + \bm{B}_{\bar{1}}$
    \\ \hline 
    $\Intr_{2L}^2\times\Intr_{2R}^2$ & 2 & I--I, $r=1$ & SG+2\,RS+15\,V+14\,H & 4 & SG+14\,V  & $\Sv + \bm{B}_3; \bSv + \bm{B}_{\bar{3}}$
    \\ 
    & & \phantom{I--I; }$r=3$ & SG+6\,RS+15\,V+10\,H & 8 & SG & $\Sv + \bm{B}_\ga$ or $\bSv + \bm{B}_{\bgb}$
    \\ \cline{3-7}
    & & II--II, $r=1$ & SG+RS+3\,V+3\,H & 3 & SG+3\,V & $\Sv + \bm{B}_1$ or $\bSv + \bm{B}_{\bar{1}}$\\ 
    & & \phantom{II--II, }$r=2$ & SG+2\,RS+3\,V+2\,H & 4 & SG+2\,V & $\Sv + \bm{B}_1, \bSv + \bm{B}_{\bar{1}}$
    \\ \cline{3-7} 
    & & II--IV; IV--II & SG+RS+3\,V+3\,H & 3 & SG+3\,V & $\Sv + \bm{B}_1; \bSv + \bm{B}_{\bar{1}}$
    \\ \cline{3-7} 
    & & III--III, $r=1$ & SG+RS+11\,V+11\,H & 3 & SG+11\,V & $\Sv + \bm{B}_1$ or $\bSv + \bm{B}_{\bar{1}}$
    \\
    & & \phantom{III--III, }$r=2$ & SG+2\,RS+7\,V+6\,H & 4 & SG+6\,V  & $\Sv + \bm{B}_1, \bSv + \bm{B}_{\bar{1}}$
    \\ \cline{3-7} 
    & & III--IV; IV--III & SG+RS+7\,V+7\,H & 3 & SG+7\,V & $\Sv + \bm{B}_1; \bSv + \bm{B}_{\bar{1}}$
   \\ \hline 
    \end{tabular}
    }
    \caption{\label{tab:EnhancedSUGRAs}
        This overview of the configurations with additional twisted Rarita--Schwinger multiplets shows that all cases where there are twisted Rarita--Schwinger multiplets, they can be enhanced to a higher extended supergravity without any remaining loose Rarita--Schwinger multiplets. The final column specifies the additional twisted supersymmetry generators. If $\Sv+\bm{B}_\ga$ or $\bSv+\bm{B}_\bgb$ is indicated there this signifies that all  three generators for $\ga, \bgb=1,2,3$ are present. Moreover, a comma between two twisted generators means that both are present; while a semicolon that the generator depends on the configurations listed in the third column. 
        }
\end{table}

Consider the models associated with the configurations of the $\Intr_{2L}^2\times\Intr_{2R}$ orbifolds that can give rise to additional Rarita--Schwinger multiplets. Because of \eqref{eq:Z2L2Z2R_RS_relation}, the models associated with configuration I--i either give $r=1$ or $5$ Rarita--Schwinger multiplets, depending on the choice of GGSO phases. Hence, they either lift to $\cN=4$ with 14 additional vector multiplets or the full $\cN=8$ supergravity. Similarly, the spectra of configurations II--i and III--i contain $r=r_1(1+2r_2)=0,1,3$ Rarita--Schwinger multiplets which either lifts to $\cN=4$ with six and ten vector multiplets, respectively, or the $\cN=6$ supergravity. 

The way the $\cN=2$ configurations i--iii--A and iii--i--A of the $\Intr_{2L}\times\Intr_{2R}\times\Intr_2$ orbifolds with a Rarita--Schwinger multiplet ($r=1$) lift $\cN=3$ is also very interesting. As can be inferred from the relations \eqref{eq:Z2LZ2RZ2_RS_relation} and \eqref{eq:Z2LZ2RZ2_h}, if $r=1$ then $h+h'=1$. Inserting this in the spectra for these configurations given in \cref{tab:NarainRotoTransClass} one finds the spectrum: SG + RS + 11\,(V+H) which lifts to $\cN=3$. 

There are seven configurations of the $\Intr_{2L}^2\times\Intr_{2R}^2$ orbifolds in which additional Rarita--Schwinger multiplets may arise, namely: I--I, II--II, II--IV, IV--II, III--III, III--IV, IV--III. The configuration I--I has $2r$ Rarita--Schwinger multiplets for $r=0,1,3$. If $r=1$, the model enhances to $\cN=4$ supergravity with 14 vector multiplets, and if $r=3$ to $\cN=8$ supergravity. The configurations II--II have a single additional Rarita--Schwinger multiplet when $r_1=1$ and $r_{\bar{1}}=0$ (or vice versa) or two when $r_1=r_{\bar{1}}=1$. In the former cases the theories lift to $\cN=3$ supergravities with 3 vector multiplets and in the latter case to $\cN=4$ supergravity with 2 vector multiplets. For configurations II--IV (or IV--II) there is enhancement if $r_1=1$ (or $r_{\bar{1}}$). This implies that $p_{1\bar{3}}=0$ (or $p_{3\bar{1}}=0$), so that the $\cN=3$ spectra contain the supergravity multiplet and three vector multiplets. The configurations III--III have a single additional Rarita--Schwinger multiplet when $r_1=1$ and $r_{\bar{1}}=0$ (or vice versa) or two when $r_1=r_{\bar{1}}=1$. In the former cases the theories lift to $\cN=3$ supergravities with 11 vector multiplets and in the latter case to $\cN=4$ supergravity with 6 vector multiplets. Given the sophisticated spectra of configuration III--III given in \cref{tab:NarainRotoTransClass} with intrinsic dependences on the various GGSO factors, these results are highly non--trivial. (Below, some details of this computation for the III--III configuration with two Rarita--Schwinger multiplets are given as an example.) Finally, for configurations III--IV (or IV--III) there is enhancement if $r_1=1$ (or $r_{\bar{1}}=1$) to $\cN=3$ supergravity coupled to seven vector multiplets. 

The fact that in all these cases the spectra of these models with twisted Rarita--Schwinger multiplets enhance to a higher extended supersymmetry can be understood in the following fashion. The basis vectors $\Sv$ and $\bSv$ generate at least $\cN=2$ supersymmetry in target space as the GGSO phases were chosen such as to preserve the maximal supersymmetry. Now, each twisted Rarita--Schwinger state has the structure of $\bgps^\gm|\Sv'\rangle$ or $\gps^\gm|\bSv'\rangle$. Hence, just like for the untwisted gravitini, see \eqref{eq:10DsugraStates}, the basis vectors $\Sv'$ or $\bSv'$ can be thought of as additional (twisted) supersymmetry generators. To see how this plays out in practice, again consider the I--I configuration of the $\Intr_{2L}^2\times\Intr_{2R}^2$ orbifold with $r=3$. Beside the two untwisted gravitini, there are now six additional twisted Rarita--Schwinger fields, 
$\bgps^\gm | \Sv_\ga'\rangle$ and $\gps^\gm | \bSv_\bgb'\rangle$, hence there are six additional generators of local supersymmety: 
\equ{ \label{eq:TwistedSUSY}
    \Sv_\ga' = \Sv + \bm{B}_\ga~, 
    \qquad 
    \bSv_\bgb' = \bSv + \bm{B}_\bgb~, 
}
with $\ga,\bgb=1,2,3$. So in total there are eight generators of local supersymmetry, so the spectrum has to organise itself in $\cN=8$ multplets; and at the massless level there is then only the full $\cN=8$ supergravity. In table \cref{tab:EnhancedSUGRAs} the additional twisted supersymmetry generators are indicated. As the $\Intr_{2L}$ and $\Intr_{2L}\times\Intr_{2R}$ orbifolds can be thought of as six dimensional models compactified on an additional two--torus, a single twisted supersymmetry generator corresponds to two supersymmetry generators in four dimensions.  

In summary, \cref{tab:EnhancedSUGRAs} shows that in all cases where additional Rarita--Schwinger multiplets arise, they can be absorbed in the supergravity multiplet of an higher extended supersymmetry. Only models with only asymmetric twist basis vectors, the enhancement can be maximal to $\cN=8$ supergravity. An additional noteworthy feature seems to be that when the enhancement is to $\cN=3$, the number of vector multiplets is always odd, while when the enhancement is to $\cN=4$ this number is always even.

\subsubsection*{Example: Supergravity enhancement of configuration III--III of the \boldsymbol{$\Intr_{2L}^2\times\Intr_{2R}^2$} orbifold with two Rarita--Schwinger multiplets}

Consider the configuration III--III of the $\Intr_{2L}^2\times\Intr_{2R}^2$ orbifold with two Rarita--Schwinger multiplets. 
In this case, the parameters determining the existence of Rarita--Schwinger multiplets are $r_1=r_{\bar{1}}=1$. According to \eqref{eq:Parameters_r},
this fixes the following GGSO phases: 
\equ{
    \CC{\bm{B}_1}{\bm{B}_{\bar{1}}} = \CC{\bm{B}_1}{\bm{B}_{\bar{2}}} = \CC{\bm{B}_2}{\bm{B}_{\bar{1}}} = +1~. 
}
By \eqref{eq:Parameters_p}, it follows that $p_{1\bar{2}}=p_{2\bar{1}}=p_{1\bar{3}}=p_{3\bar{1}}=0$. In addition, by \eqref{eq:Parameters_h}, one concludes that 
\equ{
    h_{2\bar{1}} + h_{2\bar{2}}' + h_{1\bar{2}}' + h_{2\bar{1}}' = 2~. 
}
Inserting all this in the spectrum given by this configuration in \cref{tab:NarainRotoTransClass}, the $\cN=2$ spectrum reduces to: 
\equ{
    \text{SG + 2\,RS + 7\,V + 6\,H}~.
}
This spectrum lifts to $\cN=4$ supergravity coupled to six $\cN=4$ vector multiplets. The two twisted supersymmetry generators are 
$\Sv' = \Sv + \bm{B}_1$ and $\bSv' = \bSv + \bm{B}_{\bar{1}}$, as indicated in \cref{tab:EnhancedSUGRAs}.

\subsection{Rarita--Schwinger fields within models with reduced supersymmetry}
\label{sc:ReducedSusyModels}

Similar supersymmetry enhancement may even occur in non--supersymmetric asymmetric orbifolds,
where all supersymmetries are projected out in the untwisted sector, but some novel supersymmetry is recovered by massless spin--$3/2$ states arising from twisted sectors.
An example of a model with this reintroduction of supersymmetry\footnote{Sometimes this is considered to be an unwanted feature, which can be avoided if twists were accompanied by shifts to lift such twisted gravitini~\cite{LarotondaMonteroTartaglia2026}.}
with a $\Intr_4$ orbifold action was indicated in refs.~\cite{SatohSugawaraWada2015,Sugawara_2016}.

To construct such examples, we recall that section \ref{sc:typeII4d} employed the GGSO phase choices
\equ{
    \CC{\bSv}{\bm{B}_\alpha} = \CC{\Sv}{\bm{B}_{\bga}} = \CC{\bSv}{\bm{B}_{\gb\bgb}} = \CC{\Sv}{\bm{B}_{\gb\bgb}} = -1,
}
for $\alpha,\bga, \gb=\bgb =1,2$, to preserve a maximum amount of supersymmetry for the general Narain point group $\Intr_{2L}^{p_L}\times \Intr_{2R}^{p_R}\times\Intr_2^p$ with $0\leq p_L+p, p_R+p\leq 2$. For other choices less or no supersymmetries survive. 
The amounts of supersymmetries $\cN_L$ and $\cN_R$ preserved on the left-- and right--moving sides can be written as
\equ{
    \cN_L = \frac{4}{2^{p_L+p}}\prod_{\bga \leq p_R} \cS_\bga~,  
    \qquad    
    \cN_R = \frac{4}{2^{p_R+p}} \prod_{\alpha\leq p_L} \cS_\ga~. 
}   
(If a product is empty, it is taken to be unity.) The prefactors give the maximal amount of supersymmetries~\eqref{eq:NsusyLR}  that can be preserved on either side given the point group. The other factors test whether the individual asymmetric twist basis vectors preserve these supersymmetries or not. Indeed, the values $1$ or $0$ of the parameters 
\equ{ \label{eq:TestSUSYTwists}
    \cS_\ga = \frac{1}{2}\Big( 1-\CC{\bSv}{\bm{B}_\alpha}\Big)~,
    \qquad 
    \cS_\bga = \frac{1}{2}\Big( 1-\CC{\Sv}{\bm{B}_\bga}\Big)~, 
}
indicate whether the asymmetric twists $\bm{B}_{\ga}$ or $\bm{B}_{\bga}$, preserve or break the supersymmetry generating vectors, respectively. As soon as one of the twist basis vectors is incompatible with the supersymmetries on that side, none survive there.


The total number of Rarita--Schwinger fields 
can be expressed as
\equ{ \label{eq:N_RS_comp}
    \cN_\text{RS} =  \frac{4}{2^{p_L+p}} \sum_\ga r_\ga \cS_\ga + \frac{4}{2^{p_R+p}} \sum_\bga r_\bga \cS_\bga~, 
}
where the sums are over $1\leq \ga \leq 2^{p_L}-1$, $1\leq \bga \leq 2^{p_R}-1$ (if a sum is empty, it does not contribute). Here, the following parameters are introduced:
\equ{ \label{eq:N_RS_parameters}
    r_\ga = \prod_{\bgb} \frac 12 \Big( 1-\CC{\Sv}{\bm{B}_\bgb}\CC{\bm{B}_\alpha}{\bm{B}_\bgb}\Big)~, 
    \qquad 
    r_\bga = \prod_{\gb} \frac 12 \Big( 1-\CC{\bSv}{\bm{B}_\gb}\CC{\bm{B}_\bga}{\bm{B}_\gb}\Big)~. 
}
Their values $1$ or $0$ indicate whether the sectors $\ga$ or $\bga$ produce Rarita--Schwinger fields or not, respectively. (Note that these become identical to the parameters introduced in \eqref{eq:RSmultiplet} and \eqref{eq:RSmultiplets} provided that the twists preserve supersymmetry.) The parameters \cref{eq:N_RS_parameters} show that additional Rarita--Schwinger fields may arise even when some of the twist basis vectors break supersymmetries.

The expression \eqref{eq:N_RS_comp} can be understood by working out the consequences of \eqref{eq:GGSOprojections} for a potential spin--$3/2$ state $\bgps^\gm | \bm{B}_\ga + \Sv\rangle$. The basis vectors $\bSv$, $\bm{B}_{\bgb}$ lead to projections that keep this state or project it out. The prefactor ensures that there will be 2 spin--$3/2$ states per sector if  $p_L+p=1$, while a single one for $p_L+p=2$.
Performing the same analysis for potential Rarita--Schwinger fields $\gps^\gm | \bm{B}_\bga + \bSv\rangle$ and adding the contributions gives rise to the expression above. This expression counts the number of surviving Rarita--Schwinger states irrespective of whether they are massless or not. 
In all these cases the spectra organise themselves in terms of multiplets of extended $\cN=\cN_L+\cN_R+\cN_\text{RS}$ generated by the original untwisted supersymmetry generators $\Sv$ and $\bSv$ (as long as  they are not projected out) and the  twisted supersymmetry generators \eqref{eq:TwistedSUSY}. Hence, even if all original supersymmetries are broken, supersymmetry may reintroduce itself via this mechanism. 

If there is a $\Intr_{2L}^2$ factor, the following relation holds 
\equ{
    \CC{\bSv}{\bm{B}_1} \CC{\bSv}{\bm{B}_2} \CC{\bSv}{\bm{B}_3} = -1~, 
}
as a consequence of \cref{eq:ExpansionGGSO}. This implies that either all these three phases are $-1$ (the supersymmetry preserving situation) or two of these phases are $+1$ and the remaining one $-1$ (the supersymmetry breaking cases). In terms of the parameters $\cS_\ga$, this can be stated as either they are all $1$ or two equal zero and the remaining one $1$. 

By varying these GGSO phases, a range of possibilities are obtained for the amount of left-- and right--moving spacetime supersymmetries $(\cN_L,\cN_R)$ and the possible number of Rarita-Schwinger states. \Cref{tab:RSMultipletsLowerSUSY} summarises these possibilities for all the asymmetric pure twist models with point groups $\Intr_{2L}^{p_L}\times \Intr_{2R}^{p_R}$ with $p_L, p_R \leq 2$. Since by equivalence relation E1 the labels of the twist basis vectors may be permuted, only two cases need to be considered for an $\Intr_{2L}^2$ factor: both $\bm{B}_1$ and $\bm{B}_2$ preserve or both break $\bSv$. Then in either case $\bm{B}_3$ preserves $\bSv$, so that in the case where $\bSv$ is broken, the parameter $r_3$ indicates whether spin--$3/2$ states are produced.

\Cref{tab:RSMultipletsLowerSUSY} exhibits several interesting features. The cases with maximal preserved supersymmetry with additional Rarita--Schwinger fields correspond to the cases with enhanced supergravity displayed already in \cref{tab:EnhancedSUGRAs}. Furthermore, note that the asymmetric orbifolds with point groups $\Intr_{2L}$, $\Intr_{2L}^2$ and $\Intr_{2L}\times \Intr_2$ remain supersymmetric independently of how the GGSO phases with $\Sv$ or $\bSv$ are chosen, though the amount of supersymmetry may be reduced. For the other asymmetric orbifolds with point groups $\Intr_{2L}\times\Intr_{2R}$, $\Intr_{2L}^2\times\Intr_{2R}$, $\Intr_{2L}\times\Intr_{2R}\times\Intr_2$ and $\Intr_{2L}^2\times\Intr_{2R}^2$ all spacetime supersymmetries are broken by the twist vectors. The models with the point groups $\Intr_{2L}^2\times \Intr_{2R}$ and $\Intr_{2L}^2\times \Intr_{2R}^2$ with all original supersymmetries broken, can still have $\cN=1$ or even $\cN=2$ supergravity spectra in four dimensions. Below, two examples of this are given:

\subsubsection*{A \boldsymbol{$\Intr_{2L}^2\times\Intr_{2R}$} T--fold example \boldsymbol{$\cN=0\ra 1$} supersymmetry enhancement }

The pure twist $\Intr_{2L}^2\times\Intr_{2R}$ T--fold with GGSO phases 
\equ{ \label{eq:Z2L2Z2R_SUSYbreakRS}
    \CC{\bSv}{\bm{b}_1} = \CC{\bSv}{\bm{b}_2} = \CC{\Sv}{\bm{b}_{\bar{1}}} =  +1~, 
    \qquad 
    \CC{\Sv}{\bm{b}_{1}} = \CC{\Sv}{\bm{b}_{2}} = \CC{\bSv}{\bm{b}_{\bar{1}}} = -1~, 
    \qquad 
    \CC{\bm{b}_{1}}{\bm{b}_{\bar{1}}} = -\CC{\bm{b}_{2}}{\bm{b}_{\bar{1}}}~, 
}
breaks all supersymmetries since all parameters \eqref{eq:TestSUSYTwists} are zero except $\cS_3=+1$, hence by \eqref{eq:N_RS_comp}: 
\equ{
    \cN_\text{RS}=r_3 =\frac12 \Big( 1 - \CC{\bm{b}_{1}}{\bm{b}_{\bar{1}}} \CC{\bm{b}_{2}}{\bm{b}_{\bar{1}}}\Big)~. 
}
Thus only if the two final GGSO phases in \eqref{eq:Z2L2Z2R_SUSYbreakRS} are opposite one has a single spin--$3/2$ field. The full massless four dimensional spectrum of this model consists of: 1 graviton, 1 Rarita--Schwinger field, 18 vector fields, 59 spin--$1/2$ fermions and 82 real scalars. This spectrum can be grouped into $\cN=1$ multiplets: the graviton and the Rarita--Schwinger field form the supergravity multiplet. In addition, there are 18 vector and 41 scalar multiplets. 
The fact that the spectrum reorganises itself in this case in $\cN=1$ multiplets can be understood by realising that 
\equ{
    \Sv' = \Sv+\bm{b}_1+\bm{b}_2~, 
}
acts as a twisted supersymmetry generator as it is exactly the vector that leads to the Rarita--Schwinger field in the massless twisted spectrum. So supersymmetry enhancement is even possible if all original supersymmetries were broken.

\subsubsection*{A \boldsymbol{$\Intr_{2L}^2\times\Intr_{2R}^2$} T--fold example \boldsymbol{$\cN=0\ra 2$} supersymmetry enhancement}

The pure twist $\Intr_{2L}^2\times\Intr_{2R}^2$ T--fold with GGSO phases 
\equ{
    \CC{\bSv}{\bm{b}_\ga}  = \CC{\Sv}{\bm{b}_{\bgb}}  = +1~, 
    \qquad 
    \CC{\Sv}{\bm{b}_{\ga}} = \CC{\bSv}{\bm{b}_{\bgb}}  =-1~, 
    \qquad
    \CC{\bm{b}_{\ga}}{\bm{b}_{\bgb}}  = (-1)^{\ga+\bgb}\, \CC{\bm{b}_{1}}{\bm{b}_{\bar{1}}} ~, 
}
for $\ga, \bgb=1,2$ corresponds to the last $\Intr_{2L}^2\times\Intr_{2R}^2$ configuration in \cref{tab:RSMultipletsLowerSUSY} with $r_3=r_{\bar{3}}=1$ and hence has two spin--$3/2$ fields. The full massless four dimensional spectrum of this model consists of: 1 graviton, 2 Rarita--Schwinger fields, 16 vector fields, 58 spin--$1/2$ fermions and 86 real scalars. This spectrum can be group into $\cN=2$ multiplets: the graviton and the Rarita--Schwinger fields form the supergravity multiplet. In addition, there are 15 vector and 14 hyper multiplets. 
Again, the two vectors that define the Rarita--Schwinger fields in the massless twisted string spectrum, leading to two twisted supersymmetry generators 
\equ{
    \Sv' = \Sv+\bm{b}_1+\bm{b}_2~,
    \qquad 
    \bSv' = \bSv+\bm{b}_{\bar{1}}+\bm{b}_{\bar{2}}~, 
}
producing the target space $\cN=2$ structure.

\begin{table}
\small
\centering
\begin{tabular}{|c|cccc|| c|c|}
\hline
\textbf{Point group}  & \multicolumn{4}{c||}{\textbf{SUSY GGSO phases}} & \textbf{SUSY} & \textbf{\#(Twisted spin--\boldsymbol{$3/2$})}
\\ 
& $\CC{\bSv}{\bv_1}$ & $\CC{\bSv}{\bv_2}$ & $\CC{\Sv}{\bv_{\bar{1}}}$ & $\CC{\Sv}{\bv_{\bar{2}}}$ & $(\cN_L,\cN_R)$& $\cN_\text{RS}$
\\  \hline\hline 
$\Intr_{2L}$ & $-$ & \multicolumn{3}{c||}{}& $(2,4)$ & $2$ \\ 
& $+$ &  \multicolumn{3}{c||}{} & $(2,0)$ & $0$
\\ \hline   
$\Intr_{2L}\times \Intr_{2R}$ & $-$ & & $-$ & & $(2,2)$ & $4,0$ \\ 
& $+$ & & $-$ & & $(2,0)$ & $2,0$ \\ 
& $-$ & & $+$ & & $(0,2)$ & $2,0$ \\ 
& $+$ & & $+$ & & $(0,0)$ & $0$ \\ 
\hline   
$\Intr_{2L}^2$ & $-$ & $-$ & \multicolumn{2}{c||}{} &   $(1,4)$ & $3$ \\
& $+$ & $+$ & \multicolumn{2}{c||}{} &   $(1,0)$ & $1$
\\ \hline 
$\Intr_{2L}\times \Intr_{2}$ & $-$ & \multicolumn{3}{c||}{} & $(1,2)$ & $1$ \\
& $+$ & \multicolumn{3}{c||}{} & $(1,0)$ & $0$
\\ \hline
$\Intr_{2L}^2\times \Intr_{2R}$ & $-$ & $-$ & $-$ &  &   $(1,2)$ & $5,1$  \\
& $-$ & $-$ & $+$ & &   $(0,2)$ & $2,0$ \\
& $+$ & $+$ & $-$ &  &   $(1,0)$ & $3,1,0$\\
& $+$ & $+$ & $+$ & &   $(0,0)$ & $1,0$
\\ \hline 
\hline

$\Intr_{2L}\times \Intr_{2R}\times \Intr_{2}$ & $-$ & & $-$ & & $(1,1)$ & $2,0$ \\
& $+$ & & $-$ & & $(1,0)$ & $1,0$ \\
& $-$ & & $+$ & & $(0,1)$ & $1,0$ \\
& $+$ & & $+$ & & $(0,0)$ & $0$
\\ \hline  
$\Intr_{2L}^2\times \Intr_{2R}^2$ & $-$ & $-$ & $-$ & $-$ &   $(1,1)$ & $6,2,0$ \\
& $+$ & $+$ & $-$ & $-$ &   $(1,0)$ & $3,1,0$ \\
& $-$ & $-$ & $+$ & $+$ &   $(0,1)$ & $3,1,0$ \\
& $+$ & $+$ & $+$ & $+$ &   $(0,0)$ & $2,1,0$ \\ \hline 
\end{tabular}
\caption{\label{tab:RSMultipletsLowerSUSY}
Summary of the possible left-- and right--moving spacetime supersymmetries and their corresponding possibilities for spin--$3/2$ producing sectors from 
the order--two point groups of the type II superstring. The number of Rarita--Schwinger fields in the final column is determined by the choices of GGSO phases through the dependence in \eqref{eq:N_RS_parameters}. In this table it is used that the supersymmetry breaking twists can always be chosen to be the first two twist vectors when there is a $\Intr_{2L}^2$ or $\Intr_{2R}^2$ factor.
}
\end{table}

\section{Conclusions}
\label{sc:Conclusions}

\subsection{Main results}

The main purpose of this paper was to give a classification of order--two asymmetric toroidal orbifolds on the $SO(12)$ lattice at the free fermionic point, providing a more extensive list of T--folds that have an exact string theory description existing in the literature thus far. To this end, we considered the free fermionic formulation of the type II theory associated with non--geometric backgrounds and determined the massless spectra.

A complication with any classification is that identical configurations may be represented by different input data, in the case of asymmetric orbifolds in the free fermionic formulation, the set of basis vectors that define the models. Three equivalence relations are proposed to define when two configurations should be considered to be identical.

Using these equivalence relations a full classification of all possible order--two point groups in six internal dimensions preserving some amount of supersymmetry on both sides is obtained, which is collected in \cref{tab:NarainPointGroupClass}. Interestingly, the possible point groups lead to supergravities ranging over all extended supersymmetries $\cN=2,\ldots,6,8$ in four dimensions. It should be emphasised that this classification of possible order--two point groups is complete. It does not depend on any specific choice of generalised GSO phases, nor on any specific choice of the underlying torus lattice. In particular, not on the choice of the $SO(12)$ lattice or any specific point in the (discrete) moduli space used in this paper primarily. 

Next, the equivalence relations were used to define a standard form of the twist basis vectors to obtain a classification of all order--two T--folds on the $SO(12)$ lattice up to point group mirrors. \Cref{tab:NarainRotoTransClass} gives the full list of such T--fold configurations at the free fermionic point. In addition, this table gives the resulting type IIA/B spectra in terms of the various multiplets of these extended supersymmetric theories, such as the supergravity, Rarita--Schwinger, vector and hyper multiplets. The spectra of the configurations are presented in parametric form depending on certain generalised GSO phases. 

These classification results with parametric spectra can be used for various purposes. One application is to search for configurations with small effective Hodge numbers. The minimal effective Hodge numbers were found to be $(1,1)$ for six T--fold configurations of which one was previously known in the literature. 

Another application is to investigate mirror symmetry on these order--two T--folds. Mirror symmetry of symmetric $T^6/\Intr_2^2$ orbifolds can be identified through studying how flipping the GSO phase that distinguishes the type IIA and type IIB theories can be compensated by a flip of a certain generalised GSO phase. This phase is conventionally referred as the discrete torsion phase between the two twists. Using the parametric form of the spectra of the T--fold configurations, this procedure leads to the conclusion that independent left-- and right--mirror symmetry maps can be identified. 

Not all asymmetric orbifolds admit an orientifold projection. In particular, none of the left/right--asymmetric point groups admit an orientifold projection. All orientifoldable configurations using the basic worldsheet parity were identified within our classification as left/right-symmetric T--fold constructions for which the set of twist basis vectors are left/right--symmetric. 

A particularly striking feature of these T--folds is that the Rarita--Schwinger multiplets (multiplets with spin--3/2 fermions as their highest component) may appear in twisted sectors. There has been discussion in the supersymmetric literature as to whether such multiplets can be consistently coupled to supergravity and other multiplets. The so--called ``twisting mechanism''~\cite{Dolivet:2007sz} enhances the supergravity multiplet to a higher extended supersymmetry absorbing the Rarita--Schwinger multiplets. We note that this is reminiscent of gauge symmetry enhancement in which spacetime vector boson states arising from sectors with periodic fermions always enhance the original gauge symmetry that arises from the Neveu--Schwarz sector.  The additional supersymmetry generators were identified as the vectors that define these twisted Rarita--Schwinger fields. \Cref{tab:EnhancedSUGRAs} showed that in all such cases this mechanism is at work as higher extended supergravities could always be formed absorbing all Rarita--Schwinger multiplets.

This twisting mechanism does not only act in supersymmetry preserving constructions. Relaxing the condition that the generalised GSO projections are chosen to preserve the maximal amount of supersymmetry possible (as was otherwise assumed through this work), non--supersymmetric T--folds were found that enhance to $\cN=1$ or even $2$ supergravities by twisted spin--$3/2$ states.

\subsection{Outlook}

There are various directions in which the results of this work may be extended, for example: 

The classification of the T--fold configurations assumed that the underlying six--torus lattice is $SO(12)$ at the free fermionic point. Other torus lattices can be obtained by performing additional shift orbifolds. Depending on the choices of generalised GSO phases of these shift vectors with the supersymmetry basis vectors, this may lead to models with reduced supersymmetry. 

This work was in the type II context and only briefly the topic of orientifolding was touched. Of course, that is only the starting point to investigate brane configurations on these T--folds. Only a few examples of such D--brane models exist in the literature, hence a more systematic investigation of such models is overdue and our classification provides a larger pool of left/right--symmetric T--fold constructions to start from.

Another extension is to investigate these T--folds in the heterotic string theories (possibly along the lines of~\cite{Bianchi:2012xz}). In the heterotic case, the left-- and right--moving worldsheet theories are no longer identical, hence the distinction between the properties of asymmetric twist becomes more significant. For example, an asymmetric twist may break a certain amount of supersymmetry on one side but leave the gauge group from the other side untouched, or the other way around. Moreover, on the non--supersymmetric side of the heterotic string, there are many more choices (than only the standard embedding) for how the asymmetric twist acts.

The models on the non-supersymmetric T--folds that enhanced to $\cN=1$ or $2$ supergravity spectra seem  peculiar and lead to many interesting effective field theory questions. In particular, it would be interesting to check whether interactions respect relations expected from conventional supergravity theories. To perform such a study presumably would involve computing relevant correlators in these string theory backgrounds. 
Obvious such models have vanishing cosmological constant. 
Hence, it would be interesting to revisit non-supersymmetric string vacua with vanising one--loop partition functions and examine whether twisted spin-$3/2$ exist in their spectra and therefore do in fact possess spacetime supersymmetry, similar to what we have found here.

In this paper we discussed the notion of mirror symmetry on asymmetric Type II string compactification. Another interesting direction to explore is the analogue of Spinor--Vector Duality (SVD) in the type II string \cite{fkr1,fkr2}. SVD was observed in heterotic string compactifications under the exchange of the total number of spinorial and anti--spinorial representations of $SO(10)$ with the total number of vectorial representations. It is seen to arise due to the breaking of the $(2,2)$ worldheet supersymmetry to $(2,0)$ and is induced by the spectral flow operator of the broken $N=2$ worldsheet supersymmetry \cite{fkr3, svd3}. Mirror symmetry operates in type II string as well as in the heterotic--string and arises due to a map of the internal moduli, {\it i.e.} an exchange between complex structure and K\"ahler structure moduli of the internal manifold. In the type II string, mirror symmetry arises from a map between type IIA and type IIB string. SVD on the other hand arises in the heterotic--string due to exchange of Wilson line moduli that do not exist in the type II perturbative string. It will be interesting to explore the dualities/symmetries that arise in the type II string by the breaking of the worldsheet supersymmetry from $(2,2)$ to $(2,0)$ and using the spectral flow operator to induce the transformations between dual vacua. The twisted generators that we defined herein are precisely suited to explore this operation in the case of the $SO(12)$ lattice. It is noted that mirror symmetry and SVD are the tip of the iceberg of the space of symmetries underlying the string vacua that are particularly extended in compactifications to two dimensions \cite{Dixon:1988qd,msds1, msds2}. 

\subsection{Acknowledgements}

We would like to thank Massimo Bianchi, Jose Morales and Gianfranco Pradisi for detailed communications on their pioneering works on this subject. In addition, we would like to thank Ralph Blumenhagen for an extended email correspondence. The work of AEF is supported by the STFC Consolidated Grant ST/X000699/1.

\appendix
\crefalias{section}{appendix}
\section{Extended supersymmetric multiplets in various dimensions}
\label{app:ExtendedSUSY} 

This appendix gives a short overview of the basic supersymmetric multiplets in the dimensions relevant in this work. For a more extensive review see the vast literature on supersymmetry, {\em e.g.}\ \cite{West:1990tg,deWit:1997sz}. Methods for analyzing multiplet structures for extended supersymmetry in six and four dimensions can be found in \cite{Strathdee:1986jr}. In that reference many multiplets are identified; while some others relevant for this work were omitted. Several of those can be found in \cite{Blumenhagen:2016rof}. 

The possible four dimensional multiplets of extended supersymmetry are listed in \cref{tab:4DMultiplets}. The massless multiplets are uniquely identified by their highest spin component: the SuperGravity (SG), Rarita--Schwinger (RS), Vector (V) and Hyper (H) and Scalar (S) multiplets characterised by their highest spin components of spin--2, 3/2, 1 and 1/2, respectively. In particular, the Rarita--Schwinger multiplets contain fermions of spin--3/2 but no massless spin--2 gravitons. Except for spin--0, all spin components constitute two degrees of freedom on--shell. Taking this into account shows that all multiplet have an equal number of bosonic and fermionic degrees of freedom. \Cref{tab:4DMultiplets} shows the well--known result that the supergravity multiplets of $\cN=7$ and $8$ supersymmetry are identical. Consequently, we will not distinguish these supergravities and only refer to them as $\cN=8$ supergravity. 

Contrary to four dimensions, in six dimensions the chirality of fermions is not reversed by Majorana conjugation. Consequently, the chirality of the spin--1/2 and 3/2 fermions is relevant. Moreover, this implies that six dimensional extended supersymmetric theories are characterised by the number of positive and negative chirality supercharges, denoted by $\cN_+$ and $\cN_-$, respectively. The possible extended supersymmetric theories are the non--chiral for $(\cN_+,\cN_-)=(2,2)$, $(1,1)$ theories and the chiral $(2,1)$, $(2,0)$ and $(1,0)$. The massless degrees of freedom on the light--cone  can be distinguished by their $(r_+,r_-)$ representations of the little group $Spin(4)=SU(2)_+\times SU(2)_-$ in six dimensions. Moreover, in this dimension rank two tensors may be selfdual or anti--selfdual. \Cref{tab:6DMultiplets} gives an overview of possible massless multiplets of extended supersymmetry in six dimensions.

\begin{table}[t]
\centering
\renewcommand{\arraystretch}{1.2}
\setlength{\tabcolsep}{4.75pt}
\begin{tabular}{c}
\begin{tabular}{| c c| c c c c c|}
\hline 
$\boldsymbol{\mathcal{N}}$& \textbf{Multiplet} & \multicolumn{5}{c|}{\textbf{Spin}} \\ 
& & $\boldsymbol{[0]}$ & $\boldsymbol{[\frac{1}{2}]}$ & $\boldsymbol{[1]}$ & $\boldsymbol{[\frac{3}{2}]}$ & $\boldsymbol{[2]}$    
\\ \hline\hline 
8 & SuperGravity & 70 & 56 & 28 & 8 & 1 
\\ \hline 
7 & SuperGravity & 70 & 56 & 28 & 8 & 1  
\\ \hline 
6 & SuperGravity & 30 & 26 & 16 & 6 & 1 \\  \cdashline{2-7}
  & Rarita--Schwinger & 20 & 15 & 6 & 1 &  
\\ \hline 
5 & SuperGravity & 10 & 11 & 10 & 5 & 1   \\  \cdashline{2-7}
  & Rarita--Schwinger & 20 & 15 & 6 & 1 &
\\ \hline 
4 & SuperGravity & 2 & 4 & 6 & 4 & 1 \\  \cdashline{2-7}
  & Rarita--Schwinger & 8 & 7 & 4 & 1 & \\ \cdashline{2-7}
  & Vector & 6 & 4 & 1 & &
\\ \hline 
3 & SuperGravity & & 1 & 3 & 3 & 1 \\ \cdashline{2-7}
  & Rarita--Schwinger & 2 & 3 & 3 & 1 & \\  \cdashline{2-7}
  & Vector & 6 & 4 & 1 & & 
\\ \hline
\end{tabular}
\begin{tabular}{| c c| c c c c c|}
\hline 
$\boldsymbol{\mathcal{N}}$ & \textbf{Multiplet} & \multicolumn{5}{c|}{\textbf{Spin}} \\ 
& & $\boldsymbol{[0]}$ & $\boldsymbol{[\frac{1}{2}]}$ & $\boldsymbol{[1]}$ & $\boldsymbol{[\frac{3}{2}]}$ & $\boldsymbol{[2]}$   
\\ \hline\hline 
2 & SuperGravity & & & 1 & 2 & 1 \\ \cdashline{2-7}
  & Rarita--Schwinger & & 1 & 2 & 1 & \\  \cdashline{2-7}
  & Vector & 2 & 2 & 1 & & \\  \cdashline{2-7}
  & Hyper & 4 & 2 & & & \\  \cdashline{2-7}
  & Half--Hyper & 2 & 1 & & & 
\\ \hline
1 & SuperGravity & & & & 1 & 1 \\ \cdashline{2-7}
  & Rarita--Schwinger & & & 1 &1 & \\ \cdashline{2-7}
  & Vector & & 1 & 1 & & \\ \cdashline{2-7}
  & Scalar & 2 & 1 & & & 
\\ \hline
\multicolumn{7}{c}{}
\\
\multicolumn{7}{c}{}
\\
\multicolumn{7}{c}{}
\\
\end{tabular}
\end{tabular}
\caption{\label{tab:4DMultiplets} 
Supersymmetry multiplets in four dimensions of $\mathcal{N}=1,2,\ldots,8$ extended supersymmetry~\cite{West:1990tg}.}
\end{table}

\begin{table}[t]
\centering
\renewcommand{\arraystretch}{1.2}
\setlength{\tabcolsep}{5.25pt}
\begin{tabular}{|c c| c c c c c c c c c|}
\hline 
 \multicolumn{2}{|l|}{$\boldsymbol{(\mathcal{N}_+,\mathcal{N}_-)}$\textbf{--} \qquad   \textbf{State}} & $\boldsymbol{[S]}$ &  $~~\boldsymbol{[\frac{1}{2}]^+}$ & $~~\boldsymbol{[\frac{1}{2}]}^-$ & $\boldsymbol{[V]}$ & $~~\boldsymbol{[T]^+}$ & $~~\boldsymbol{[T]^-}$ & $~~\boldsymbol{[\frac{3}{2}]}^+$ & $~~\boldsymbol{[\frac{3}{2}]}^-$ & $\boldsymbol{[G]}$  \\ 
\multicolumn{2}{|c|}{\textbf{Multiplet} \qquad  $\boldsymbol{(r_+,r_-)}$} & $(1,1)$ & $(2,1)$ & $(1,2)$ & $(2,2)$ & $(3,1)$ & $(1,3)$ & $(3,2)$ & $(2,3)$ & $(3,3)$
\\ \hline\hline
(2,2) & SuperGravity & 25 & 20 & 20 & 16 & 5 & 5 & 4 & 4 & 1
\\ \hline 
(2,1) & SuperGravity & 5 & 4 & 10 & 8 & 1 & 5 & 2 & 4 & 1 \\ \cdashline{2-11} 
      & Rarita--Schwinger & 10 & 8 & 5 & 4 & 2 & & 1 &  &
\\ \hline 
(2,0) & SuperGravity &  &  &  &  &  & 5 &  & 4 & 1 \\ \cdashline{2-11}
      & Rarita--Schwinger &  &  & 5 & 4 &  & & 1 &  & \\ \cdashline{2-11}
      & Self--dual Tensor & 5 & 4 &  &  & 1 &  &  &  &
\\ \hline
(1,1) & SuperGravity & 1 & 2 & 2 & 4 & 1 & 1 & 2 & 2 & 1 \\ \cdashline{2-11}
      & Rarita--Schwinger$^-$ & 2 & 1 & 4 & 2 &  & 2 &  & 1 & \\ \cdashline{2-11}
      & Rarita--Schwinger$^+$ & 2 & 4 & 1 & 2 & 2 &  & 1 &  & \\ \cdashline{2-11}
      & Vector & 4 & 2 & 2 & 1 & & & & &
\\ \hline
(1,0) & SuperGravity & & & & & & 1 & & 2 & 1 \\ \cdashline{2-11}
      & RS--Vector & & & 1 & 2 & & & 1 & & \\  \cdashline{2-11}
      & RS--Tensor & & & & & & 2 & & 1 & \\ \cdashline{2-11}
      & Self--Dual Tensor & 1 & 2 & & & 1 & & & & \\ \cdashline{2-11}
      & Vector &  &  & 2 & 1 & & & & & \\ \cdashline{2-11}
      & Hyper & 4 & 2 & & & & & & & \\ \cdashline{2-11}
      & Half--Hyper & 2 & 1 & & & & & & & 
\\ \hline
\end{tabular}
\caption{\label{tab:6DMultiplets} 
Supersymmetry multiplets in six dimensions of $\mathcal{N}=(1,0); (1,1); (2,0); (2,1); (2,2)$ extended supersymmetry. 
The states, scalars $[0]$, chiral fermions $[\sfrac12]^\pm$, vector $[1]$, (anti--)selfdual tensors $[T]^\pm$, chiral Rarita-Schwinger fermions $[\sfrac 32]^\pm$ and gravitons $[G]$, are characterised by $Spin(4)=SU(2)_+\times SU(2)_-$ little group representations $(r_+,r_-)$. 
Notation and many multiplets taken in part from \cite{Strathdee:1986jr}. }
\end{table}

\subsection[Type II realisation of six dimensional $(2,1)$ Rarita--Schwinger multiplets]{Type II realisation of six dimensional $\boldsymbol{(2,1)}$ Rarita--Schwinger multiplets}
\label{app:21RSstates}

The string states \eqref{eq:RS} correspond to $(2,1)$ Rarita--Schwinger multiplet in six dimensions given in \cref{tab:6DMultiplets}: 
 
First of all, observe that the GGSO associated with the basis vector $\bm{E}$ only acts on the two complexified $y$'s or $w$'s in $\bm{B}_\ga$, leading to an overall multiplicity of 2. Ignoring this factor in the rest of the arguments (but of course keeping it in mind for the final result), note that the ten dimensional vector in light--cone gauge reduces to $8_v = 4\cdot (1,1)+(2,2)$. 

Now, consider the target space fermions first. By the GGSO projections, $\big|\bm{B}_\ga + \bm{S}\big\rangle$ defines a positive chiral spinor in six dimensions, {\em i.e.}\ lives in the $(2,1)$ representation, since $(2,2) \times (2,1) = (3,2) + (1,2)$. Hence $\bgps^M \big|\bm{B}_\ga + \bm{S}\big\rangle$ gives rise to the representation $4\cdot (2,1) + (3,2) + (1,2)$. Note that this includes a spin--3/2 Rarita--Schwinger fermion. Next, note both chiralities survive from the states $\big|\bm{B}_\ga + \bar{\bm{S}}\big\rangle$, because $\bm{B}_\ga$ does not overlap with $\bar{\bm{S}}$, the ten dimensional spinor $8_s=2\cdot (2,1) + 2\cdot (1,2)$. The GGSO associated with $\bm{S}$ reduces the internal spinor components in $\bm{b}_\ga$ by two, hence $\big|\bm{B}_\ga + \bar{\bm{S}}\big\rangle$ gives rise to the representation $4\cdot (2,1) + 4\cdot (1,2)$. Collecting things together, this analysis gives rise to the states: $(3,2) + 8\cdot (2,1) + 5\cdot (1,2)$, precisely the fermionic states of the $(2,1)$ Rarita--Schwinger multiplet in \cref{tab:6DMultiplets}. 

Next, consider the target space bosons. $\bgps^M \big|\bm{B}_\ga \big\rangle$ gives rise to $2\cdot 8_v = 8\cdot (1,1) + 2\cdot (2,2)$, {\em i.e.}\ eight scalar and two vectors. On the other hand, $\big|\bm{B}_\ga + \bm{S}+\bar{\bm{S}}\big\rangle$ generates $2\cdot (2,1)\times [ (2,1)+(1,2)]$ since $\bm{S}$ gives a positive chiral fermion by the GGSO projections, while $\bar{\bm{S}}$ is non--chiral. Working out the tensor products one finds $2\cdot[(3,1) + (1,1) + (2,2)]$, {\em i.e.}\ two self--dual tensors, two scalars and two additional vectors. Combined, this results in $4\cdot (2,2) + 2\cdot (3,1) + 10\cdot (1,1)$, which matches precisely the bosonic states of the $(2,1)$ Rarita--Schwinger multiplet in \cref{tab:6DMultiplets}.

Taking everything together, this shows that the supersector \eqref{eq:RS} gives rise to two Rarita--Schwinger multiplets of $(2,1)$ supersymmetry.

\subsection[Type II realisation of six dimensional $(1,0)$ hyper multiplets]{Type II realisation of six dimensional $\boldsymbol{(1,0)}$ hyper multiplets}
\label{app:10Hstates}

The string states \eqref{eq:Hyper} correspond to sixteen $(1,0)$ hyper  multiplets in six dimensions given in \cref{tab:6DMultiplets}:
$|\bm{B}_{\ga}'\rangle$ contains 16 real (thus eight complex) worldsheet fermions, hence represents $2^{8}=256$ states. 
Because $\bm{S}$ and $\bm{E}$ have overlaps with $\bm{B}_{\ga}'$, they both halve the number of degrees, $64$ of them are independent. They constitute 16 sets of two complex scalars. 
$|\bm{B}_{\ga}'+\bm{S}\rangle$ represents target space fermions. They contain $2^2/2=2$ helicities. Each of these two helicities come with $2^6=64$ copies. The projection defined by $\bm{E}$ reduces this number to $32$. These thus give 16 fermion doublets.

\subsection[Type II realisation of six dimensional $(2,0)$ tensor and $(1,1)$ vector multiplets]{Type II realisation of six dimensional $\boldsymbol{(2,0)}$ tensor and $\boldsymbol{(1,1)}$ vector multiplets}
\label{app:TVstates}

The string states \eqref{eq:TensorVector} in the type IIB theory correspond to eight $(2,0)$ tensor multiplets in six dimensions given in \cref{tab:6DMultiplets}. First of all, because $|\bm{B}_{\ga\bgb} \rangle$ contains four $y$'s or $w$'s and four $\byy$'s and $\bw$'s, the GGSO projection associated to $\bm{E}$ leads to a multiplicity $2^4/2=8$. Now, because of the $\bm{S}$ and $\bar{\bm{S}}$ projection the first string states in \eqref{eq:TensorVector} correspond to $2\cdot 2 =4$ scalars. The fermionic states both live in the $(2,1)$ representation in terms of $Spin(4)=SU(2)_+\times SU(2)_-$ light--cone representations in six dimensions for the type IIB theory and come with a multiplicity of 2. Consequently, the final states form $(2,1)\times (2,1) = (3,1) + (1,1)$, {\em i.e.} one self--dual tensor and one additional scalar. These are the states of a $(2,0)$ self--dual tensor multiplet given in \cref{tab:6DMultiplets}.

In the type IIA theory the chiralities of $\bm{S}$ and $\bar{\bm{S}}$ are opposite. Consequently, the final states in \cref{eq:TensorVector} form a vector: $(2,1)\times (1,2) = (2,2)$. Hence, these states result in eight $(1,1)$ vector multiplets given in \cref{tab:6DMultiplets}.

\section{One loop partition functions and generalised GSO phases}
\label{app:PartitionFunctions}

The general form of the partition function for a free fermionic model in $d$ non--compact lightcone dimensions is given by 
\equ{ \label{eq:FullPartitionFun}
Z(\gt,\bgt) = Z_d(\gt,\bgt) \sum_{\Bga,\Bga'\in \gX} \frac{1}{2^{|\mathcal{B}|}}\, \CC{\mathbf{\Bga}}{\Bga'} 
\, 
\frac{ \gTh\brkt{\Bga_L}{\Bga_L'}(\gt)}{\get^{D_L/2}(\gt)} \frac{\bgTh\brkt{\Bga_R}{\Bga_R'}(\bgt)}{\bget^{D_R/2}(\bgt)}~,
}
where the sums are over all vectors in the additive set $\gX$. 
Here $D_L$ and $D_R$ count the number of real holomorphic and anti--holomorphic fermions, respectively. Here the partition function of $d$ non--compact real bosons is given by 
\equ{
Z_d(\gt,\bgt) = 
\frac{1}{\gt_2^{d/2} \get^{d}(\gt)\bget^{d}(\bgt)}~. 
}

Here the Mumford form of the genus $g$ theta functions with vector valued characteristics $\Bga$ and $\Bga'$ is given by
\equ{
\gTh\brkt{\Bga}{\Bga'}(\gt) = 
\sum_{\boldsymbol{n}\in\Intr^g} q^{\sfrac12 \big(\boldsymbol{n}+\sfrac12\Bga\big)^2}\,
e^{\pi i\, \boldsymbol{n}^T\Bga'}~, 
\qquad 
q = e^{2\pi i\, \gt}~. 
}
This form of the theta functions is convenient since the characteristic $\Bga$ only affects the masses of the states whereas the other characteristic $\Bga'$ only affects the projection conditions on the integral vectors $\boldsymbol{n}$. 
The characteristics are defined modulo even integral vectors $\Bgd$ and $\Bgd'$ 
\equ{
\gTh\brkt{\Bga+\Bgd}{\Bga'+\Bgd'}(\gt) =
e^{-\sfrac 12 \gp i\, \Bgd^T\Bga'}
\gTh\brkt{\Bga}{\Bga'}(\gt)~. 
}
The Dedekind function reads
\equ{
\get(\gt) = q^{\sfrac1{24}} \prod_{n\geq 1} \big( 1 - q^n \big)~. 
}
Their modular transformation properties are 
\equ{
\gTh\brkt{\Bga}{\Bga'}(\gt+1) =
e^{\sfrac 14 \gp i\, \Bga^2}
\gTh\brkt{\Bga}{\Bga'+\Bga}(\gt)~, 
\qquad
\gTh\brkt{\Bga}{\Bga'}\big(\sfrac{-1}\gt\big) =
e^{-\sfrac 12 \gp i\, \Bga^T\Bga'}
(-i\gt)^{\sfrac d2}\, 
\gTh\brkt{\Bga'}{-\Bga}(\gt)~, 
}
and
\equ{
\get(\gt+1)  = e^{2\pi i\, \sfrac 1{24}}\, \get(\gt)~, 
\qquad
\get\big( \sfrac{-1}\gt\big) = (-i\gt)^{\sfrac 12}\, \get(\gt)~. 
}

\subsection{Some properties of generalised GGSO phases}

The generalised GSO phases are subject to a number of conditions, see~\cite{ABK1, ABK2} for details. The relations 
\equ{ \label{eq:CCinterchange}
\CC{\Bga}{\Bga} = e^{\sfrac 14\gp i\,\Bga^2} \CC{\Bga}{\bm{1}}~, 
\qquad 
\CC{\Bga}{\Bga'} = e^{\sfrac 12 \gp i\, \Bga\cdot \Bga'}\CC{\Bga'}{\Bga}~, 
}
shows that only the generalised GSO phase $\CC{\Bgb_a}{\Bgb_b}$ for $a>b$ and $\CC{\bm{1}}{\bm{1}}$ are independent. 
A useful splitting formula,
\equ{ \label{eq:CCsplit}
\CC{\Bga}{\Bgb+\Bgg} = \gd_\Bga\CC{\Bga}{\Bgb}\,\CC{\Bga}{\Bgg}~, 
}
follows from two--loop modular invariance~\cite{ABK1, ABK2}.
The general form of the generalised GSO phases may be expanded as~\cite{ABK1, ABK2}
\equ{ \label{eq:ExpansionGGSO}
\CC{\Bga}{\Bga'} = 
\gd_\Bga^{\sum_{a}n_a'-1} \gd_{\Bga'}^{\sum_{a}n_a-1}\, 
e^{-\gp i\, r(\Bga)\cdot \Bga'}\, 
\prod_{a,b} \CC{\Bgb_a}{\Bgb_b}^{n_an_b'}~, 
}
where $\Bga=\sum_a n_a \Bgb_a$, $\Bga'=\sum_a n_a' \Bgb_a$ and $r(\Bga) = \Bga - [\Bga]$. 

In this paper we used the basis $\bm{S}$, $\bar{\bm{S}}$, $\bm{E}$ and $n$ twist basis vectors $\Bgb_a$.  In particular, the diagonal GGSO phase of any basis vector $\Bga$ is given by
\equ{ \label{eq:MixedPhasesE}
 \CC{\Bga}{\Bga} = e^{\sfrac 14\gp i\, \Bga^2}\, \CC{\Bga}{\bm{S}} \CC{\Bga}{\bar{\bm{S}}}\, \CC{\Bga}{\bm{E}}~. 
}
If $\Bga = \sum_a n_a\, \Bgb_a$, the diagonal phase $\CC{\Bga}{\Bga}$ can be evaluated in two ways: by applying \eqref{eq:MixedPhasesE} to $\Bga$ and the individual $\Bgb_a$ one finds 
\begin{subequations}
\equ{
\CC{\Bga}{\Bga} =  e^{\sfrac 12 \gp i \sum\limits_{a>b}n_an_b\, \Bgb_a\cdot \Bgb_b}\, \prod_a \CC{\Bgb_a}{\Bgb_a}^{n_a}
~, 
}
but by expanding it using \eqref{eq:ExpansionGGSO} one concludes that 
\equ{
\CC{\Bga}{\Bga} = \prod_{a,b}  \CC{\Bgb_a}{\Bgb_b}^{n_an_b} = 
\prod_{a>b}  \Bigg( \CC{\Bgb_a}{\Bgb_b}\CC{\Bgb_b}{\Bgb_b} \Bigg)^{n_an_b} \prod_a \CC{\Bgb_a}{\Bgb_a}^{n_a}~. 
}    
\end{subequations}
These two results are identical by \eqref{eq:CCinterchange} which can be expressed as 
\equ{ \label{eq:CrossedPhases}
    \CC{\Bgb_a}{\Bgb_b}\CC{\Bgb_b}{\Bgb_a} = e^{\sfrac 12 \pi i\, \Bgb_a\cdot \Bgb_b}~.  
}

\subsection{Ten dimensional type II partition functions}

The one loop type II partition functions reads 
\equ{
Z(\gt,\bgt) = Z_8(\gt,\bgt) \sum_{s,s',\bs,\bs'} \frac{1}{4}\, (-)^{s+s'+\bs+\bs'} 
\, \CC{\Sv}{\Sv}^{s's}\CC{\bar{\Sv}}{\bar{\Sv}}^{\bs'\bs}\,
\frac{ \gTh\brkt{s\, 1^4}{s'1^4}(\gt)}{\get^{4}(\gt)} \frac{\bgTh\brkt{\bs\, 1^4}{\bs' 1^4}(\bgt)}{\bget^{4}(\bgt)}~.
}
This is compatible with the general expression \eqref{eq:ExpansionGGSO} setting the mixed phases to unity
\equ{ \label{eq:AbsenceMixedPhases}
\CC{\Sv}{\bar{\Sv}} = \CC{\bar{\Sv}}{\Sv} = 1~. 
}
For \eqref{eq:GSOtypeIIB} both the holomorphic and anti--holomorphic sides of the theory select target space spinors of the same (positive) chirality and hence defines the type IIB theory. (If the sign was opposite, negative chirality spinors are selected on both side, leading to the same theory in ten dimensions.) For the type IIA theory the selected chiralities in target space of the holomorphic and anti--holomorphic sides are opposite, see {\em e.g.}\ \eqref{eq:GSOtypeIIA}. The relation between the generalised GSO phases in the conventional basis $\{\mathbf{1},\Sv\}$ and the basis $\{\Sv,\bar{\Sv}\}$ is given by 
\equ{ \label{eq:Conversion}
\CC{\mathbf{1}}{\mathbf{1}} = \CC{\Sv}{\Sv}\CC{\bar{\Sv}}{\bar{\Sv}}\CC{\Sv}{\bar{\Sv}}\CC{\bar{\Sv}}{\Sv}~, 
\qquad 
\CC{\mathbf{1}}{\Sv} = - \CC{\Sv}{\Sv} \CC{\bar{\Sv}}{\Sv}~, 
}
in general ({\em i.e.}\ without the assumption that the model is supersymmetric in ten dimensions).

\clearpage
\printbibliography[heading=bibintoc]

\end{document}